\documentclass[a4paper,10pt, prd, twocolumn, superscriptaddress, amsmath,amssymb, aps, nofootinbib, floatfix]{revtex4-2}

\usepackage{graphicx}
\usepackage{amsfonts}
\usepackage{amsmath}
\usepackage{float}
\usepackage{amssymb,amsbsy}
\usepackage{epstopdf}
\usepackage{xcolor}
\usepackage[normalem]{ulem}
\usepackage{braket}
\usepackage{combelow}

\usepackage[colorlinks=True,linkcolor=blue,citecolor=blue,urlcolor=blue]{hyperref}
\usepackage{soul}

\def\bk{{\mathbf{k}}}

\def\fk{f_\bk}
\def\fS{f_{{\rm S}\bk}}

\renewcommand{\d}{\mathrm{d}}

\allowdisplaybreaks

\begin{document}

\title{High-order Shakhov-like extension of the relaxation time approximation\texorpdfstring{\\}{}
in relativistic kinetic theory}

\author{Victor E. Ambru\cb{s}}
\email{victor.ambrus@e-uvt.ro}
\affiliation{Department of Physics, West University of Timi\cb{s}oara, \\
Bd.~Vasile P\^arvan 4, Timi\cb{s}oara 300223, Romania}
\author{David Wagner}
    \affiliation{
Institut f\"ur Theoretische Physik, Johann Wolfgang Goethe-Universit\"at, Max-von-Laue-Straße 1, D-60438 Frankfurt am Main, Germany}%
\affiliation{Department of Physics, West University of Timi\cb{s}oara, \\
Bd.~Vasile P\^arvan 4, Timi\cb{s}oara 300223, Romania}


\begin{abstract}
In this paper we present a relativistic Shakhov-type generalization of the Anderson-Witting relaxation time model for the Boltzmann collision integral. The extension is performed by modifying the path on which the distribution function $f_{\bk}$ is taken towards local equilibrium $f_{0\bk}$, by replacing $f_{\bk} - f_{0\bk}$ via $f_{\bk} - f_{{\rm S}\bk}$. The Shakhov-like distribution $f_{{\rm S} \bk}$ is constructed using $f_{0\bk}$ and the irreducible moments $\rho_r^{\mu_1 \cdots \mu_\ell}$ of $f_\bk$ and reduces to $f_{0\bk}$ in local equilibrium. Employing the method of moments, we derive systematic high-order Shakhov extensions that allow both the first- and the second-order transport coefficients to be controlled independently of each other. We illustrate the capabilities of the formalism by tweaking the shear-bulk coupling coefficient $\lambda_{\Pi \pi}$ in the frame of the Bjorken flow of massive particles, as well as the diffusion-shear transport coefficients $\ell_{V\pi}$, $\ell_{\pi V}$ in the frame of sound wave propagation in an ultrarelativistic gas. Finally, we illustrate the importance of second-order transport coefficients by comparison with the results of the stochastic BAMPS method in the context of the one-dimensional Riemann problem.
\end{abstract}

\markboth{left head}{\today \; - \; 2021 v4.0}

\date{\today }
\maketitle
\section{Introduction}

In the frame of the relativistic Boltzmann equation,
the computation of the collision term $C[\fk]$ even for the simplest case 
of binary collisions remains the most expensive step. 
For this reason, models which approximate the main features of $C[\fk]$ 
are highly desirable. One such approximation is the relaxation 
time approximation (RTA) introduced by Anderson and Witting \cite{Anderson:1974a,Anderson:1974b}. 

Historically, the Anderson-Witting (AW) approximation came as an extension of the model proposed by Marle \cite{Marle.1969}, which focussed on the case of massive constituents. Both these models reduce in the non-relativistic limit to the  Bhatnagar-Gross-Krook (BGK) single-relaxation time approximation of the collision integral appearing in the nonrelativistic Boltzmann equation \cite{Bhatnagar.1954}. 

Much like its nonrelativistic cousin, the AW model has gained popularity due to its relatively simple structure which allows analytical calculations to be performed in simplified $0+1$-D setups, such as the Bjorken flow \cite{Florkowski:2013lya,Florkowski:2014sfa,Giacalone:2019ldn} or Gubser flow \cite{Denicol:2014xca,Denicol:2014tha,Behtash:2019qtk}, as well as numerical calculations in setups such as the Riemann problem \cite{Romatschke:2011hm,Ambrus:2016aub,Gabbana:2019ydb,Ambrus:2022adp} or the $2+1$-D Bjorken flow with transverse expansion \cite{Kurkela:2018qeb,Kurkela:2019kip,Ambrus:2022qya}. Especially for conformal (massless) uncharged (vanishing chemical potential) fluids, which are characterised only by tensor moments such as the stress-energy tensor $T^{\mu\nu}$ (i.e., no charge current), the single relaxation time $\tau_R$ of the Anderson-Witting model can be used to control the shear viscosity $\eta$, which is sufficient to achieve agreement with the solution of the full Boltzmann equation \cite{Ambrus:2021sjg}. 

In general, these single-relaxation-time models share the caveat that the transport coefficients governing dissipation
within the fluid are all derived from a single function -- the relaxation
time $\tau_R$. In the case of the BGK model, this had the unpleasant
consequence that the Prandtl number $\mathrm{Pr} = c_p \eta / \lambda$,
where $c_p$ is the heat capacity of the gas at constant pressure, $\eta$ is
the dynamical or shear viscosity and $\lambda$ is the heat conductivity, is
fixed at the value $1$. Most ideal gases are known to have $\mathrm{Pr} \simeq
2/3$ \cite{Sharipov:2016book}. This limitation was overcome in the extension proposed by Shakhov \cite{shakhov68a,shakhov68b}, which introduces a new parameter that allows $\mathrm{Pr}$ to be controlled independently. There is still some controversy regarding the well-posedness of the
Shakhov modification of the BGK collision model \cite{wang_wu_ho_li_li_zhang_2020}. For example, the second law of thermodynamics was proven only in the linear regime of small departures from equilibrium. Also, because the model relies on a polynomial extension of the equilibrium distribution function, it may lead to negative values of the distribution function in the case of flows with are sufficiently far from equilibrium. Finally, as is the case for the single-relaxation time models, the Shakhov model lacks a fundamental justification, being in essence an effective model. Despite these drawbacks, the Shakhov model has been highly successful at describing fluids far from equilibrium, i.e. deep into the transition regime \cite{sharipov1998data}, in the strongly non-linear regime \cite{graur2009comparison}, as well as in the case of non-ideal (dense) gases \cite{wang_wu_ho_li_li_zhang_2020,busuioc2023dense}.

Recently, a Shakhov-like extension of the AW model was proposed in the frame of relativistic kinetic theory \cite{Ambrus:2023ilm}. Much like its non-relativistic counterpart, this extension provides new free parameters that allow the first-order transport coefficients, $\zeta$, $\kappa$, and $\eta$, to be controlled separately. We shall refer to this model as {\it the first-order Shakhov model}. In this paper, we introduce a systematic procedure to extend the Shakhov model beyond first order, in a manner allowing a selection or all of the second-order transport coefficients to be separately controlled. 

As in the case of the first-order Shakhov model, the idea is to replace the relaxation term $f_\bk - f_{0\bk}$ of the original AW model with $f_\bk - f_{{\rm S}\bk}$, where $f_{{\rm S}\bk}$ is the Shakhov distribution. For the high-order Shakhov models that we discuss in this paper, we construct $f_{{\rm S}\bk}$ using a finite polynomial basis similar to the one employed in Grad's method of moments (see Ref.~\cite{Denicol:2012cn} for a rigorous discussion in the frame of relativistic kinetic theory). The basis involves a finite range $-s_\ell \le r \le N_\ell$ of irreducible moments $\rho^{\mu_1 \cdots \mu_\ell}_{{\rm S};r}$ of the deviation $\delta f_{{\rm S}\bk} \equiv f_{{\rm S}\bk} - f_{0\bk}$ of the Shakhov distribution from equilibrium. Here, $N_\ell$ represents the usual truncation order of the polynomial basis, while $s_\ell$ represents a downwards shift, allowing the basis to directly incorporate irreducible moments of negative energy index. Such a shifted basis was shown in Ref.~\cite{Ambrus:2022vif} to play an important role in deriving the hydrodynamic limit of the AW model and we exploit this feature also in the Shakhov model.

We denote the elements of collision matrix that are controlled by the Shakhov model by $\mathcal{A}^{(\ell)}_{{\rm S};rn}$, where $-s_\ell \le r,n \le N_\ell$. Of course, the collision model has an infinite-dimensional collision matrix, which we are able to derive analytically in a basis-free manner, as described in Ref.~\cite{Ambrus:2022vif} (see Ref.~\cite{Ambrus:2023ilm} for an application to the case of the first-order Shakhov model). The functions $\mathcal{A}^{(\ell)}_{{\rm S};rn}$ represent the direct degrees of freedom of the high-order Shakhov model. On the other hand, our ultimate goal is for the kinetic model to achieve prescribed transport coefficients in its hydrodynamic limit. It is known that these transport coefficients are ultimately governed by the elements of the inverse collision matrices $\tau^{(\ell)}_{rn} = [\mathcal{A}^{(\ell)}]^{-1}_{rn}$ and we derive their exact expressions in the basis-free manner of Ref.~\cite{Ambrus:2022vif}, using the Inverse-Reynolds Dominance (IReD) approach of Ref.~\cite{Wagner:2022ayd}. Then, the problem of constructing the Shakhov collision matrix boils down to solving a set of algebraic equations that involve the elements $\tau^{(\ell)}_{{\rm S};rn}$ of the inverse collision matrix, allowing an appropriate subset of transport coefficients to be set as essentially arbitrary thermodynamic functions.
An extension to third order is possible within our moment-based approach, employing the developments in Ref.~\cite{deBrito:2023tgb}, however we leave this avenue open for future research on the topic.
We illustrate the capabilities of our proposed model by considering three examples, described below. 

The first example that we consider is the $0+1$-D Bjorken flow of massive ideal particles, where we aim to separately tune the first-order bulk and shear viscosities $\zeta$ and $\eta$ (which can be tuned also by the first-order Shakhov model, see Ref.~\cite{Ambrus:2023ilm}), as well as the second-order bulk-shear coupling coefficient, $\lambda_{\Pi \pi}$. By increasing $\lambda_{\Pi \pi}$, we allow for an enhanced bulk viscous pressure in the early and intermediate times of the Bjorken expansion, even when the particle mass is not so large.

The second example involves controlling the diffusion-shear coupling coefficients, $\ell_{V\pi}$ and $\ell_{\pi V}$, which we discuss in the context of longitudinal waves propagating through a massless, ideal gas. Both these coefficients vanish in the original RTA by Anderson and Witting \cite{Ambrus:2022vif}, unlike in the more realistic case of hard-sphere interactions \cite{Denicol:2012cn,Wagner:2022ayd,Wagner:2023joq} or the interacting $\lambda \varphi^4$ scalar field theory \cite{Denicol:2022bsq,Rocha:2023hts}. 

The third problem consists of the time-honored Riemann problem \cite{sod1978survey} for a dissipative, ideal gas of massless particles. Besides providing a benchmark test for solvers of perfect fluid dynamics, the flow around the shock front is dominated by strong non-equilibrium effects. Our goal is to derive a Shakhov model that is able to reproduce the results obtained using the Boltzmann Approach to Multi-Parton Scattering (BAMPS) method \cite{Xu:2004mz}, simulating hard-sphere interactions via the test-particle method. We demonstrate that using a Shakhov model able to tune all first- and second-order transport coefficients for both diffusion and shear leads to excellent agreement with the BAMPS data reported in Refs.~\cite{Bouras:2009nn,Bouras:2010hm,Denicol:2012vq}. 
We also demonstrate the importance of the second-order transport coefficients, which differ depending on the computational method employed to derive them.
While Ref.~\cite{Denicol:2012vq} found several discrepancies between the BAMPS data and a formulation of second-order hydrodynamics using the DNMR coefficients \cite{Denicol:2012cn}, the Shakhov model tuned to recover the transport coefficients computed within the IReD (inverse Reynolds dominance \cite{Wagner:2022ayd}) approach gives an excellent agreement with the BAMPS data.

For all of the above examples, we show numerical results of the kinetic Shakhov model obtained using a discrete-velocity method (DVM) implementation derived from the relativistic lattice Boltzmann \cite{Ambrus:2022adp} method using finite-difference techniques for the advection and time-stepping. Our approach employs the so-called rapidity-based moments \cite{Ambrus:2023qcl}, allowing the momentum magnitude to be integrated out exactly. Taking into account azimuthal symmetry, the momentum space complexity is reduced to a single degree of freedom, namely the projection of the particle velocity on the propagation axis ($v_z$). The $v_z$ degree of freedom is discretized using the Gauss-Legendre quadrature, as described in Refs.~\cite{Romatschke:2011hm,Ambrus:2016aub}. The algorithm is highly efficient and its accuracy has been tested in previous publications for the Riemann problem \cite{Ambrus:2016aub}, longitudinal waves \cite{Ambrus:2017keg,Ambrus:2022vif} and Bjorken flow \cite{Ambrus:2016aub,Ambrus:2021sjg,Ambrus:2021fej,Ambrus:2022koq,Ambrus:2022qya,Ambrus:2023qcl,Ambrus:2023ilm} setups. For validation purposes, we also solve the equations of second-order hydrodynamics in the $0+1$-D Bjorken flow setup, as well as in the linearized limit of the longitudinal wave damping problem, using Runge-Kutta time integration. The code is available online as a CodeOcean capsule \cite{codeoceanshk2} (see Sec.~\ref{app:numsch:code} for more details on code availability).

The paper is structured as follows. In Sec.~\ref{sec:shk1}, we review the first-order Shakhov model introduced in Ref.~\cite{Ambrus:2023ilm}. We then introduce the higher-order extension of the Shakhov model in Sec.~\ref{sec:shk2}, where we also discuss how to extract its corresponding first- and second-order transport coefficients arising in its hydrodynamic limit. Sections~\ref{sec:bjork}, \ref{sec:long}, and \ref{sec:riemann} illustrate the capabilities of the Shakhov model in the context of Bjorken flow, longitudinal waves, and the Riemann problem, respectively. Section~\ref{sec:conc} concludes this paper. We also include two appendices: Appendix \ref{app:numsch} describes our DVM solver employing rapidity-based moments, while Appendix~\ref{app:shk_higher} summarizes the high-order Shakhov models considered in Sec.~\ref{sec:riemann}, taylored to capture a selection of the second-order transport coefficients of an ultrarelativistic gas of hard-sphere particles.

\section{First-order Shakhov model}\label{sec:shk1}

The purpose of this section is to review the relaxation-time approximation (RTA) introduced by Anderson and Witting \cite{Anderson:1974a,Anderson:1974b} (in Sec.~\ref{sec:shk1:AW}) and the first-order Shakhov model introduced in Ref.~\cite{Ambrus:2023ilm} (see Sec.~\ref{sec:shk1:shk}). This section also serves to introduce much of the notation used later on in this paper. 

\subsection{The Anderson-Witting model}\label{sec:shk1:AW}

The starting point of this model is the relativistic Boltzmann equation,
\begin{equation}
 k^{\mu }\partial _{\mu }f_\bk = C[f], 
 \label{eq:boltz}
\end{equation}
where $f_\bk$ is the one-particle distribution function, $k^\mu$ is the on-shell particle four-momentum with $k^2 = (k^0)^2 - \bk^2 = m^2$, while $C[f]$ is the 
Boltzmann collision term. The Anderson-Witting (AW) approximation for the collision term $C[f]$ reads
\begin{equation}
 C_{\rm AW}[ f] \equiv -%
 \frac{E_\bk}{\tau_R} \delta f_\bk,
 \label{eq:CAW}
\end{equation}
where $\delta f_\bk = f_\bk - f_{0\bk}$ represents the deviation of the distribution function $f_\bk$ from local thermodynamic equilibrium. In this paper, we focus on ideal gases, for which $f_{0\bk}$ is given by
\begin{equation}
 f_{0\bk}=\left(e^{\beta E_\bk - \alpha} + a\right)^{-1},  \label{eq:feq}
\end{equation}
where $\beta = T^{-1}$ is the inverse temperature, $\alpha = \beta \mu$ is the ratio between the chemical potential and the temperature, $u^\mu$ is the fluid four-velocity, $E_\bk = u_\mu k^\mu$ is the particle energy in the fluid-rest frame, while $a = 1$, $-1$ and $0$ for Fermi-Dirac, Bose-Einstein, and Boltzmann statistics, respectively. 

The distribution $f_\bk$ can be used to compute the macroscopic particle four-flow $N^\mu$ and stress-energy tensor $T^{\mu\nu}$ as 
\begin{align}
 N^\mu &= \int \d K\, k^\mu f_{\bk}, &
 T^{\mu\nu} &= \int \d K\, k^\mu k^\nu f_{\bk},
 \label{eq:macro_def}
\end{align}
where $\d K \equiv  g \d^3k / [(2\pi)^3 k^0]$ and $g$ is the degeneracy factor. 
The equilibrium contributions to the above quantities read
\begin{equation}
 N^\mu_0 = n u^\mu, \qquad 
 T^{\mu\nu}_0 = e u^\mu u^\nu - P \Delta^{\mu\nu},
 \label{eq:hydro_eq}
\end{equation}
where $\Delta^{\mu\nu} \equiv g^{\mu\nu} - u^\mu u^\nu$ is the projector on the hypersurface orthogonal to $u^\mu$, while the particle number density $n$, energy density $e$ and hydrostatic pressure $P$ are given by
\begin{equation}
 n = I_{10}, \quad e = I_{20}, \quad P = I_{21},
 \label{eq:I_ex}
\end{equation}
where $I_{nq}$ are thermodynamic integrals defined as
\begin{equation}
 I_{nq} \equiv \frac{1}{(2q+1)!!} \int \d K\, E_\bk^{n-2q} (-\Delta^{\alpha\beta} k_\alpha k_\beta)^q f_{0\bk}.
 \label{eq:Inq}
\end{equation}
For future use, we introduce the associated integrals $J_{nq}$, defined as
\begin{align}
 J_{nq} &\equiv \left(\frac{\partial I_{nq}}{\partial \alpha}\right)_\beta 
 = -\left(\frac{\partial I_{n-1,q}}{\partial \beta}\right)_\alpha \nonumber\\
 &= \frac{1}{(2q+1)!!} \int \d K E_\bk^{n-2q} (-\Delta^{\alpha\beta} k_\alpha k_\beta)^q 
 f_{0\bk} \tilde{f}_{0\bk} \nonumber\\
 &= \beta^{-1}[I_{n-1,q-1} + (n - 2q) I_{n-1,q}],
 \label{eq:Jnq}
\end{align}
with $\tilde{f}_{0\bk}\equiv 1-a f_{0\bk}$,
as well as the recurrence relations:
\begin{subequations}
\begin{align}
 I_{nq} &= \frac{1}{2q+1} (I_{n,q-1} - m^2 I_{n-2,q-1}),\\
 J_{nq} &= \frac{1}{2q+1} (J_{n,q-1} - m^2 J_{n-2,q-1}).
\end{align}
\end{subequations}
Taking into account the thermodynamic relations
\begin{equation}
 \left(\frac{\partial P}{\partial T}\right)_{\mu} = s, \qquad 
 \left(\frac{\partial P}{\partial \mu}\right)_{T} = n,
\end{equation}
where $s = (e + P - \mu n) / T$ is the entropy density, one can establish:
\begin{align}
 J_{21} &= \left(\frac{\partial P}{\partial \alpha}\right)_\beta = n T,\nonumber\\
 J_{31} &= -\left(\frac{\partial P}{\partial \beta}\right)_\alpha = T(e + P).
 \label{eq:J_ex}
\end{align}

Multiplying Eq.~\eqref{eq:boltz} by $1$ and $k^\mu$ and integrating with respect to $\d K$ leads to
\begin{align}
\partial_\mu N^\mu &= -\frac{u_\mu}{\tau _R} (N^\mu - N^\mu_0), \nonumber\\
\partial_\nu T^{\mu\nu} &= -\frac{u_\nu}{\tau _R} (T^{\mu\nu} - T_0^{\mu\nu}).\label{eq:AW_conservation_1}
\end{align}
Imposing the conservation equations $\partial_\mu N^\mu = 0$ and $\partial_\nu T^{\mu\nu} = 0$, we arrive at the Landau matching conditions,
\begin{equation}
 u_\mu N^\mu = u_\mu N^\mu_0 = n, \quad 
 u_\nu T^{\mu\nu} = u_\nu T^{\mu\nu}_0 = e u^\mu.
 \label{eq:Landau_matching}
\end{equation}
The eigenvalue equation implied in the second relation, $T^\mu{}_\nu u^\nu = eu^\mu$, corresponds to the Landau (energy) frame definition of the fluid four-velocity $u^\mu$.
In the Landau frame, the deviations from equilibrium $\delta N^\mu = N^\mu - N^\mu_0$ and $\delta T^{\mu\nu} = T^{\mu\nu} - T^{\mu\nu}_0$ can be decomposed with respect to $u^\mu$ as
\begin{equation}
 \delta N^\mu = V^\mu, \qquad 
 \delta T^{\mu\nu} = -\Pi \Delta^{\mu\nu} + \pi^{\mu\nu},
 \label{eq:Landau_deviations}
\end{equation}
where $\Pi$ is the bulk viscous pressure, $V^\mu$ is the diffusion current, and $\pi^{\mu\nu}$ is the shear-stress tensor. The conservation equations for $N^\mu$ and $T^{\mu\nu}$ give rise to the following evolution equations for $n$, $e$ and $u^\mu$:
\begin{subequations}\label{eq:hydro_cons}
\begin{align}
 \dot{n} + n \theta + \partial_\mu V^\mu &= 0, \label{eq:hydro_dotn}\\
 \dot{e} + (e + P + \Pi) \theta - \pi^{\mu\nu} \sigma_{\mu\nu} &= 0, \label{eq:hydro_dote}\\
 (e + P + \Pi) \dot{u}^\mu - \nabla^\mu (P + \Pi) + \Delta^\mu{}_\lambda \partial_\nu \pi^{\lambda \nu} &= 0, \label{eq:hydro_dotu}
\end{align}
\end{subequations}
where the dot denotes the comoving derivative, $\dot{f} \equiv u^\mu \partial_\mu f$, while $\nabla_\mu f \equiv \Delta_\mu^\nu \partial_\nu f = \partial_\mu f - u_\mu \dot{f}$ becomes the spatial gradient in the fluid rest frame. 
Furthermore, $\theta \equiv \partial_\mu u^\mu$ is the expansion scalar and $\sigma_{\mu\nu} \equiv \nabla_{\langle \mu} u_{\nu \rangle}$ is the shear tensor, while the angular brackets denote symmetrization and orthogonality with respect to $u^\mu$ in all indices. In the case of 
rank-one and rank-two tensors, $V^\mu$ and $A^{\mu\nu}$, we have $V^{\langle \mu \rangle} = \Delta^\mu_\nu V^\nu = V^\mu - u^\mu (u \cdot V)$ and
\begin{equation}
 A^{\langle \mu\nu \rangle} =  \left[\frac{1}{2}\left(\Delta^\mu_\alpha \Delta^\nu_\beta + 
 \Delta^\nu_\alpha \Delta^\mu_\beta\right) - \frac{1}{3} \Delta^{\mu\nu} \Delta_{\alpha\beta}\right] A^{\alpha\beta}.
\end{equation}

Considering now that the deviation $\delta f_\bk$ from equilibrium is small, the Chapman-Enskog method allows one to estimate $\delta f_\bk$ in the AW model as follows:
\begin{equation}
    \delta f_\bk \simeq -\frac{\tau_R}{E_\bk} k^\mu \partial_\mu f_{0\bk},
    \label{eq:AW_CE}
\end{equation}
where $\tau_R$ is assumed to be of the same order as $\delta f_\bk$. Using the expression \eqref{eq:feq} for $f_{0\bk}$, the right-hand side of the above equation evaluates to 
\begin{align}
 k^\mu \partial_\mu f_{0\bk} &= -f_{0 \bk} \tilde{f}_{0\bk} [k^\mu (E_\bk \partial_\mu \beta - \partial_\mu \alpha) + \beta k^\mu k^\nu \partial_\mu u_\nu] \nonumber\\
 &= - f_{0\bk} \tilde{f}_{0\bk} \Big[E_\bk^2 \dot{\beta} - E_\bk \dot{\alpha} + \frac{\beta}{3} (m^2 - E_\bk^2) \theta \nonumber\\
 & + k^{\langle \mu \rangle} (\beta E_\bk \dot{u}_\mu + E_\bk \nabla_\mu \beta - I_\mu) + \beta k^{\langle \mu} k^{\nu \rangle} \sigma_{\mu\nu}\Big],
\end{align}
where $I^\mu = \nabla^\mu \alpha$ and we used the properties $k^\mu = k^{\langle \mu \rangle} + u^\mu E_\bk$ and
\begin{multline}
 k^\mu k^\nu = E_\bk^2 u^\mu u^\nu + E_\bk (u^\mu k^{\langle \nu \rangle} + u^\nu k^{\langle \mu \rangle})  \\
 +k^{\langle \mu} k^{\nu \rangle}+ \frac{1}{3} \Delta^{\mu\nu} \Delta_{\alpha\beta} k^\alpha k^\beta.
 \label{eq:kmunu}
\end{multline}

We now seek to compute the diffusive quantities $\Pi$, $V^\mu$, and $\pi^{\mu\nu}$, expressed as
\begin{equation}
 \Pi = -\frac{m^2}{3} \rho_0, \quad 
 V^\mu = \rho_0^\mu, \quad 
 \pi^{\mu\nu} = \rho_0^{\mu\nu},
\end{equation}
where the irreducible moments $\rho^{\mu_1 \cdots \mu_\ell}_r$ of tensor-rank $\ell$ and energy index $r$ of $\delta f_\bk$ are defined as
\begin{equation}
 \rho_r^{\mu_1 \cdots \mu_\ell} \equiv \int \d K E_\bk^r
 k^{\langle \mu_1} \cdots k^{\mu_\ell \rangle}\delta f_\bk.
 \label{eq:rho_def}
\end{equation}
In the above, $k^{\langle \mu_1} \cdots k^{\mu_\ell \rangle}
\equiv \Delta_{\nu_1 \cdots \nu_\ell}^{\mu_1 \cdots \mu_\ell} k^{\nu_1} \cdots k^{\nu_\ell}$ represents a complete set of basis vectors \cite{Groot.1980,Denicol:2012cn}.

Using the integration formula given in Eq.~(20) of Ref.~\cite{Denicol:2012cn}, 
\begin{multline}
 \int \d K\, F_\bk k^{\langle \mu_1} \cdots k^{\mu_m \rangle} k_{\langle \nu_1} \cdots k_{\nu_n \rangle} \\ 
 = \frac{m! \delta_{mn}}{(2m+1)!!} \Delta^{\mu_1 \cdots \mu_m}_{\nu_1 \cdots \nu_m} 
 \int \d K\, F_\bk (\Delta^{\alpha\beta} k_\alpha k_\beta)^m,
\end{multline}
with $F_\bk \equiv F_\bk(E_\bk)$ being a function that depends on $k^\mu$ only through the combination $E_\bk = k^\mu u_\mu$, we obtain
\begin{subequations}\label{eq:AW_CE_aux}
\begin{align}
 \Pi &\simeq -\tau_R \frac{m^2}{3} \left[J_{10} \dot{\beta} - J_{00} \dot{\alpha} - \beta J_{11} \theta\right],\label{eq:AW_CE_aux_0}\\
 V^\mu &\simeq \tau_R \left[J_{11} I^\mu - J_{21} (\beta \dot{u}^\mu + \nabla^\mu \beta)\right],\label{eq:AW_CE_aux_1}\\
 \pi^{\mu\nu} &\simeq 2\tau_R \beta J_{32} \sigma^{\mu\nu}.\label{eq:AW_CE_aux_2}
\end{align}
\end{subequations}

Equation~\eqref{eq:AW_CE_aux_2} for $\pi^{\mu\nu}$ is already in its familiar Navier-Stokes form, $\pi^{\mu\nu} \simeq 2 \eta_{\rm AW} \sigma^{\mu\nu}$, where the shear viscosity of the Anderson-Witting model is
\begin{equation}
 \eta_{\rm AW} = \tau_R \alpha^{(2)}_0, \quad \alpha^{(2)}_r = \beta J_{3+r, 2}.
 \label{eq:eta_alpha2}
\end{equation}
In order to obtain similar constitutive relations for $\Pi$ and $V^\mu$, we must employ the conservation equations \eqref{eq:hydro_cons} to eliminate the comoving derivatives $\dot{\alpha}$, $\dot{\beta}$, and $\dot{u}^\mu$. 
We start with the case of $V^\mu$, when $\d P = J_{21} \d\alpha - J_{31} \d\beta$ can be used to replace $\nabla^\mu \beta$ in favor of $F^\mu \equiv \nabla^\mu P$ and $I^\mu \equiv \nabla^\mu \alpha$:
\begin{equation}
 \nabla^\mu \beta = \frac{J_{21}}{J_{31}} I^\mu - \frac{F^\mu}{J_{31}}
 = \frac{I^\mu}{h} - \frac{\beta F^\mu}{e+P},
\end{equation}
where $h \equiv (e + P) / n$ is the specific enthalpy per particle.
Using Eq.~\eqref{eq:hydro_dotu} to express $\dot{u}^\mu$ in terms of $F^\mu$ and higher-order terms, Eq.~\eqref{eq:AW_CE_aux_1} leads to $V^\mu \simeq \kappa_{\rm AW} I^\mu$, where the diffusion coefficient $\kappa_{\rm AW}$ reads
\begin{equation}
 \kappa_{\rm AW} = \tau_R \alpha^{(1)}_0, \quad 
 \alpha^{(1)}_r = J_{r+1,1} - \frac{1}{h} J_{r+2, 1}.
 \label{eq:kappa_alpha1}
\end{equation}
Finally, $\dot{n}$ and $\dot{e}$ can be related to $\dot{\alpha}$ and $\dot{\beta}$ using the following relations:
\begin{align}
 \d n &= J_{10} \d\alpha - J_{20} \d\beta, \nonumber\\
 \d e &= J_{20} \d\alpha - J_{30} \d\beta.
\end{align}
This leads to
\begin{equation}
 \dot{\alpha} = \frac{1}{G_{22}}(J_{20} \dot{e} - J_{30} \dot{n}), \quad 
 \dot{\beta} = \frac{1}{G_{22}}(J_{10} \dot{e} - J_{20} \dot{n}),
\end{equation}
where we introduced the notation:
\begin{equation}
 G_{nm} \equiv J_{n0} J_{m0} - J_{n-1,0} J_{m+1,0}.
 \label{eq:Gnm}
\end{equation}
Using now the conservation Eqs.~\eqref{eq:hydro_dotn} and \eqref{eq:hydro_dote}, we have 
\begin{subequations}\label{eq:dotab_charged}
\begin{align}
 \dot{\alpha} &= \mathcal{H}_\alpha \theta + \frac{J_{20}}{G_{22}} (\pi^{\mu\nu} \sigma_{\mu\nu} -\Pi \theta) + \frac{J_{30}}{G_{22}} \partial_\mu V^\mu,\label{eq:dota_charged}\\
 \dot{\beta} &= \mathcal{H}_\beta \theta + \frac{J_{10}}{G_{22}} (\pi^{\mu\nu} \sigma_{\mu\nu} -\Pi \theta) + \frac{J_{20}}{G_{22}} \partial_\mu V^\mu,\label{eq:dotb_charged}
\end{align}
\end{subequations}
where $\mathcal{H}_\alpha$ and $\mathcal{H}_\beta$ are given by
\begin{subequations}\label{eq:mathcalH}
\begin{align}
 \mathcal{H}_\alpha &\equiv \frac{1}{G_{22}}[J_{30} n - J_{20} (e + P)],\label{eq:mathcalHa}\\
 \mathcal{H}_\beta &\equiv \frac{1}{G_{22}}[J_{20} n - J_{10} (e + P)].\label{eq:mathcalHb}
\end{align}
\end{subequations}
Substituting now Eqs.~\eqref{eq:dotab_charged} into Eq.~\eqref{eq:AW_CE_aux_0}, we find $\Pi \simeq -\zeta_{\rm AW} \theta$, where the coefficient of bulk viscosity $\zeta_{\rm AW}$ reads
\begin{equation}
 \zeta_{\rm AW} = \tau_R \frac{m^2}{3} \alpha^{(0)}_0, \quad 
 \alpha^{(0)}_r = \mathcal{H}_\beta J_{r+1,0} - \mathcal{H}_\alpha J_{r0} - \beta J_{r+1,1}.
 \label{eq:alpha0_charged}
\end{equation}

We take a moment here to remark that the above relations arise in a fluid with one conserved charge. If there is no conserved charge, or if there are multiple conserved charges, then the above discussion must be generalized, as discussed in Ref.~\cite{Fotakis:2022usk}. For the purpose of this paper, we also consider the case when the fluid possesses no such conserved charge. This case can be obtained as the limit when $\alpha = 0$, leading to
\begin{equation} \label{eq:dotab_uncharged}
 \dot{\alpha} = 0, \quad 
 \dot{\beta} = -\frac{\dot{e}}{J_{30}} = \frac{e + P}{J_{30}} \theta + \frac{\Pi \theta - \pi^{\mu\nu} \sigma_{\mu\nu}}{J_{30}}.
\end{equation}
Comparing the above relations to Eq.~\eqref{eq:dotab_charged}, we see that for the uncharged fluid we can formally identify 
\begin{equation}
 \mathcal{H}_\alpha \rightarrow \overline{\mathcal{H}}_\alpha \equiv 0, \quad 
 \mathcal{H}_\beta \rightarrow \overline{\mathcal{H}}_\beta \equiv \frac{e + P}{J_{30}}.
 \label{eq:Hs_nocharge}
\end{equation}
Ultimately, this leads to a modification of the constitutive equation for the bulk viscous pressure and diffusion current, which now read $\Pi \simeq -\bar{\zeta}_{\rm AW} \theta$ and $V^\mu \simeq 0$, where
\begin{equation}
 \bar{\zeta}_{\rm AW} = \tau_R \frac{m^2}{3} \bar{\alpha}^{(0)}_0, \quad 
 \bar{\alpha}^{(0)}_r = \overline{\mathcal{H}}_\beta J_{r+1,0} - \beta J_{r+1,1}.
 \label{eq:alpha0_uncharged}
\end{equation}

\subsection{The first-order Shakhov-like extension}\label{sec:shk1:shk}

In the first-order Shakhov model introduced in Ref.~\cite{Ambrus:2023ilm}, Eq.~\eqref{eq:CAW} is replaced by
\begin{equation}
 C_{\rm S}[\fk] = -\frac{E_\bk}{\tau_R}(\fk - \fS) = -\frac{E_\bk}{\tau_R}(\delta \fk - \delta \fS),
 \label{eq:CS}
\end{equation}
where $\fS = f_{0\bk} + \delta \fS$ drives $f_\bk$ towards $f_{0\bk}$ on a modified path compared to the AW model. Multiplying Eq.~\eqref{eq:boltz} by $1$ and $k^\mu$ and integrating with respect to $\d K$ leads to
\begin{align}
\partial _{\mu }N^{\mu } =& -\frac{1}{\tau _R} (n - n_{\rm S}), \nonumber\\
\partial _{\mu }T^{\mu \nu } =& -\frac{1}{\tau _R} [(e - e_{\rm S})u^\nu + W^\nu - W^\nu_{\rm S}].\label{eq:AW_conservation}
\end{align}
where $W^\mu \equiv \Delta^\mu{}_\alpha T^{\alpha\beta} u_\beta$ is the heat flux ($W^\mu = 0$ in the Landau frame). 
The conservation of the particle four-flow $N^\mu$ and stress-energy
tensor is achieved when
\begin{align}
 n &= n_{\rm S}, & e &= e_{\rm S}, &
 W^\mu &= W_{\rm S}^\mu.
 \label{eq:shk_matching}
\end{align}
In the AW model, the velocity is taken in the Landau frame, such that $W^\mu =0$ and thus $W_{\rm S}^\mu = 0$.
The extension proposed in Ref.~\cite{Pennisi.2018} allows for a different frame to be used, e.g., the Eckart frame,  however 
we do not pursue this freedom in the remainder of this manuscript. We further assume the Landau matching conditions, such that $n$, $e$, and $u^\mu$ define the local equilibrium distribution $f_{0\bk}$.

Considering now that the  deviations $\delta f_\bk$ and $\delta f_{{\rm S}\bk}$ from equilibrium are small, the Chapman-Enskog procedure can be applied just as in the previous subsection, leading to
\begin{equation}
    \delta f_\bk - \delta f_{{\rm S}\bk} \simeq -\frac{\tau_R}{E_\bk} k^\mu \partial_\mu f_{0\bk}.
    \label{eq:shk_CE}
\end{equation}
Taking the off-equilibrium moments of the above equation gives 
\begin{subequations}\label{eq:shk_CE_hydro}
\begin{align}
 \Pi - \Pi_{\rm S} &= -\zeta_{\rm AW} \theta, \\
 V^\mu - V^\mu_{\rm S} &= \kappa_{\rm AW} I^\mu, \\ 
 \pi^{\mu\nu} - \pi^{\mu\nu}_{\rm S} &= 2\eta_{\rm AW} \sigma^{\mu\nu}.
\end{align}
\end{subequations}
As originally proposed by Shakhov \cite{shakhov68a}, the non-equilibrium moments $\Pi_{\rm S}$, $V_{\rm S}^\mu$ and $\pi_{\rm S}^{\mu\nu}$ are chosen as
\begin{gather}
 \Pi_{\rm S} \equiv \Pi\left(1 - \frac{\tau_R}{\tau_\Pi}\right), \nonumber\\
 V_{\rm S}^\mu \equiv V^\mu\left(1 - \frac{\tau_R}{\tau_V}\right), \quad 
 \pi_{\rm S}^{\mu\nu} \equiv \pi^{\mu\nu}\left(1 - \frac{\tau_R}{\tau_\pi}\right),
 \label{eq:shk_dissipative}
\end{gather}
where the new relaxation times $\tau_\Pi$, $\tau_V$, and $\tau_\pi$ are thermodynamic functions representing new model parameters. Substituting the above equalities in Eqs.~\eqref{eq:shk_CE_hydro} leads to the modified constitutive relations 
\begin{align}
 \Pi &= -\zeta_{\rm S} \theta, &
 V^\mu &= \kappa_{\rm S} I^\mu, &
 \pi^{\mu\nu} &= 2\eta_{\rm S} \sigma^{\mu\nu},
\end{align}
where the first-order transport coefficients of the Shakhov model read:
\begin{align}
 \zeta_{\rm S} &= \frac{m^2}{3} \alpha^{(0)}_0 \tau_\Pi, &
 \kappa_{\rm S} &= \alpha^{(1)}_0 \tau_V, &
 \eta_{\rm S} &= \alpha^{(2)}_0 \tau_\pi,
\end{align}
or equivalently, $\zeta_{\rm S} = (\tau_\Pi / \tau_R) \zeta_{\rm AW}$, $\kappa_{\rm S} = (\tau_V / \tau_R) \kappa_{\rm AW}$, and $\eta_{\rm S} = (\tau_\pi / \tau_R) \eta_{\rm AW}$.

Following Shakhov's prescription, the simplest way to achieve the relations in Eq.~\eqref{eq:shk_dissipative} is to construct the Shakhov distribution as 
\begin{subequations}\label{eq:shk}
\begin{equation}
 f_{{\rm S}\bk} = f_{0\bk} + f_{0\bk} \tilde{f}_{0\bk} \mathbb{S}_\bk,\label{eq:Sk_def}
\end{equation}
where
\begin{multline}
 \mathbb{S}_\bk \equiv - 
 \frac{3\Pi}{m^2} \left(1 - \frac{\tau_R}{\tau_\Pi}\right)\mathcal{H}^{(0)}_{\bk 0}
 + k_\mu V^\mu 
 \left(1 - \frac{\tau_R}{\tau_V}\right)
 \mathcal{H}^{(1)}_{\bk 0} \\
 + k_\mu k_\nu \pi^{\mu\nu} \left(1 - \frac{\tau_R}{\tau_\pi}\right)\mathcal{H}^{(2)}_{\bk 0},
 \label{eq:shk1}
\end{multline}
where $\mathcal{H}^{(\ell)}_{\bk 0}$ are polynomials that ensure the recovery of the matching conditions in Eq.~\eqref{eq:shk_matching} and the relations \eqref{eq:shk_dissipative}, such that
\begin{align}
 \begin{pmatrix}
  \rho_{{\rm S};0} \\ \rho_{{\rm S};1} \\ \rho_{{\rm S};2}
 \end{pmatrix}
  &= \int \d K
 \begin{pmatrix}
  1 \\ E_\bk \\ E_\bk^2
 \end{pmatrix}
 \delta f_{{\rm S}\bk} = -\frac{3}{m_0^2} \Pi_{\rm S}
 \begin{pmatrix}
  1 \\ 0 \\ 0
 \end{pmatrix}, \nonumber\\
 \begin{pmatrix}
  \rho^\mu_{{\rm S};0} \\ \rho^\mu_{{\rm S};1}
 \end{pmatrix} &= \int \d K
 \begin{pmatrix}
  1 \\ E_\bk
 \end{pmatrix} k^{\langle \mu \rangle} \delta f_{{\rm S}\bk} = V^\mu_{\rm S} \begin{pmatrix}
  1 \\ 0
 \end{pmatrix}, \nonumber\\
 \rho_{{\rm S};0}^{\mu\nu} &= \int \d K
 k^{\langle \mu} k^{\nu \rangle} \delta f_{{\rm S}\bk} = \pi_{\rm S}^{\mu\nu}.
 \label{eq:Heqs}
\end{align}
Taking $\mathcal{H}^{(0)}_{\bk 0}$, $\mathcal{H}^{(1)}_{\bk 0}$, and $\mathcal{H}^{(2)}_{\bk 0}$ as second-, first- and zeroth-order polynomials, 
their exact expressions can be obtained as \cite{Ambrus:2022vif} 
\begin{align}
\mathcal{H}_{\bk 0}^{(0)}=& \frac{G_{33} - G_{23} E_\bk + G_{22} E_\bk^2}{J_{00} G_{33} - J_{10} G_{23} + J_{20} G_{22}}, \nonumber\\
\mathcal{H}_{\bk 0}^{(1)} =& \frac{J_{31} E_\bk - J_{41}}{J_{21} J_{41} - J_{31}^2}, \quad 
\mathcal{H}_{\bk 0}^{(2)} = \frac{1}{2 J_{42}},
\label{eq:shk1_Hfunctions}
\end{align}%
\end{subequations}
where $J_{nq}$ and $G_{nm}$ were introduced in Eqs.~\eqref{eq:Jnq} and \eqref{eq:Gnm}. The above functions coincide with those appearing in the 14-moment approximation of $\delta f_\bk$ in Ref.~\cite{Denicol:2012cn}.


\section{Higher-order Shakhov-like extension}\label{sec:shk2}

As mentioned in the introduction, the Shakhov-like model as introduced in Eq.~\eqref{eq:CS} can be used to control also second-order transport coefficients. This requires an extension of the first-order Shakhov model summarized in the previous section by adding extra terms on the right-hand side of Eq.~\eqref{eq:shk1}. Anticipating material that will be reviewed in this section, our proposed extension effectively amounts to modifying an increasingly large set of entries in the collision matrix $\mathcal{A}^{(\ell)}_{rn}$ associated with the Shakhov collision term, $C_{\rm S}[f]$. 

We begin this section with a discussion on the equations of motion for the irreducible moments $\rho^{\mu_1 \cdots \mu_\ell}_r$ of $\delta f_\bk$, derived from the Boltzmann equation in Ref.~\cite{Denicol:2012cn} and summarized in Subsec.~\ref{sec:shk2:eoms} of the present section. We then discuss the collision matrix of the extended Shakhov model and its inverse in Subsec.~\ref{sec:shk2:tau}. The equations of second-order hydrodynamics and the corresponding transport coefficients arising from the Shakhov model are discussed in Subsec.~\ref{sec:shk2:tcoeffs}. These results are specialized to the case of a fluid without conserved charges in Subsec.~\ref{sec:shk2:uncharged} and to the case of an ultrarelativistic ideal gas in Subsec.~\ref{sec:shk2:UR}. Subsection \ref{sec:shk2:dofs} discusses the degrees of freedom that can be fixed by the Shakhov model.

\subsection{Equations of motion for the irreducible moments} \label{sec:shk2:eoms}

The central objects of the analysis are the irreducible moments $\rho^{\mu_1 \cdots \mu_\ell}_r$ of $\delta f_\bk$, introduced in Eq.~\eqref{eq:rho_def}.
Their equations of motion can be derived starting from the Boltzmann Eq.~\eqref{eq:boltz}, by substituting $f_\bk = f_{0\bk} + \delta f_{\bk}$:
\begin{equation}
 \dot{\delta f_\bk} = -\frac{1}{E_\bk} k^\mu \partial_\mu f_{0\bk} - 
 \frac{1}{E_\bk} k^{\langle \mu \rangle} \nabla_\mu \delta f_\bk + 
 \frac{C[f]}{E_\bk}.
 \label{eq:boltz_df}
\end{equation}
Similar to $\delta f_\bk$, the collision term $C[f]$ can also be characterized by its irreducible moments, defined as
\begin{equation}
 C_r^{\mu_1 \cdots \mu_\ell} \equiv \int \d K\, E_\bk^r k^{\langle \mu_1} \cdots k^{\mu_\ell \rangle} C[f].
\end{equation}
Multiplying Eq.~\eqref{eq:boltz_df} by $E_\bk^r$, $E_\bk^r k^{\langle \mu \rangle}$, and $E_\bk^r k^{\langle \mu} k^{\nu \rangle}$ and integrating with respect to $\d K$ leads to \cite{Denicol:2012cn}:
\begin{widetext}
\begin{subequations}\label{eq:rhodot_all}
\begin{align}
 \dot{\rho}_r - C_{r-1} &= \alpha_r^{(0)} \theta +
 \frac{G_{2r}}{G_{22}} (\Pi \theta - \pi^{\mu\nu} \sigma_{\mu\nu}) + 
 \frac{G_{3r}}{G_{22}} (V^\mu \dot{u}_\mu - \nabla_\mu V^\mu)+r\dot{u}_\mu \rho_{r-1}^\mu \nonumber\\
 & - \rho_r \theta - \frac{r-1}{3} (\rho_r - m^2 \rho_{r-2}) \theta 
 - \nabla_\mu \rho^\mu_{r-1} + (r - 1) \rho^{\mu\nu}_{r-2} \sigma_{\mu\nu}\;, \label{eq:rhodot_l0}\\
 \dot{\rho}^{\langle\mu\rangle}_r - C_{r-1}^{\langle \mu \rangle} &= \alpha_r^{(1)} I^\mu + 
 \rho^\nu_r \omega^\mu{}_\nu + 
 \frac{1}{3}[(r-1) m^2 \rho^\mu_{r-2} - (r+ 3) \rho^\mu_r] \theta 
 - \Delta^\mu_\lambda \nabla_\nu \rho^{\lambda \nu}_{r-1} 
 + r \rho^{\mu\nu}_{r-1} \dot{u}_\nu \nonumber\\
 &+ \frac{1}{5} \left[(2r-2) m^2 \rho^\nu_{r-2} - (2r+3) \rho^\nu_r\right] \sigma^\mu_\nu 
 + \frac{1}{3}\left[m^2 r \rho_{r-1} - (r+3) \rho_{r+1}\right] \dot{u}^\mu \nonumber\\
 &+ \frac{\beta J_{r+2,1}}{e + P} 
 (\Pi \dot{u}^\mu - \nabla^\mu \Pi + \Delta^\mu_\nu \partial_\lambda \pi^{\lambda \nu}) 
 -\frac{1}{3} \nabla^\mu (m^2 \rho_{r-1} - \rho_{r+1}) 
 + (r -1 ) \rho^{\mu\nu\lambda}_{r-2} \sigma_{\lambda\nu}\;,
 \label{eq:rhodot_l1} \\
 \dot{\rho}^{\langle \mu\nu\rangle}_r - C^{\langle \mu\nu \rangle}_{r-1} &=
 2 \alpha^{(2)}_r \sigma^{\mu\nu} - \frac{2}{7} \left[(2r+5) \rho^{\lambda\langle \mu}_r - 2m^2(r-1) \rho^{\lambda\langle \mu}_{r-2}\right] \sigma^{\nu\rangle}_\lambda + 
 2\rho^{\lambda\langle \mu}_r \omega^{\nu\rangle}{}_\lambda \nonumber\\
 &+ \frac{2}{15}[(r+4)\rho_{r+2} - (2r+3) m^2 \rho_r + 
 (r - 1)m^4 \rho_{r-2}] \sigma^{\mu\nu} + 
 \frac{2}{5} \nabla^{\langle \mu} (\rho^{\nu \rangle}_{r+1} - m^2 \rho^{\nu\rangle}_{r-1}) \nonumber\\
 &- \frac{2}{5} \left[(r+5) \rho^{\langle \mu}_{r+1} - r m^2 \rho^{\langle \mu}_{r-1}\right] \dot{u}^{\nu \rangle} 
 - \frac{1}{3} \left[(r + 4) \rho^{\mu\nu}_r - m^2 (r-1) \rho^{\mu\nu}_{r-2}\right] \theta 
 \nonumber\\
 & + (r - 1) \rho^{\mu\nu\lambda\rho}_{r-2} \sigma_{\lambda \rho} - \Delta^{\mu\nu}_{\alpha\beta} 
 \nabla_\lambda \rho^{\alpha\beta\lambda}_{r-1} + 
 r \rho^{\mu\nu\lambda}_{r-1} \dot{u}_\lambda\;,
 \label{eq:rhodot_l2}
\end{align}
\end{subequations}
\end{widetext}
where we considered that the fluid possesses a conserved charge, characterized by $\alpha = \mu / T \neq 0$. In the case when no such charge is present, $\alpha = 0$ and the above equations are modified as discussed in Sec.~\ref{sec:shk1:AW}. The modifications to the equation for the scalar moments can be summarized by
\begin{equation}
 \alpha^{(0)}_r \rightarrow \bar{\alpha}^{(0)}_r, \quad 
 \frac{G_{2r}}{G_{22}} \rightarrow \frac{J_{r+1,0}}{J_{30}}, \qquad 
 \frac{G_{3r}}{G_{22}} \rightarrow 0,
 \label{eq:uncharged_repl}
\end{equation}
such that Eq.~\eqref{eq:rhodot_l0} becomes
\begin{multline}
 \dot{\rho}_r - C_{r-1} = \bar{\alpha}_r^{(0)} \theta + \frac{J_{r+1,0}}{J_{30}} (\Pi \theta - \pi^{\mu\nu} \sigma_{\mu\nu}) - \rho_r \theta +r \dot{u}_\mu \rho_{r-1}^\mu \\
 - \frac{r-1}{3} (\rho_r - m^2 \rho_{r-2}) \theta - \nabla_\mu \rho^\mu_{r-1} 
 + (r - 1) \rho^{\mu\nu}_{r-2} \sigma_{\mu\nu} 
 \;.
\end{multline}
From a structural point of view, the equations of motion for the vector and tensor moments remain unchanged, with the observation that in this case, $I^\mu = 0$ and all vector moments become of second order.

\subsection{Collision matrix of the extended Shakhov model} \label{sec:shk2:tau}

In the approximation when the fluid is not far from equilibrium, $C_r^{\mu_1 \cdots \mu_\ell}$
can be linearized with respect to the irreducible moments $\rho^{\mu_1 \cdots \mu_\ell}_r$ of the non-equilibrium part of the distribution function $\delta f_\bk = f_\bk - f_{0\bk}$,
\begin{equation}
 C_{r-1}^{\mu_1 \cdots \mu_\ell} = -\sum_n \mathcal{A}^{(\ell)}_{rn} \rho^{\mu_1 \cdots \mu_\ell}_n,
\end{equation}
where the summation over $n$ goes in principle from $-\infty$ to $\infty$. In the case of a general collision term, the collision matrix $\mathcal{A}^{(\ell)}_{rn}$ must be computed with respect to a finite basis (cf. Ref.~\cite{Denicol:2012cn}), which accounts only for the moments with indices $-s_\ell \le r \le N_\ell$, where $s_\ell$ is a shift parameter allowing for negative-order moments to be represented \cite{Ambrus:2022vif}, while $N_\ell$ is the expansion order. Here we employ the basis-free approach introduced for the Anderson-Witting collision model in Ref.~\cite{Ambrus:2022vif}, which takes advantage of the simplicity of the relaxation-time approximation to bypass the use of any basis.

We now compute the collision matrix corresponding to the Shakhov model:
\begin{equation}
 C_{{\rm S};r - 1}^{\mu_1 \cdots \mu_\ell} = -\frac{1}{\tau_R} \rho^{\mu_1 \cdots \mu_\ell}_r + \frac{1}{\tau_R} \rho^{\mu_1 \cdots \mu_\ell}_{\rm S;r},
 \label{eq:CSr}
\end{equation}
where the first term originates from the original AW model, while the second one involves the irreducible moments 
$\rho^{\mu_1 \cdots \mu_\ell}_{\rm S;r}$ of $\delta f_{S\bk}$. Since $\delta f_{S\bk}$ vanishes in equilibrium, its irreducible moments can be written in terms of those of $\delta f_\bk$. For simplicity, we restrict the discussion in this paper to the case of a linear dependence and leave the extension to a quadratic dependence (as is the case for a generic $2 \rightarrow 2$ collision term) for future work. 

Naturally, we may ask which moments $\rho^{\mu_1 \cdots \mu_\ell}_{{\rm S};r}$ of the Shakhov model are required and what should their relation to the moments $\rho^{\mu_1 \cdots \mu_\ell}_r$ of $\delta f_\bk$ be? For the first-order model presented in Eq.~\eqref{eq:shk1}, we have 
$\rho_{{\rm S};1} = \rho_{{\rm S}; 2} = \rho^\mu_{{\rm S};1} = 0$, while Eq.~\eqref{eq:shk_dissipative} provides relations for $\rho_0$, $\rho^\mu_0$, and $\rho^{\mu\nu}_0$. Inspection of the right-hand sides of Eqs.~\eqref{eq:rhodot_all} shows that the natural extension of this set of moments should include both negative-order and positive-order moments. We therefore consider
a subset of moments with indices $-s_\ell \le r \le N_\ell$ and write
\begin{equation}
 \rho^{\mu_1 \cdots \mu_\ell}_{{\rm S};r} = \sum_{n = -s_\ell}^{N_\ell} \left(\delta_{rn} - \tau_R \mathcal{A}^{(\ell)}_{{\rm S};rn}\right) \rho^{\mu_1 \cdots \mu_\ell}_n\;,
 \label{eq:connection_rhoS_rho}
\end{equation}
where the reduced collision matrix $\mathcal{A}^{(\ell)}_{{\rm S};rn}$ is in principle arbitrary, representing the degrees of freedom of the extended Shakhov model. 
Substituting the above into Eq.~\eqref{eq:CSr} gives
\begin{equation}
 C_{{\rm S};r - 1}^{\mu_1 \cdots \mu_\ell} = -\sum_{n = -s_\ell}^{N_\ell} \mathcal{A}^{(\ell)}_{{\rm S};rn} \rho^{\mu_1 \cdots \mu_\ell}_n.
\end{equation}
In what follows, we demand that the above relation is satisfied for $-s_\ell \le r \le N_\ell$.

Before proceeding further, we must first give an explicit expression for $\delta f_{{\rm S}\bk}$. As already discussed for the first-order Shakhov model, the deviation $\delta f_{{\rm S}\bk} = f_{{\rm S}\bk} - f_{0\bk} = f_{0\bk} \tilde{f}_{0\bk} \mathbb{S}_\bk$ considered in Eq.~\eqref{eq:shk} must be constructed using an orthogonal basis which ensures the exact recovery of the irreducible moments of $\delta f_{{\rm S}\bk}$ that we require, namely $\rho^{\mu_1 \cdots \mu_\ell}_{{\rm S};r}$ for $-s_\ell \le r \le N_\ell$. For this purpose, we employ the shifted basis introduced in Ref.~\cite{Ambrus:2022vif} and write
\begin{align}
 \delta f_{{\rm S}\bk} &= f_{0\bk} + f_{0\bk} \tilde{f}_{0\bk} \mathbb{S}_\bk,\nonumber\\
 \mathbb{S}_\mathbf{k}
 &= \sum_{\ell =0}^{L}\sum_{n=-s_\ell}^{N_\ell}
 \rho_{{\rm S};n}^{\mu_1 \cdots \mu_\ell} E_\bk^{-s_\ell} k_{\langle \mu_1} \cdots
 k_{\mu_\ell \rangle} \widetilde{\mathcal{H}}_{\mathbf{k},n+s_\ell}^{(\ell)},
 \label{eq:dfS}
\end{align}%
where $L$ is a finite but otherwise arbitrary tensor-rank threshold, while $\widetilde{\mathcal{H}}_{\mathbf{k}n}^{(\ell )}$ are polynomials in
energy $E_\bk$ of order $N_\ell + s_\ell$, defined as
\begin{equation}
 \widetilde{\mathcal{H}}_{\bk n}^{(\ell)}\equiv
 \frac{(-1)^\ell}{\ell! J_{2\ell -2s_\ell,\ell}}
 \sum_{m=n}^{N_\ell + s_\ell} \widetilde{a}_{mn}^{(\ell)}
 \widetilde{P}_{\bk m}^{(\ell)}.
 \label{eq:Hfunctions_def}
\end{equation}%
In the above, $\widetilde{P}_{\bk m}^{(\ell)} = \sum_{r=0}^{m} \widetilde{a}_{mr}^{(\ell)} E_{\bk}^{r}$ are polynomials of order $m$, satisfying the orthogonality relation
\begin{equation}
 \int \d K\, \widetilde{\omega}^{(\ell)} 
 \widetilde{P}_{\bk m}^{(\ell)} \widetilde{P}_{\bk n}^{(\ell)} 
 = \delta_{mn},
 \label{eq:P_ortho}
\end{equation}
where the weight function $\widetilde{\omega}^{(\ell)}$ is defined as
\begin{equation}
 \widetilde{\omega}^{(\ell)} \equiv\frac{(-1)^\ell}{(2\ell + 1)!!} 
 \frac{E_\bk^{-2s_\ell}}{J_{2\ell - 2s_\ell,\ell}}
 (\Delta^{\alpha\beta} k_\alpha k_\beta)^\ell 
 f_{0\bk} \widetilde{f}_{0\bk},\label{eq:omega_def}
\end{equation}
such that $\widetilde{P}^{(\ell)}_{\bk 0} = \tilde{a}^{(\ell)}_{00} = 1$.

The finite-dimensional basis employed in Eq.~\eqref{eq:dfS} ensures that $\rho^{\mu_1 \cdots \mu_\ell}_{{\rm S};r} = \int \d K\, \delta f_{{\rm S};\bk} k^{\langle \mu_1} \cdots k^{\mu_\ell \rangle} E_\bk^r$ for all $-s_\ell \le r \le N_\ell$. When considering either $r < -s_\ell$ or $r > N_\ell$, the corresponding irreducible moments will be expressed in terms of those appearing in Eq.~\eqref{eq:dfS} via
\begin{equation}
 \rho^{\mu_1 \cdots \mu_\ell}_{{\rm S};r} = \sum_{n = -s_\ell}^{N_\ell} \rho^{\mu_1 \cdots \mu_\ell}_{{\rm S}; n} \widetilde{\mathcal{F}}^{(\ell)}_{-(r + s_\ell),n + s_\ell}\;,
\end{equation}
where the functions $\widetilde{\mathcal{F}}^{(\ell)}_{rn}$ are given by
\begin{equation}
\widetilde{\mathcal{F}}^{(\ell)}_{rn} \equiv (-1)^\ell \ell! J_{2\ell - 2s_\ell,\ell} 
 \int \d K\, \widetilde{\omega}^{(\ell)} E_\bk^{-r} 
\widetilde{\mathcal{H}}^{(\ell)}_{\bk n}.
 \label{eq:F_def}
\end{equation}
For this reason, the collision matrix $\mathcal{A}^{(\ell)}_{{\rm S};rn}$ will contain non-trivial entries when $r < -s_\ell$ and $r > N_\ell$, as  illustrated below:
\begin{align}
 \mathcal{A}^{(\ell)}_{rn} = 
 \begin{pmatrix}
  \frac{1}{\tau_R} \delta_{rn} & \mathcal{A}^{(\ell)}_{<; rn} & 0 \\
  0 & \mathcal{A}^{(\ell)}_{{\rm S}; rn} & 0 \\
  0 & \mathcal{A}^{(\ell)}_{>; rn} & \frac{1}{\tau_R} \delta_{rn}
 \end{pmatrix},
 \label{eq:Arn}
\end{align}
where the first and last lines correspond to row indices satisfying $r < -s_\ell$ and $r > N_\ell$, respectively. The elements on the first and third columns, having column index satisfying $n < -s_\ell$ and $n > N_\ell$, respectively, have entries that are identical to those of the AW model. On the contrary, the middle column (where $-s_\ell \le n \le N_\ell$) exhibits non-vanishing entries for $r < -s_\ell$ and $r > N_\ell$, given by
\begin{multline}
 \mathcal{A}^{(\ell)}_{</>;rn} = -\frac{1}{\tau_R} \widetilde{\mathcal{F}}^{(\ell)}_{-(r+s_\ell),n+s_\ell} \\ 
 + \sum_{j = -s_\ell}^{N_\ell} \widetilde{\mathcal{F}}^{(\ell)}_{-(r+s_\ell),j+s_\ell} \mathcal{A}^{(\ell)}_{{\rm S}; jn}.
 \label{eq:A_><}
\end{multline}


The inverse of the collision matrix $\mathcal{A}^{(\ell)}_{rn}$ in Eq.~\eqref{eq:Arn} can be obtained analytically, as follows:
\begin{equation}
 \tau^{(\ell)}_{rn} = 
 \begin{pmatrix}
  \tau_R \delta_{rn} & \tau^{(\ell)}_{<; rn} & 0 \\
  0 & \tau^{(\ell)}_{{\rm S}; rn} & 0 \\
  0 & \tau^{(\ell)}_{>; rn} & \tau_R \delta_{rn}
 \end{pmatrix}\;,
 \label{eq:taurn}
\end{equation}
where the reduced inverse matrix $\tau^{(\ell)}_{{\rm S}; rn} \equiv (\mathcal{A}^{(\ell)}_{\rm S})^{-1}_{rn}$ is the inverse of the $(N_\ell + s_\ell+1) \times (N_\ell + s_\ell+1)$ reduced collision matrix $\mathcal{A}^{(\ell)}_{{\rm S};rn}$. As in the case of $\mathcal{A}^{(\ell)}_{rn}$, the inverse matrix $\tau^{(\ell)}_{rn}$ exhibits the same elements as the AW model when the column index satisfies $n < -s_\ell$ or $n > N_\ell$. However, for $-s_\ell \le n \le N_\ell$, the rows with indices $r < -s_\ell$ and $r> N_\ell$ exhibit non-vanishing entries given by
\begin{multline}
 \tau^{(\ell)}_{<,>; rn} = -\tau_R \sum_{j = -s_\ell}^{N_\ell} \mathcal{A}^{(\ell)}_{<,>; rj} \tau^{(\ell)}_{{\rm S};jn}
 = -\tau_R \widetilde{\mathcal{F}}^{(\ell)}_{-(r+s_\ell),n+s_\ell} \\ 
 + \sum_{j = -s_\ell}^{N_\ell} \widetilde{\mathcal{F}}^{(\ell)}_{-(r+s_\ell),j+s_\ell} \tau^{(\ell)}_{{\rm S}; jn}.
\end{multline}

\subsection{Transport coefficients of the extended Shakhov model}
\label{sec:shk2:tcoeffs}

We now employ the method of moments introduced in Ref.~\cite{Denicol:2012cn} using the basis-free approach of Ref.~\cite{Ambrus:2022vif} to derive the second-order equations of M\"uller-Israel-Stewart-type hydrodynamics and the corresponding first- and second-order transport coefficients arising from our higher-order Shakhov collision model. In order to ensure that the system of equations is hyperbolic, we employ the Inverse-Reynolds Dominance (IReD) approach introduced in Ref.~\cite{Wagner:2022ayd}. To achieve second-order accuracy, we employ as book-keeping parameters the inverse Reynolds number ${\rm Re}^{-1}$ and the Knudsen number ${\rm Kn}$. The former is proportional to the ratio between the dissipative quantities and a corresponding equilibrium quantity, i.e., ${\rm Re}^{-1} \sim |\Pi| / P$, $|V^\mu| / n$ or $|\pi^{\mu\nu}| / P$. The latter is proportional to the ratio of microscopic to macroscopic scales, ${\rm Kn} \sim \lambda_{\rm mfp} \nabla_\mu f$, where $\lambda_{\rm mfp}$ is related to the particle mean free path, while $f$ is a function of the local thermodynamic parameters. As a rule of thumb, the irreducible moments $\rho^{\mu_1 \cdots \mu_\ell}_r$ are of order ${\rm Re}^{-1}$ (or higher), while gradient terms such as $\theta$, $I^\mu$ or $\sigma^{\mu\nu}$ will appear in combinations making $O({\rm Kn})$ contributions.

The second-order equations of fluid dynamics can be obtained from the moment equations~\eqref{eq:rhodot_all}, which we summarize as:
\begin{subequations} \label{eq:rhoeqs_short}
\begin{align}
 \dot{\rho}_n + \sum_r \mathcal{A}^{(0)}_{nr} \rho_r &= \alpha^{(0)}_n \theta + O({\rm Re}^{-1}{\rm Kn}),\\
 \dot{\rho}^{\langle\mu\rangle}_n + \sum_r \mathcal{A}^{(1)}_{nr} \rho_r^\mu &= \alpha^{(1)}_n I^\mu + O({\rm Re}^{-1}{\rm Kn}),\\
 \dot{\rho}_n^{\langle\mu\nu\rangle} + \sum_r \mathcal{A}^{(2)}_{nr} \rho_r^{\mu\nu} &= 2\alpha^{(2)}_n \sigma^{\mu\nu} + O({\rm Re}^{-1}{\rm Kn}).
\end{align}
\end{subequations}
For brevity, we omitted the terms which are of second or higher order on the right hand side of the above relations. Note that, in the spirit of the basis-free paradigm, we did not specify fixed limits for the summation over $r$. However, the structure of $\mathcal{A}^{(\ell)}_{rn}$ shown in Eq.~\eqref{eq:Arn} guarantees that this summation is both finite and well defined. Concretely, $r$ is restricted to the interval $\{-s_\ell, -s_\ell + 1, \ldots, N_\ell\} \cup \{n\}$, i.e., the middle column shown in Eq.~\eqref{eq:taurn} and the element on the diagonal (implied by the $\tau_R^{-1} \delta_{rn}$ entries). Explicitly:
\begin{align}
 n < -s_\ell:& & 
 r &\in \{n\} \cup \{-s_\ell, \dots, N_\ell\}, \nonumber\\
 -s_\ell \le n \le N_\ell:& & 
 r &\in \{-s_\ell, \dots ,N_\ell\}, \nonumber\\
 n > N_\ell:& & 
 r &\in \{-s_\ell, \dots, N_\ell\} \cup \{n\}.
 \label{eq:shk_summation_range}
\end{align}

We now multiply the relations in Eq.~\eqref{eq:rhoeqs_short} by $\tau_{rn}^{(\ell)}$ and sum over $n$. At leading order, we obtain the first-order Navier-Stokes-like relationship between the $O({\rm Re}^{-1})$ irreducible moments $\rho_r^{\mu_1 \cdots \mu_\ell}$ and the $O({\rm Kn})$ thermodynamic forces,
\begin{equation}
 \rho_r \simeq \frac{3}{m^2} \zeta_r \theta, \quad 
 \rho^\mu_r \simeq \kappa_r I^\mu, \quad 
 \rho^{\mu\nu}_r \simeq 2\eta_r \sigma^{\mu\nu},
 \label{eq:NS_r}
\end{equation}
where the first-order transport coefficients are 
\begin{gather}
 \zeta_r = \frac{m^2}{3} \sum_{n \neq 1,2} \tau_{rn}^{(0)} \alpha^{(0)}_n, \quad 
 \kappa_r = \sum_{n \neq 1} \tau_{rn}^{(1)} \alpha^{(1)}_n, \nonumber\\
 \eta_r = \sum_n \tau_{rn}^{(2)} \alpha^{(2)}_n.\label{eq:first_order}
\end{gather}
As before, we did not specify the summation limits for the index $n$. Since the structure of $\tau^{(\ell)}_{rn}$, shown in Eq.~\eqref{eq:taurn}, is identical to that of $\mathcal{A}^{(\ell)}_{nr}$, the values allowed for $n$ in the above expressions for a given $r$ are identical to the values of $r$ allowed by Eq.~\eqref{eq:shk_summation_range} for a given of $n$.
Throughout this section, we will continue to omit the summation limits, with the implicit understanding that the summation is performed as described above.
Coming back to the computation of the first-order transport coefficients, when $-s_\ell \le r \le N_\ell$ we have:
\begin{gather}
 \zeta_r = \frac{m^2}{3} \sum_{n = -s_0,\neq 1,2}^{N_0} \tau^{(0)}_{{\rm S};rn} \alpha^{(0)}_n,\nonumber\\
 \kappa_r = \sum_{n = -s_1,\neq 1}^{N_1} \tau^{(1)}_{{\rm S};rn} \alpha^{(1)}_n,\quad 
 \eta_r = \sum_{n = -s_2}^{N_2} \tau^{(2)}_{{\rm S};rn} \alpha^{(2)}_n.
 \label{eq:shk_tcoeffsr_in}
\end{gather}
When $r < -s_\ell$ or $r > N_\ell$, one has
\begin{subequations}\label{eq:shk_tcoeffsr_out}
\begin{align}
 \zeta_r &= \frac{\tau_R m^2}{3} \alpha^{(0)}_r +\hspace{-14pt} \sum_{n = -s_0,\neq 1,2}^{N_0} \hspace{-10pt} \widetilde{\mathcal{F}}^{(0)}_{-r-s_0, n+s_0} \left(\zeta_n - \frac{\tau_R m^2}{3} \alpha^{(0)}_n\right),\label{eq:shk_zetar}\\
 \kappa_r &= \tau_R \alpha^{(1)}_r + \hspace{-8pt} \sum_{n = -s_1,\neq 1}^{N_1} \hspace{-6pt} \widetilde{\mathcal{F}}^{(1)}_{-r-s_1, n+s_1} (\kappa_n - \tau_R \alpha^{(1)}_n),\label{eq:shk_kappar}\\
 \eta_r &= \tau_R \alpha^{(2)}_r + \sum_{n = -s_2}^{N_2} \widetilde{\mathcal{F}}^{(2)}_{-r-s_2, n+s_2} (\eta_n - \tau_R \alpha^{(2)}_n).
 \label{eq:shk_etar}
\end{align}
\end{subequations}

Next, we employ the IReD approach, by which Eq.~\eqref{eq:NS_r} is employed at $r = 0$ to approximate, up to first order in ${\rm Kn}$ and ${\rm Re}^{-1}$, the thermodynamic forces $\theta$, $I^\mu$ and $\sigma^{\mu\nu}$ in terms of the main dissipative quantities,
\begin{equation}
 \theta \simeq -\frac{\Pi}{\zeta}, \quad 
 I^\mu \simeq \frac{V^\mu}{\kappa}, \quad 
 \sigma^{\mu\nu} \simeq \frac{\pi^{\mu\nu}}{2\eta},
 \label{eq:NS_inv}
\end{equation}
with $\zeta \equiv \zeta_0$, $\kappa \equiv \kappa_0$ and $\eta \equiv \eta_0$. Then, the irreducible moments with $r \neq 0$ are approximated by replacing Eq.~\eqref{eq:NS_inv} in Eq.~\eqref{eq:NS_r}:
\begin{equation}
 \rho_r \simeq -\frac{3}{m^2} \mathcal{C}^{(0)}_r \Pi, \quad 
 \rho_r^\mu \simeq \mathcal{C}^{(1)}_r V^\mu, \quad 
 \rho_r^{\mu\nu} \simeq \mathcal{C}^{(2)}_r \pi^{\mu\nu},
\end{equation}
where the coefficients $\mathcal{C}^{(\ell)}_r$ are given by
\begin{equation}
 \mathcal{C}^{(0)}_r = \frac{\zeta_r}{\zeta}, \quad 
 \mathcal{C}^{(1)}_r = \frac{\kappa_r}{\kappa}, \quad 
 \mathcal{C}^{(2)}_r = \frac{\eta_r}{\eta}.
\end{equation}
Using the above approximations, we arrive at the hydrodynamical equations in the IReD approach:
\begin{subequations}\label{eq:hydro}
\begin{align}
 \tau_\Pi \dot{\Pi} + \Pi &=  -\zeta \theta + \mathcal{J} + \mathcal{R},\label{eq:Pidot}\\
 \tau_V \dot{V}^{\langle \mu \rangle} + V^\mu &= \kappa I^\mu + 
 \mathcal{J}^\mu + \mathcal{R}^\mu, \label{eq:Vdot}\\
 \tau_\pi \dot{\pi}^{\langle \mu \nu \rangle} + \pi^{\mu\nu} &= 
 2 \eta \sigma^{\mu\nu} + \mathcal{J}^{\mu\nu} + \mathcal{R}^{\mu\nu},
 \label{eq:pidot}
\end{align}
\end{subequations}
where the relaxation times $\tau_\Pi$, $\tau_V$, and $\tau_\pi$ are given by
\begin{gather}
 \tau_\Pi = \sum_{r \neq 1,2} \tau^{(0)}_{0r} \mathcal{C}^{(0)}_r, \quad 
 \tau_V = \sum_{r \neq 1} \tau^{(1)}_{0r} \mathcal{C}^{(1)}_r, \nonumber\\
 \tau_\pi = \sum_{r} \tau^{(2)}_{0r} \mathcal{C}^{(2)}_r.
 \label{eq:rtimes}
\end{gather}
The terms $\mathcal{R}$, $\mathcal{R}^\mu$, and $\mathcal{R}^{\mu\nu}$ are of order $O({\rm Re}^{-2})$ and originate from contributions to the collision term which are quadratic with respect to $\delta f_\bk$, or equivalently, with respect to its irreducible moments $\rho^{\mu_1 \cdots \mu_\ell}_r$. As already mentioned before, in this paper we do not include such terms in the Shakhov model and therefore these terms vanish identically: $\mathcal{R} = \mathcal{R}^\mu = \mathcal{R}^{\mu\nu} = 0$. We note that for a realistic collision kernel, such as the $2 \rightarrow 2$ binary scattering kernel, these terms do not vanish, see e.g.~Refs.~\cite{Molnar.2014,Bazow.2016} for a discussion of such quadratic terms in the case of hard-sphere interactions.

The $\mathcal{J}, \mathcal{J}^{\mu}$ and $\mathcal{J}^{\mu\nu}$ collect terms of first order in $\rm{Kn}$ and ${\rm Re}^{-1}$:
\begin{subequations}\label{eq:J}
\begin{align}
	\mathcal{J} &= -\ell_{\Pi V} \nabla_\mu V^\mu - 
	\tau_{\Pi V} V_\mu \dot{u}^\mu - \delta_{\Pi\Pi} \Pi \theta \nonumber\\
	& - \lambda_{\Pi V} V_\mu I^\mu + 
	\lambda_{\Pi \pi} \pi^{\mu\nu} \sigma_{\mu\nu},\label{eq:J0}\\
	\mathcal{J}^\mu &= -\tau_V V_\nu \omega^{\nu\mu} - \delta_{VV} V^\mu \theta 
	- \ell_{V\Pi} \nabla^\mu \Pi \nonumber\\
	& + \ell_{V\pi} \Delta^{\mu\nu} \nabla_\lambda \pi^\lambda{}_\nu + \tau_{V\Pi} \Pi \dot{u}^\mu - 
	\tau_{V\pi} \pi^{\mu\nu} \dot{u}_\nu \nonumber\\
	& -\lambda_{VV} V_\nu \sigma^{\mu\nu} + 
	\lambda_{V \Pi} \Pi I^\mu - \lambda_{V\pi} \pi^{\mu\nu} I_\nu,\label{eq:J1}\\
	\mathcal{J}^{\mu\nu} &= 2\tau_\pi \pi^{\langle\mu}_\lambda \omega^{\nu\rangle \lambda} - 
	\delta_{\pi\pi} \pi^{\mu\nu} \theta - 
	\tau_{\pi\pi} \pi^{\lambda\langle \mu}\sigma^{\nu\rangle}_\lambda + 
	\lambda_{\pi \Pi} \Pi \sigma^{\mu\nu} \nonumber\\
	& - \tau_{\pi V} V^{\langle \mu} \dot{u}^{\nu \rangle} + \ell_{\pi V} \nabla^{\langle \mu} V^{\nu \rangle} 
	+ \lambda_{\pi V} V^{\langle \mu} I^{\nu \rangle}.
	\label{eq:J2}
\end{align}
\end{subequations}
The transport coefficients appearing above can be computed based on the expression for the inverse of the collision matrix $\tau^{(\ell)}_{rn}$, given in Eq.~\eqref{eq:taurn}. In particular, we will need the elements on the $r = 0$ line, where $\tau^{(\ell)}_{0n} = \tau^{(\ell)}_{{\rm S};0n}$. For the terms appearing in $\mathcal{J}$, corresponding to the equation of motion for $\Pi$, we have:
\begin{subequations}\label{eq:tcoeffs0}
\begin{align}
    \ell_{\Pi V} =& -\frac{m^2}{3}\sum_{r \neq 1,2} \tau_{0r}^{(0)}\left(\mathcal{C}^{(1)}_{r-1}-\frac{G_{3r}}{D_{20}}\right)\;,\\
    \tau_{\Pi V} =& \sum_{r\neq 1,2} \frac{m^2 \tau_{0r}^{(0)}}{3} \left( r \mathcal{C}^{(1)}_{r-1}+ \frac{\partial\mathcal{C}_{r-1}^{(1)}}{\partial\ln \beta}-\frac{G_{3r}}{D_{20}}\right)\;,\\
    \delta_{\Pi\Pi} =&\sum_{r\neq 1,2} \tau_{0r}^{(0)}\Bigg[ \frac{r+2}{3} \mathcal{C}^{(0)}_r + \mathcal{H}_\alpha \frac{\partial\mathcal{C}^{(0)}_r}{\partial\alpha} + \mathcal{H}_\beta \frac{\partial\mathcal{C}_r^{(0)}}{\partial\beta}\nonumber\\
    & -\frac{m^2}{3}(r-1)\mathcal{C}^{(0)}_{r-2}-\frac{m^2}{3} \frac{G_{2r}}{D_{20}} \Bigg]\;,\\
    \lambda_{\Pi V} = &-\frac{m^2}{3}\sum_{r\neq 1,2} \tau_{0r}^{(0)}\left(\frac{\partial\mathcal{C}_{r-1}^{(1)}}{\partial\alpha}+\frac{1}{h}\frac{\partial\mathcal{C}_{r-1}^{(1)}}{\partial\beta}\right)\;,\\
    \lambda_{\Pi \pi} = &\frac{m^2}{3}\sum_{r \neq 1,2} \tau_{0r}^{(0)}\left[\frac{G_{2r}}{G_{22}}+(1-r)\mathcal{C}_{r-2}^{(2)}\right]\;. 
\end{align}
\end{subequations}
The transport coefficients appearing in $\mathcal{J}^\mu$ read:
\begin{subequations}\label{eq:tcoeffs1}
\begin{align}
 \delta_{VV} &= 
 \frac{1}{3} \sum_{r \neq 1} \tau^{(1)}_{0r} [(r+3) \mathcal{C}^{(1)}_r - m^2(r-1) \mathcal{C}^{(1)}_{r-2}]\nonumber\\
 &+ \sum_{r \neq 1} \tau^{(1)}_{0r} \left(
 \mathcal{H}_\alpha \frac{\partial \mathcal{C}^{(1)}_{r}}{\partial \alpha} + \mathcal{H}_\beta \frac{\partial \mathcal{C}^{(1)}_{r}}{\partial \beta}\right),\\
 \ell_{V\Pi} &= \sum_{r \neq 1} \tau^{(1)}_{0r} 
 \left(\frac{\beta J_{r+2,1}}{e + P} - \mathcal{C}^{(0)}_{r-1}+\frac{\mathcal{C}^{(0)}_{r+1}}{m^2}\right),\\
 \tau_{V\Pi} &= \sum_{r \neq 1} \tau^{(1)}_{0r} 
 \left(\frac{\beta J_{r+2,1}}{e + P} + 
 \frac{r+3}{m^2} \mathcal{C}^{(0)}_{r+1} - r \mathcal{C}^{(0)}_{r-1}\right)\nonumber\\
 &+ \frac{1}{m^2} \sum_{r \neq 1} \tau^{(1)}_{0r} 
 \frac{\partial}{\partial \ln \beta} (\mathcal{C}^{(0)}_{r+1} - m^2 \mathcal{C}^{(0)}_{r-1})
 ,\\
 \ell_{V\pi} &= 
 \sum_{r \neq 1} \tau^{(1)}_{0r} \left(
 \frac{\beta J_{r+2,1}}{e + P} - \mathcal{C}^{(2)}_{r-1}\right),\\
 \tau_{V\pi} &= \sum_{r \neq 1} \tau^{(1)}_{0r} \left(
 \frac{\beta J_{r+2,1}}{e + P} -
 \frac{\partial \mathcal{C}^{(2)}_{r-1}}{\partial \ln \beta} -
 r \mathcal{C}^{(2)}_{r-1}\right),\\
 \lambda_{VV} &= \frac{1}{5} \sum_{r \neq 1} \tau^{(1)}_{0r} \left[(2r+3) \mathcal{C}^{(1)}_{r} - 2m^2 (r-1) \mathcal{C}^{(1)}_{r-2}\right],\\
 \lambda_{V\Pi} &= -\sum_{r \neq 1} \frac{\tau^{(1)}_{0r}}{m^2} \left(\frac{1}{h} \frac{\partial}{\partial \beta} + \frac{\partial}{\partial\alpha} \right) (\mathcal{C}^{(0)}_{r+1} - m^2 \mathcal{C}^{(0)}_{r-1}), \\
%
 \lambda_{V\pi} &= \sum_{r \neq 1} \tau^{(1)}_{0r} 
 \left(\frac{1}{h} \frac{\partial \mathcal{C}^{(2)}_{r-1}}{\partial \beta} + \frac{\partial \mathcal{C}^{(2)}_{r-1}}{\partial \alpha}\right).
 \label{eq:DNMR_tcoeff_n}
\end{align}
\end{subequations}
Finally, the transport coefficients from $\mathcal{J}^{\mu\nu}$ are:
\begin{subequations}\label{eq:tcoeffs2}
\begin{align}
 \delta_{\pi\pi} &= \sum_r \tau_{0r}^{(2)}
 \left[\frac{r+4}{3} \mathcal{C}^{(2)}_r - \frac{m^2}{3} (r-1) \mathcal{C}^{(2)}_{r-2}\right] \nonumber\\
 & + \sum_r \tau_{0r}^{(2)} \left(\mathcal{H}_\alpha \frac{\partial \mathcal{C}^{(2)}_r}{\partial\alpha} + 
 \mathcal{H}_\beta \frac{\partial \mathcal{C}^{(2)}_r}{\partial\beta}\right),\label{eq:delta_pi_pi}\\
 \tau_{\pi\pi} &= \frac{2}{7}\sum_{r} \tau^{(2)}_{0r} \left[(2r+5) \mathcal{C}^{(2)}_r - 2m^2(r-1) \mathcal{C}^{(2)}_{r-2}\right], \label{eq:tau_pi_pi}\\ 
 \lambda_{\pi\Pi} &= \frac{2}{5m^2} \sum_{r} \tau^{(2)}_{0r} \left[-(r+4)\mathcal{C}_{r+2}^{(0)}+m^2(2r+3) \mathcal{C}^{(0)}_r \right.\nonumber\\
 &\left.- m^4 (r-1) \mathcal{C}^{(0)}_{r-2}\right], \\
 \tau_{\pi V} &= \frac{2}{5}\bigg\{ \sum_r \tau^{(2)}_{0r} \left[(r+5) \mathcal{C}^{(1)}_{r+1} - m^2 r \mathcal{C}^{(1)}_{r-1}\right] \nonumber\\
 & + \sum_{r} \tau^{(2)}_{0r} \left(
 \frac{\partial \mathcal{C}^{(1)}_{r+1}}{\partial \ln \beta} - m^2 \frac{\partial \mathcal{C}^{(1)}_{r-1}}{\partial \ln \beta}\right)\bigg\},\\
 \ell_{\pi V} &= \frac{2}{5}\sum_{r} \tau_{0r}^{(2)}\left(\mathcal{C}^{(1)}_{r+1} - m^2 \mathcal{C}^{(1)}_{r-1}\right), \\
 \lambda_{\pi V} &= \frac{2}{5}\sum_{r}\tau_{0r}^{(2)}\left(\frac{\partial \mathcal{C}^{(1)}_{r+1}}{\partial\alpha}+\frac{1}{h}\frac{\partial \mathcal{C}^{(1)}_{r+1}}{\partial\beta}\right) \nonumber\\
 & - \frac{2m^2}{5} \sum_{r} \tau_{0r}^{(2)}\left(\frac{\partial \mathcal{C}^{(1)}_{r-1}}{\partial\alpha}+\frac{1}{h}\frac{\partial \mathcal{C}^{(1)}_{r-1}}{\partial\beta}\right).
\end{align}
\end{subequations}

The above equations give the tools to extract the transport coefficients of a fluid with one conserved charge, corresponding to a given implementation of the Shakhov model. In the following subsections, we will discuss the transport coefficients arising in a fluid with no conserved charges, as well as for an ultrarelativistic ideal gas.

\subsection{Transport coefficients for the fluid without conserved charges} \label{sec:shk2:uncharged} 

The case when the fluid bears no conserved charge will be considered when discussing applications to the Bjorken flow in Sec.~\ref{sec:bjork} and requires formally to set $\alpha = 0$, $\dot{\alpha} = 0$, and $I^\mu = \nabla^\mu \alpha = 0$ [see the discussion around Eqs.~\eqref{eq:dotab_uncharged} and \eqref{eq:Hs_nocharge}]. Thus, all vector moments $\rho^\mu_r$ are of second order with respect to ${\rm Kn}$ and ${\rm Re}^{-1}$, such that the equations of second-order hydrodynamics \eqref{eq:hydro} reduce to:
\begin{subequations}\label{eq:bjorken_hydro}
\begin{align}
 \bar{\tau}_\Pi \dot{\Pi} + \Pi &= -\bar{\zeta} \theta - \bar{\delta}_{\Pi\Pi} \Pi \theta + 
 \bar{\lambda}_{\Pi \pi} \pi^{\mu\nu} \sigma_{\mu\nu},\label{eq:bjorken_Pidot}\\
 \tau_\pi \dot{\pi}^{\langle \mu \nu \rangle} + \pi^{\mu\nu} &= 
 2 \eta \sigma^{\mu\nu} + 2\tau_\pi \pi^{\langle\mu}_\lambda \omega^{\nu\rangle \lambda} - 
 \delta_{\pi\pi} \pi^{\mu\nu} \theta \nonumber\\ 
 & -  \tau_{\pi\pi} \pi^{\lambda\langle \mu}\sigma^{\nu\rangle}_\lambda + 
 \bar{\lambda}_{\pi \Pi} \Pi \sigma^{\mu\nu},
 \label{eq:bjorken_pidot}
\end{align}
\end{subequations}
where the overhead bar indicates transport coefficients which are different from the charged case discussed in the previous subsection. The transport coefficients appearing in the equation for $\Pi$ differ structurally from those reported in Sec.~\ref{sec:shk2:tcoeffs}. They can be obtained from the latter under the replacements summarized in Eq.~\eqref{eq:uncharged_repl}. Specifically, we list below the expressions for all barred transport coefficients:
\begin{subequations}
\begin{align}
 \bar{\zeta}_n &= \frac{m^2}{3} \sum_{r \neq 1,2} \tau^{(0)}_{nr} \bar{\alpha}^{(0)}_r, \quad 
%
 \bar{\tau}_\Pi = \sum_{r \neq 1,2} \tau^{(0)}_{0r} \overline{\mathcal{C}}^{(0)}_r, 
 \\
 \bar{\delta}_{\Pi\Pi} &= \sum_{r\neq 1,2} \tau_{0r}^{(0)}\Bigg[ \frac{r+2}{3} \overline{\mathcal{C}}^{(0)}_r + \frac{e + P}{J_{30}} \frac{\partial\overline{\mathcal{C}}_r^{(0)}}{\partial\beta}\nonumber\\
 & -\frac{m^2}{3}(r-1)\overline{\mathcal{C}}^{(0)}_{r-2} + \frac{m^2}{3} \frac{J_{r+1,0}}{J_{30}} \Bigg]\;,\\
 \bar{\lambda}_{\Pi \pi} &= \frac{m^2}{3}\sum_{r \neq 1,2} \tau_{0r}^{(0)}\left[(1-r)\mathcal{C}_{r-2}^{(2)} +\frac{J_{r+1,0}}{J_{30}}\right]\;,\\
 \bar{\lambda}_{\pi\Pi} &= \frac{2}{5m^2} \sum_{r} \tau^{(2)}_{0r} \left[-(r+4)\overline{\mathcal{C}}^{(0)}_{r+2}+m^2(2r+3) \overline{\mathcal{C}}^{(0)}_r\right.\nonumber\\
 &\left.- m^4 (r-1) \overline{\mathcal{C}}^{(0)}_{r-2}\right],
\end{align}
\end{subequations}
where $\overline{\mathcal{C}}^{(0)}_r = \bar{\zeta}_r / \bar{\zeta}_0$. In addition, $\eta$, $\tau_\pi$, $\delta_{\pi\pi}$, and $\tau_{\pi\pi}$ are given in Eqs.~\eqref{eq:shk_tcoeffsr_in}, \eqref{eq:rtimes}, \eqref{eq:delta_pi_pi}, and \eqref{eq:tau_pi_pi}.

\subsection{Transport coefficients for the ultrarelativistic classical gas}
\label{sec:shk2:UR}

Another limit of the system discussed in Sec.~\ref{sec:shk2:tcoeffs} is that of the ultrarelativistic classical gas, which we will refer to in Sections~\ref{sec:long} and \ref{sec:riemann}.
The equilibrium distribution for the classical gas can be obtained by setting $a = 0$ in Eq.~\eqref{eq:feq}, leading to the Maxwell-J\"uttner distribution function:
\begin{equation}
 f_{0\bk} = e^{\alpha - \beta E_\bk}.\label{eq:feq_MJ}
\end{equation}
Since $\partial f_{0\bk} / \partial \alpha = f_{0\bk}$, we have $J_{nq} = I_{nq}$, with
\begin{equation}
 I_{nq} = \frac{P \beta^{2 - n}}{2(2q+1)!!} (n+1)!\;.
 \label{eq:Inq_m0}
\end{equation}
The functions $\alpha^{(1)}_r$ and $\alpha^{(2)}_r$ appearing in Eqs.~\eqref{eq:kappa_alpha1} and \eqref{eq:eta_alpha2}, respectively, reduce to
\begin{align}
 \alpha^{(1)}_r =& \frac{P(r+2)!(1-r)}{24 \beta^{r-1}}, &
 \alpha^{(2)}_r =& \frac{P(r+4)!}{30\beta^r}.
 \label{eq:UR_alpha}
\end{align}
In particular, we find
\begin{equation}
 G_{22} = -3P^2, \quad 
 J_{20} - h J_{10} = -P, \quad 
 J_{30} - h J_{20} = 0,
\end{equation}
such that $\mathcal{H}_\alpha = 0$ and $\mathcal{H}_\beta = \beta / 3$.

For an ultrarelativistic fluid, $\Pi = 0$ by the tracelessness condition $T^\mu{}_\mu = 0$, such that the scalar moments do not play any role. The equations of second-order hydrodynamics in the IReD approach \cite{Wagner:2022ayd} read:
\begin{subequations}\label{eq:UR_hydro}
\begin{align}
 \tau_V \dot{V}^{\langle \mu \rangle} + V^\mu &= \kappa I^\mu + \mathcal{J}^\mu, \label{eq:UR_Vdot}\\ 
 \tau_\pi \dot{\pi}^{\langle \mu \nu \rangle} + \pi^{\mu\nu} &= 2 \eta \sigma^{\mu\nu} + \mathcal{J}^{\mu\nu}.
\label{eq:UR_pidot}
\end{align}
\end{subequations}
The tensors $\mathcal{J}^\mu$ and $\mathcal{J}^{\mu\nu}$ containing second-order terms of order $O({\rm Re}^{-1}  {\rm Kn})$ are given by
\begin{subequations}
\begin{align}
	\mathcal{J}^\mu &= -\tau_V V_\nu \omega^{\nu\mu} - \delta_{VV} V^\mu \theta 
	+ \ell_{V\pi} \Delta^{\mu\nu} \nabla_\lambda \pi^\lambda{}_\nu \nonumber\\
	& - 
	\tau_{V\pi} \pi^{\mu\nu} \dot{u}_\nu - \lambda_{VV} V_\nu \sigma^{\mu\nu} - \lambda_{V\pi} \pi^{\mu\nu} I_\nu,\label{J_mu}\\
	\mathcal{J}^{\mu\nu} &= 2\tau_\pi \pi^{\langle\mu}_\lambda \omega^{\nu\rangle \lambda} - 
	\delta_{\pi\pi} \pi^{\mu\nu} \theta - 
	\tau_{\pi\pi} \pi^{\lambda\langle \mu}\sigma^{\nu\rangle}_\lambda \nonumber\\
	& - \tau_{\pi V} V^{\langle \mu} \dot{u}^{\nu \rangle} + \ell_{\pi V} \nabla^{\langle \mu} V^{\nu \rangle} 
	+ \lambda_{\pi V} V^{\langle \mu} I^{\nu \rangle}.
	\label{J_munu}
\end{align}
\end{subequations}
As usual, the first-order transport coefficients $\kappa_n$ and $\eta_n$ are obtained from
Eq. \eqref{eq:first_order}, while the relaxation times $\tau_V$ and $\tau_\pi$ are computed as in Eq.~\eqref{eq:rtimes}.
Using now the relations
\begin{equation}
 \beta \frac{\partial \mathcal{C}^{(\ell)}_r}{\partial \beta} = -r \mathcal{C}^{(\ell)}_r, \qquad 
 \frac{\partial \mathcal{C}^{(\ell)}_r}{\partial \alpha} = 0,
\end{equation}
it is not difficult to establish that
\begin{gather}
 \delta_{VV} = \tau_V, \quad \tau_{V\pi} = \ell_{V\pi}, \quad
 \delta_{\pi\pi} = \frac{4}{3} \tau_\pi, \quad 
 \tau_{\pi V} = 4 \ell_{\pi V},
 \label{eq:UR_tcoeffs_fixed}
\end{gather}
as can be seen after setting $m = 0$ in Eqs.~\eqref{eq:tcoeffs1}--\eqref{eq:tcoeffs2}.
The other second-order transport coefficients reduce to
\begin{subequations} \label{eq:UR_tcoeff}
\begin{align}
 \ell_{V\pi} &= 
 \sum_{r \neq 1} \tau^{(1)}_{0r} \left(
 \frac{\beta J_{r+2,1}}{e + P} - \mathcal{C}^{(2)}_{r-1}\right),\\
 \lambda_{VV} &= \frac{1}{5} \sum_{r \neq 1} (2r+3) \tau^{(1)}_{0r}  \mathcal{C}^{(1)}_{r},\\
 \lambda_{V\pi} &= -\frac{1}{4}\sum_{r \neq 1} (r - 1) \tau^{(1)}_{0r} \mathcal{C}^{(2)}_{r-1}, \\
 \tau_{\pi\pi} &= \frac{2}{7}\sum_{r} (2r+5) \tau^{(2)}_{0r} \mathcal{C}^{(2)}_r, \\ %
 \ell_{\pi V} &= \frac{2}{5}\sum_{r}\tau_{0r}^{(2)} \mathcal{C}^{(1)}_{r+1},\\
 \lambda_{\pi V} &= -\frac{1}{10}\sum_{r} (r + 1) \tau_{0r}^{(2)} \mathcal{C}^{(1)}_{r+1}\;.
\end{align}
\end{subequations}

In evaluating the above expressions, it is instructive to consider first that the weight function $\tilde{\omega}^{(\ell)}$ for the basis shifted by $s_\ell$, introduced in Eq.~\eqref{eq:omega_def}, can be written in terms of the one corresponding to $s_\ell = 0$, denoted without the overhead tilde, $\omega^{(\ell)}$, as
\begin{equation}
 \tilde{\omega}^{(\ell)} = \frac{(2\ell + 1 - 2s_\ell)!!}{(2\ell+1)!!} \frac{J_{2\ell-2s_\ell,\ell-s_\ell}}{J_{2\ell-2s_\ell,\ell}} \omega^{(\ell - s_\ell)}.
\end{equation}
Enforcing $\int \d K \tilde{\omega}^{(\ell)} \widetilde{P}^{(\ell)}_{\bk m} \widetilde{P}^{(\ell)}_{\bk n} = \delta_{mn}$ shows that the polynomials in the shifted basis ($s_\ell > 0$) can be written in terms of those in the non-shifted basis ($s_\ell = 0$) as
\begin{equation}
 \widetilde{P}^{(\ell)}_{\bk m} = \sqrt{\frac{(2\ell + 1)!!}{(2\ell+1 - 2s_\ell)!!}} \sqrt{\frac{J_{2\ell-2s_\ell,\ell}}{J_{2\ell-2s_\ell,\ell-s_\ell}}} P^{(\ell - s_\ell)}_{\bk m}.
\end{equation}
Assuming now that $N_\ell + s_\ell$ in the shifted basis is equal to $N_{\ell - s_\ell}$ in the non-shifted basis, it is possible to express the $\widetilde{\mathcal{H}}_{\bk n}$ polynomials in terms of the non-shifted ones via
\begin{multline}
 \widetilde{\mathcal{H}}^{(\ell)}_{\bk n}(N_\ell, s_\ell) = 
 \frac{(-1)^{s_\ell} (\ell-s_\ell)! (2\ell + 1)!!}{\ell! (2\ell + 1 - 2s_\ell)!!} \\\times
 \mathcal{H}^{(\ell-s_\ell)}_{\bk n}(N_\ell + s_\ell, 0).
\end{multline}
The function $\widetilde{\mathcal{F}}^{(\ell)}_{rn}$, defined via Eq.~\eqref{eq:F_def}, can be shown to satisfy
\begin{equation}
 \widetilde{\mathcal{F}}^{(\ell)}_{rn} = \mathcal{F}^{(\ell - s_\ell)}_{rn}.
\end{equation}
Using the exact expression derived in Eq.~(142) of Ref.~\cite{Ambrus:2022vif}, we find
\begin{multline}
 \widetilde{\mathcal{F}}^{(\ell)}_{rn} = \frac{\beta^{r+n}(-1)^n}{(r + n)(r-1)! n!} \\\times 
 \frac{(2\ell - 2s_\ell + 1 - r)! (N_\ell + s_\ell + r)!}{(2\ell - 2s_\ell + 1 + n)! (N_\ell + s_\ell - n)!},
 \label{eq:UR_Flrn}
\end{multline}
where the results corresponding to $r > 2 \ell - 2s_\ell + 1$ diverge due to the infrared behaviour of the integrand in $\widetilde{\mathcal{F}}^{(\ell)}_{rn}$. When $0 \le r \le N_\ell + s_\ell$, it can be seen that $\widetilde{\mathcal{F}}^{(\ell)}_{-r,n} = \delta_{rn}$. For $r = -N_\ell -s_\ell - q$ and $q > 0$, it can be seen that 
\begin{multline}
    \widetilde{\mathcal{F}}^{(\ell)}_{-N_\ell-s_\ell-q,n} = 
    \frac{(-\beta)^{n - N_\ell - s_\ell}}{\beta^q n! (q-1)!} \\\times 
    \frac{(N_\ell + s_\ell + q)! (N_\ell + 2\ell -s_\ell + q  + 1)!}{(2\ell - 2s_\ell + n + 1)! (N_\ell + s_\ell - n)! (N_\ell + s_\ell + q- n)},
    \label{eq:F_m0}
\end{multline}
where the following relation was employed:
\begin{equation}
 \frac{(-q)!}{(-N_\ell -s_\ell -q - 1)!} = \frac{(-1)^{N_\ell + s_\ell + 1}}{(q-1)!} (N_\ell + s_\ell + q)!.
\end{equation}

The knowledge of the functions $\widetilde{F}^{(\ell)}_{rn}$ allows one to calculate the transport coefficients $\kappa_r$ and $\eta_r$ for indices $r$ lying outside the basis, cf.~Eqs.~\eqref{eq:shk_kappar} and \eqref{eq:shk_etar}. In particular, we will compute the terms $\sum_{n = -s_\ell}^{N_\ell} \widetilde{\mathcal{F}}^{(\ell)}_{-r-s_1, n+s_1} \alpha^{(\ell)}_n$ exactly, as follows. In the case when $r < -s_\ell$, the first index $-r-s_\ell$ of $\widetilde{\mathcal{F}}^{(\ell)}_{-r-s_\ell, n+s_\ell}$ is larger than $0$, hence we can use the representation \eqref{eq:UR_Flrn}, such that:
\begin{align}
 \sum_{n = -s_1}^{N_1} \widetilde{\mathcal{F}}^{(1)}_{-r-s_1, n+s_1} \alpha^{(1)}_n &= 
 \frac{(r + 3 - s_1)! (N_1 - r)! \alpha^{(1)}_r }{(r+2)!(1-r)(-r-s_1-1)!} \nonumber\\
  & \times \frac{1}{ (N_1 + s_1)!} S^{(1)}_{N_1+s_1,s_1}, \nonumber\\
 \sum_{n = -s_2}^{N_2} \widetilde{\mathcal{F}}^{(2)}_{-r-s_2, n+s_2} \alpha^{(2)}_n &= 
 \frac{(r + 5 - s_2)! (N_2 - r)! \alpha^{(2)}_r }{(r+4)!(-r-s_2-1)!} \nonumber\\
 & \times \frac{1}{ (N_2 + s_2)!} S^{(2)}_{N_2+s_2,s_2},
\end{align}
where the functions $\alpha^{(\ell)}_r$ were taken from Eq.~\eqref{eq:UR_alpha}. We also introduced the following functions:
\begin{align}
 S^{(1)}_{N,s_1} &= \sum_{n = 0}^N \binom{N}{n} \frac{(-1)^n (n + 2 - s_1)! (1 + s_1 - n)}{(n - r - s_1)(n + 3 - 2s_1)!},\nonumber\\
 S^{(2)}_{N,s_2} &= \sum_{n = 0}^N \binom{N}{n} \frac{(-1)^n (n + 4 - s_2)!}{(n-r-s_2) (n + 5 - 2s_2)!}.
\end{align}
In order to simplify the calculation, we replace $s_1$ and $s_2$ by the allowed values, $0 \le s_1 \le 1$ and $0 \le s_2 \le 2$:
\begin{align}
 S^{(1)}_{N,0} &= \sum_{n = 0}^N \binom{N}{n} \frac{(-1)^n}{r+3} \left(\frac{1-r}{n-r} - \frac{4}{n+3}\right), \nonumber\\
 S^{(1)}_{N,1} &= \sum_{n = 0}^N \binom{N}{n} (-1)^n \left(\frac{1-r}{n-r-1} - 1\right), \nonumber\\
 S^{(2)}_{N,0} &= \sum_{n = 0}^N \binom{N}{n} \frac{(-1)^n}{r+5} \left(\frac{1}{n-r} - \frac{1}{n + 5}\right),\nonumber\\
 S^{(2)}_{N,1} &= \sum_{n = 0}^N \binom{N}{n} \frac{(-1)^n}{n - r - 1},\nonumber\\
 S^{(2)}_{N,2} &= \sum_{n = 0}^N \binom{N}{n} (-1)^n\left(\frac{r+4}{n-r-2} + 1\right).
\end{align}
The factors $1 / (n + \alpha + 1)$ appearing inside the summation over $n$ can be replaced using the formula
\begin{equation}
 \frac{1}{n + \alpha + 1} = 
 \begin{cases}
  {\displaystyle \int_0^1 \d x\, x^{n + \alpha}}, & n + \alpha > -1, \\
  {\displaystyle -\int_0^1 \d x\, x^{-n-\alpha-2}}, & n + \alpha < -1.
 \end{cases} 
\end{equation}
Then, the sum over $n$ can be performed using the binomial theorem,
\begin{align}
 \sum_{n = 0}^N \binom{N}{n} (-1)^n x^{n + \alpha} &= x^\alpha (1 - x)^N, \nonumber\\
 \sum_{n = 0}^N \binom{N}{n} (-1)^n x^{-n - \alpha - 2} &= (-1)^N x^{-\alpha - 2 - N} (1 - x)^N.
\end{align}
The limit $x = 1$ shows that for $N > 0$, we always have $\sum_{n = 0}^N \binom{N}{n} (-1)^n = 0$. This is always the case when $s_\ell > 0$, since $N = N_\ell + s_\ell > 0$. 
Finally, the integration with respect to the auxiliary variable $x$ can be performed in terms of the Beta function \cite{olver2010nist},
\begin{equation}
 B(a,b) = \int_0^1 \d x\, x^{a-1} (1 - x)^{b-1} = \frac{\Gamma(a) \Gamma(b)}{\Gamma(a + b)},
\end{equation}
where $\Gamma(n + 1) = n!$ is the Gamma (factorial) function \cite{olver2010nist}. 
Without further ado, we find
\begin{align}
 S^{(1)}_{N,0} &= \frac{N!}{r+3} \left[\frac{(1-r)(-r-1)!}{(N-r)!} - \frac{8}{(N+3)!}\right],\nonumber\\
 S^{(1)}_{N,1} &= \frac{(1-r)(-r-2)!N!}{(N - r - 1)!}, \nonumber\\
 S^{(2)}_{N,0} &= \frac{N!}{r+5}\left[\frac{(-r-1)!}{(N-r)!} - \frac{24}{(N+5)!}\right],\nonumber\\
 S^{(2)}_{N,1} &= \frac{(-r-2)!N!}{(N-1-r)!},\nonumber\\
 S^{(2)}_{N,2} &= \frac{(r+4)(-r-3)!N!}{(N - r -2)!}.
\end{align}
We thus find for the unshifted bases ($s_1 = s_2 = 0$):
\begin{align}
 \left.\sum_{n = -s_1}^{N_1} \widetilde{\mathcal{F}}^{(1)}_{-r-s_1, n + s_1} \alpha^{(1)}_n\right\rfloor_{s_1 = 0} \hspace{-10pt} &= 
 \alpha^{(1)}_r \nonumber\\ 
 &\hspace{-50pt} \times \left[1 - \frac{8 (N_1 - r)!}{(1-r)(-r-1)!(N_1 + 3)!}\right],\nonumber\\
 \left.\sum_{n = -s_2}^{N_2} \widetilde{\mathcal{F}}^{(2)}_{-r-s_2, n + s_2} \alpha^{(2)}_n\right\rfloor_{s_2 = 0} \hspace{-10pt} &= 
 \alpha^{(2)}_r \nonumber\\
 &\hspace{-50pt} \times \left[1 - \frac{24 (N_2 - r)!}{(-r-1)!(N_2 + 5)!}\right].
\end{align}
The second term in the square brackets provides a non-vanishing correction depending on both $r$ and $N_2$ to the intuitive result, $\alpha^{(\ell)}_r$. 
These results exactly agree with those in Eqs.~(161)--(162) of Ref.~\cite{Ambrus:2022vif}. 
Remarkably, for any non-vanishing shift $s_\ell \neq 0$, the second term completely disapears:
\begin{equation}
 \left.\sum_{n = -s_\ell}^{N_\ell} \widetilde{\mathcal{F}}^{(\ell)}_{-r-s_\ell, n + s_\ell} \alpha^{(\ell)}_n\right\rfloor_{s_\ell \neq 0} \hspace{-10pt} = \alpha^{(\ell)}_r.
\end{equation}

The above calculation can be repeated for the case when $r > N_\ell$, when the representation \eqref{eq:F_m0} of $\widetilde{\mathcal{F}}^{(\ell)}_{rn}$ is appropriate. In this case, we have 
\begin{align}
 \sum_{n = -s_1}^{N_1} \widetilde{\mathcal{F}}^{(1)}_{-r-s_1, n+s_1} \alpha^{(1)}_n &= 
 \frac{(r + 3 - s_1)! (r + s_1)! \alpha^{(1)}_r }{(r+2)!(1-r)(r-1 - N_1)!} \nonumber\\
  & \times \frac{1}{ (N_1 + s_1)!} S^{(1)\prime}_{N_1+s_1,s_1}, \nonumber\\
 \sum_{n = -s_2}^{N_2} \widetilde{\mathcal{F}}^{(2)}_{-r-s_2, n+s_2} \alpha^{(2)}_n &= 
 \frac{(r + 5 - s_2)! (r + s_2)! \alpha^{(2)}_r }{(r+4)!(r-1 - N_2)!} \nonumber\\
 & \times \frac{1}{ (N_2 + s_2)!} S^{(2)\prime}_{N_2+s_2,s_2},
\end{align}
where
\begin{align}
 S^{(1)\prime}_{N,s_1} &= \sum_{n = 0}^N \binom{N}{n} \frac{(-1)^{n+N} (n + 2 - s_1)! (1 + s_1 - n)}{(r + s_1 - n)(n + 3 - 2s_1)!},\nonumber\\
 S^{(2)\prime}_{N,s_2} &= \sum_{n = 0}^N \binom{N}{n} \frac{(-1)^{n+N} (n + 4 - s_2)!}{(r+s_2 - n) (n + 5 - 2s_2)!}.
\end{align}
Repeating the same steps as above, we arrive at similar results:
\begin{align}
 \left.\sum_{n = -s_1}^{N_1} \widetilde{\mathcal{F}}^{(1)}_{-r-s_1, n + s_1} \alpha^{(1)}_n\right\rfloor_{s_1 = 0} \hspace{-10pt} &= 
 \alpha^{(1)}_r \nonumber\\ 
 &\hspace{-50pt} \times \left[1 + \frac{8 (-1)^{N_1} r!}{(1-r)(r-1 - N_1)!(N_1 + 3)!}\right],\nonumber\\
 \left.\sum_{n = -s_2}^{N_2} \widetilde{\mathcal{F}}^{(2)}_{-r-s_2, n + s_2} \alpha^{(2)}_n\right\rfloor_{s_2 = 0} \hspace{-10pt} &= 
 \alpha^{(2)}_r \nonumber\\
 &\hspace{-50pt} \times \left[1 + \frac{24 (-1)^{N_2} r!}{(r-N_2 - 1)!(N_2 + 5)!}\right], \nonumber\\
 \left.\sum_{n = -s_\ell}^{N_\ell} \widetilde{\mathcal{F}}^{(\ell)}_{-r-s_\ell, n + s_\ell} \alpha^{(\ell)}_n\right\rfloor_{s_\ell \neq 0} \hspace{-10pt} &= \alpha^{(\ell)}_r.
\end{align}

We thus conclude that for any non-vanishing shift of the basis, i.e. $s_\ell > 0$, the transport coefficients $\kappa_r$ and $\eta_r$ are obtained from Eqs.~\eqref{eq:shk_kappar} and \eqref{eq:shk_etar} as follows:
\begin{align}
 \kappa_r &= \sum_{n = -s_1,\neq 1}^{N_1} \hspace{-6pt} \widetilde{\mathcal{F}}^{(1)}_{-r-s_1, n+s_1} \kappa_n,\label{eq:UR_kappar_shift}\\
 \eta_r &= \sum_{n = -s_2}^{N_2} \widetilde{\mathcal{F}}^{(2)}_{-r-s_2, n+s_2} \eta_n,
 \label{eq:UR_etar_shift}
\end{align}
where we keep in mind that for $-s_\ell \le r \le N_\ell$, the functions $\widetilde{\mathcal{F}}^{(\ell)}_{-r-s_\ell, n + s_\ell} = \delta_{rn}$ reduce to the Kronecker symbol.
In the case of the unshifted basis, $s_1 = 0$ or $s_2 = 0$, we have for $r < 0$:
\begin{align}
 \kappa_{r < 0} &= \frac{8 (N_1 - r)! \tau_R \alpha^{(1)}_r}{(1-r)(-r-1)!(N_1 + 3)!} + 
 \sum_{n = 0,\neq 1}^{N_1} \widetilde{\mathcal{F}}^{(1)}_{-r,n} \kappa_n, \nonumber\\
 \eta_{r < 0} &= \frac{24 (N_2 - r)! \tau_R \alpha^{(2)}_r}{(-r-1)!(N_2 + 5)!} + 
 \sum_{n = 0}^{N_2} \widetilde{\mathcal{F}}^{(2)}_{-r,n} \eta_n.
\end{align}
In the case when $r > N_\ell$, we have:
\begin{align}
 \kappa_{r > N_1} &= -\frac{8 (-1)^{N_1} r! \tau_R \alpha^{(1)}_r}{(1-r)(r-1-N_1)!(N_1 + 3)!} + \hspace{-5pt}
 \sum_{n = 0,\neq 1}^{N_1} \hspace{-5pt} \widetilde{\mathcal{F}}^{(1)}_{-r,n} \kappa_n, \nonumber\\
 \eta_{r > N_2} &= -\frac{24 (-1)^{N_2} r! \tau_R \alpha^{(2)}_r}{(r-1-N_2)!(N_2 + 5)!} + 
 \sum_{n = 0}^{N_2} \widetilde{\mathcal{F}}^{(2)}_{-r,n} \eta_n.
\end{align}

\subsection{Degrees of freedom in the extended Shakhov model}\label{sec:shk2:dofs}

Let us consider a Shakhov model employing the expansion orders $(N_0, N_1, N_2)$, together with the shifts $(s_0, s_1, s_2)$. In order to enforce particle number and energy-momentum conservation, $N_0 \ge 2$ and $N_1 \ge 1$. Since the $r = 1,2$ and $r = 1$ lines of the scalar $\mathcal{A}^{(0)}_{rn}$ and vector $\mathcal{A}^{(1)}_{rn}$ matrices are ignored, the resulting Shakhov matrices have the following effective sizes:
\begin{align}
 \mathcal{A}^{(0)}:& & (N_0 + s_0 - 1) \times (N_0 + s_0 - 1), \nonumber\\ 
 \mathcal{A}^{(1)}:& & (N_1 + s_1) \times (N_1 + s_1), \nonumber\\ 
 \mathcal{A}^{(2)}:& & (N_2 + s_2 + 1) \times (N_2 + s_2 + 1).
\end{align}
Generically, the relevant size of the matrix $\mathcal{A}^{(\ell)}$ is $(N_\ell + s_\ell + \ell - 1)^2$.
For the purpose of fixing the transport coefficients of second-order fluid dynamics, we remark that Eqs.~\eqref{eq:tcoeffs0}--\eqref{eq:tcoeffs2} show that only the elements on the zeroth line of the inverse collision matrices, $\tau^{(\ell)}_{0n}$, and the coefficients $\mathcal{C}^{(\ell)}_r$ representing the ratios of first-order transport coefficients play a role. As can be seen from Eqs.~\eqref{eq:shk_tcoeffsr_out}, the transport coefficients $\zeta_r$, $\kappa_r$ and $\eta_r$ corresponding to indices $r$ lying outside the range $[-s_\ell, N_\ell]$ can be expressed completely in terms of those with indices satisfying $-s_\ell \le r \le N_\ell$. Skipping those corresponding to $r = 0,1$ and $r = 0$ for the scalar and vector sectors, we have $\sum_\ell(N_\ell + s_\ell + \ell - 1)$ first-order transport coefficients that can be independently fixed. As shown in Eq.~\eqref{eq:shk_tcoeffsr_in}, the first-order transport coefficients corresponding to $r = 0$ are fixed solely by the zeroth line of the corresponding inverse collision matrix, $\tau^{(\ell)}_{0n}$ and therefore do not represent additional independent degrees of freedom. Thus, the $\mathcal{C}^{(\ell)}_r$ coefficients provide an extra $\sum_{\ell} (N_\ell + s_\ell + \ell - 2)$ independent parameters. We conclude that the total number of relevant degrees of freedom of the extended Shakhov model is
\begin{equation}
 \text{general:} \quad \sum_{\ell = 0}^2 [2(N_\ell + s_\ell + \ell) - 3] = 2 \sum_{\ell = 0}^2 (N_\ell + s_\ell) - 3.
 \label{eq:dofs_general}
\end{equation}
In the case when there is no conserved charge, the vector moments are not relevant, such that one can safely skip the $\ell = 1$ case and the total number of degrees of freedom reads
\begin{equation}
 \text{No conserved charge:} \quad 2(N_0 + N_2 + s_0 + s_2) - 2.
 \label{eq:dofs_bjork}
\end{equation}
In the case of a gas of ultrarelativistic particles with a conserved charge, the total number of degrees of freedom becomes
\begin{equation}
 \text{UR gas:} \quad 2(N_1 + N_2 + s_1 + s_2).
 \label{eq:dofs_UR}
\end{equation}
We will denote in Sections~\ref{sec:long}, \ref{sec:riemann}, and Appendix~\ref{app:shk_higher} such Shakhov models for ultrarelativistic particles using the quartet $(N_1, N_2, s_1, s_2)$.

\section{Shear-bulk coupling: Bjorken flow}\label{sec:bjork}

We now consider one of the standard problems for heavy-ion collisions: the Bjorken flow. In Subsec.~\ref{sec:bjork:intro}, we discuss the equations of motion of second-order hydrodynamics and of kinetic theory in the Bjorken flow setup. In Subsections~\ref{sec:bjork:shk} and \ref{sec:bjork:shk_coll}, we discuss the Shakhov model that allows the cross-coupling coefficient $\bar{\lambda}_{\Pi \pi}$ to be controlled. The capabilities of the model are demonstrated in Subsec.~\ref{sec:bjork:res}. 

While in the main text, we employed an overbar for the transport coefficients computed in the absence of a conserved charge in order to avoid confusion with similar coefficients for a fluid with conserved charges, within this section we will omit the bar for notational convenience, keeping in mind that all transport coefficients correspond to the case without a conserved charge.

\subsection{Problem description} \label{sec:bjork:intro}

The Bjorken flow represents an idealization of the dynamics of the quark-gluon plasma created in a heavy-ion collision, based on the experimental observation that the system properties are independent of the space-time rapidity $\eta = {\rm artanh}(z / t)$ around mid-rapidity (when $\eta \simeq 0$). This property translates into the requirement of invariance with respect to Lorentz boosts along the longitudinal (beam) direction, which greatly restricts the possible structure of four-tensors. Ignoring the dynamics in the transverse plane, the velocity field is uniquely identified as
\begin{equation}
 u^\mu \partial_\mu = \frac{t}{\tau} \partial_t + \frac{z}{\tau} \partial_z = \partial_\tau,
\end{equation}
where $\tau = \sqrt{t^2 - z^2}$ is the Bjorken time. With respect to the Bjorken coordinates $(\tau, x, y, \eta)$, the stress-energy tensor becomes diagonal, $T^{\mu\nu} = {\rm diag}(e, P_T, P_T, P_L)$, where the transverse and longitudinal pressures can be related to the bulk viscous pressure $\Pi$ and the shear stress tensor coefficient $\pi_d$ via 
\begin{equation}
 P_T = P + \Pi - \frac{\pi_d}{2}, \quad 
 P_L = P + \Pi + \pi_d.
\end{equation}

The conservation equation for $T^{\mu\nu}$ reduces to
\begin{subequations}\label{eq:bjork_hydro}
\begin{equation}
 \tau \frac{\d e}{\d\tau} + e +P_L = 0.
 \label{eq:bjork_e}
\end{equation}
Taking into account that $\theta = 1/\tau$ and $\sigma^{\mu\nu} = {\rm diag}(0, 1/3\tau, 1/3\tau, -2/3\tau^3)$, the equations for $\Pi$ and $\pi_d$ read
\begin{align}
 \tau_\Pi \frac{\d \Pi}{\d\tau} + \Pi &= -\frac{1}{\tau}\left(\zeta + \delta_{\Pi \Pi} \Pi + \lambda_{\Pi \pi} \pi_d\right)\;,\\
 \tau_\pi \frac{\d \pi_d}{\d\tau} + \pi_d &= -\frac{1}{\tau} \left[\frac{4\eta}{3} + \left(\delta_{\pi\pi} + \frac{\tau_{\pi\pi}}{3}\right) \pi_d + \frac{2\lambda_{\pi\Pi}}{3} \Pi \right]. 
\end{align}
\end{subequations}
In the above equations, we can identify $9$ transport coefficients, out of which two appear in the combination $\lambda = \delta_{\pi\pi} + \tau_{\pi\pi} / 3$. A Shakhov-like model that allows all of these coefficients to be controlled should provide $8$ free parameters. According to Eq.~\eqref{eq:dofs_bjork}, this requires $(N_0 + s_0 -2) + (N_2 + s_2) = 3$, which is achievable employing, e.g., one $3 \times 3$ matrix and another $2 \times 2$ matrix for the scalar and the tensor sectors, respectively. 

In this section, we will focus only on the first-order transport coefficients $\zeta$ and $\eta$ and the cross-coupling coefficient  $\lambda_{\Pi\pi}$, for which we can use $(N_0, N_2, s_0, s_2) = (2,0,0,2)$. We choose $s_2 = 2$ instead of $s_2 = 1$ because in the Bjorken flow, the moments with even energy index and those with odd energy index are decoupled. Since $T^{\mu\nu}$ and its evolution can be characterized exclusively in terms of even moments, we will ignore odd ones in what follows.

\subsection{Shakhov matrices for the Bjorken flow}\label{sec:bjork:shk}

We seek to achieve
\begin{equation}
 \frac{\lambda_{\Pi \pi}}{\tau_\Pi} = A \frac{\lambda^R_{\Pi \pi}}{\tau_R},\quad 
 \eta = H \eta_R,\quad 
 \zeta = \zeta_R,
\end{equation}
where $\lambda^R_{\Pi \pi}$ corresponds to the equivalent Anderson-Witting model with relaxation time $\tau_R$, 
\begin{align}
 \lambda^R_{\Pi \pi} &= \frac{m^2 \tau_R}{3} \left(\mathcal{R}^{(2)}_{-2} + \frac{J_{10}}{J_{30}}\right)\;,
\end{align}
while $\eta_R = \tau_R \alpha^{(2)}_0$ and $\zeta_R = (m^2/3) \tau_R \alpha_0^{(0)}$. In the above, we introduced the notation 
\begin{equation}
 \mathcal{R}^{(\ell)}_{r} = \frac{\alpha^{(\ell)}_r}{\alpha^{(\ell)}_0}.
\end{equation}

Our strategy is to shift the basis for the tensor moments down by two units, while ignoring the contributions from the moments of energy-rank $-1$. In other words, we set
\begin{equation}
 N_0 = 2, \quad N_2 = 0, \quad
 s_0 = 0, \quad s_2 = 2.
\end{equation}
The relevant submatrices and their inverses then read
\begin{equation}
 \mathcal{A}^{(0)}_{{\rm S};rn} = 
  \mathcal{A}^{(0)}_{{\rm S};0,0}
 \;, \qquad
 \tau^{(0)}_{{\rm S};rn} =
  \tau^{(0)}_{{\rm S};0,0}\;,
\end{equation}
and
\begin{align}
 \mathcal{A}^{(2)}_{{\rm S};rn} =& 
 \begin{pmatrix}
  \mathcal{A}^{(2)}_{{\rm S};-2,-2} & 
  \mathcal{A}^{(2)}_{{\rm S};-2,0} \\
  \mathcal{A}^{(2)}_{{\rm S};0,-2} & 
  \mathcal{A}^{(2)}_{{\rm S};0,0}
 \end{pmatrix}, \nonumber\\
 \tau^{(2)}_{{\rm S};rn} =& 
 \begin{pmatrix}
  \tau^{(2)}_{{\rm S};-2,-2} & 
  \tau^{(2)}_{{\rm S};-2,0} \\
  \tau^{(2)}_{{\rm S};0,-2} & 
  \tau^{(2)}_{{\rm S};0,0}
 \end{pmatrix}\;.
\end{align}
Note that we did not include the rows and columns corresponding to $r=1,2$ in the matrices $\mathcal{A}^{(0)}_{\rm S}$ and $\tau^{(0)}_{\rm S}$, as these do not enter the transport coefficients.

The first-order transport coefficients are given by
\begin{subequations}
\begin{align}
 \zeta_0 &= \frac{m^2}{3} \tau^{(0)}_{{\rm S};0,0} \alpha_{0}^{(0)}\;,\label{eq:bjork_zeta}\\
\eta_0 &= \tau^{(2)}_{{\rm S};0,-2}\alpha_{-2}^{(2)}+\tau^{(2)}_{{\rm S};0,0}\alpha_{0}^{(2)}\;,\label{eq:bjork_eta}\\
\eta_{-2} &= \tau^{(2)}_{{\rm S};-2,-2}\alpha_{-2}^{(2)}+\tau^{(2)}_{{\rm S};-2,0}\alpha_{0}^{(2)}\;,\label{eq:bjork_etam2}
\end{align}
\end{subequations}
while the relaxation times read
\begin{align}
    \tau_\Pi&= \tau^{(0)}_{{\rm S};0,0}\;,&
    \tau_\pi&= \tau^{(2)}_{{\rm S};0,-2} \mathcal{C}^{(2)}_{-2}+\tau^{(2)}_{{\rm S};0,0}\label{eq:tau_pi_bjorken}\;.
\end{align}
For simplicity, we set $\tau_\Pi = \tau^{(0)}_{{\rm S};0,0} = \tau_R$.
From the bulk-shear coupling
\begin{equation}
    \lambda_{\Pi\pi}= \frac{m^2 \tau_\Pi}{3} \left( \mathcal{C}_{-2}^{(2)}+\frac{J_{10}}{J_{30}}    \right)
\end{equation}
we obtain the coefficient $\mathcal{C}^{(2)}_{-2}$ as
\begin{align}
    \mathcal{C}^{(2)}_{-2} &= \frac{3}{m^2} \frac{\lambda_{\Pi\pi}}{\tau_\Pi}-\frac{J_{10}}{J_{30}} \nonumber\\
    &= A \mathcal{R}^{(2)}_{-2} + (A - 1) \frac{J_{10}}{J_{30}}\;.
    \label{eq:bjork_C2m2}
\end{align}

For simplicity, we set $\tau^{(2)}_{{\rm S};0,-2}=0$. From Eq.~\eqref{eq:tau_pi_bjorken}, we find $\tau^{(2)}_{{\rm S};0,0} = \tau_\pi$. Substituting the above into Eq.~\eqref{eq:bjork_eta}, we find $\tau_\pi = H \tau_R$. Summarizing, we have
\begin{equation}
 \tau^{(2)}_{{\rm S};0,-2} = 0, \quad 
 \tau^{(2)}_{{\rm S};0,0} = \tau_\pi = H \tau_R.
\end{equation}
Noting that, when $\tau^{(0)}_{{\rm S};0,0} = \tau_\Pi = \tau_R$, we have $\mathcal{C}^{(0)}_r = \mathcal{R}^{(0)}_r = \alpha^{(0)}_r / \alpha^{(0)}_0$,
the other second-order transport coefficients become
\begin{align}
    \lambda_{\pi\Pi} &= \frac25 H \tau_R \left(3+m^2 \mathcal{R}_{-2}^{(0)}\right)\;,\nonumber\\
      \delta_{\pi\pi} + \frac{\tau_{\pi\pi}}{3} &= \frac{H\tau_R }{21}\left\{38 + 11 m^2 \left[A \mathcal{R}^{(2)}_{-2} + (A - 1) \frac{J_{10}}{J_{30}}\right]\right\}\;.
\end{align}

From Eq.~\eqref{eq:bjork_C2m2}, the entries in the $r = -2$ line of $\tau^{(2)}_{{\rm S};rn}$ matrix are constrained by
\begin{equation}
    \tau^{(2)}_{{\rm S};-2,-2} \mathcal{R}_{-2}^{(2)}+\tau^{(2)}_{{\rm S};-2,0} = \tau_R H \left[A \mathcal{R}_{-2}^{(2)}+(A-1)\frac{J_{10}}{J_{30}}\right]\;.
\end{equation}
For simplicity, we demand that $\tau^{(2)}_{{\rm S};-2,-2} =\tau_\pi= H\tau_R$, such that the last unknown entry $\tau^{(2)}_{{\rm S};-2,0}$ becomes
\begin{equation} 
 \tau^{(2)}_{{\rm S};-2,0} = \tau_R H(A - 1)\left(\mathcal{R}_{-2}^{(2)}+  \frac{J_{10}}{J_{30}}\right)\;.
\end{equation}
The resulting Shakhov inverse matrix thus reads 
\begin{equation}
 \tau^{(2)}_{\rm S} = \tau_R H 
 \begin{pmatrix}
  1 & (A - 1) \left(\mathcal{R}_{-2}^{(2)}+  \frac{J_{10}}{J_{30}}\right) \\
  0 & 1
 \end{pmatrix}.
\end{equation}
Consequently, the Shakhov collision matrix $\mathcal{A}^{(2)}_{{\rm S};rn}$ is given by
\begin{equation}
 \mathcal{A}^{(2)}_{\rm S} = \frac{1}{\tau_R H} 
 \begin{pmatrix}
     1 & (1 - A) \left(\mathcal{R}_{-2}^{(2)}+  \frac{J_{10}}{J_{30}}\right)  \\
     0 & 1
 \end{pmatrix}\;.\label{eq:bjorken_A2}
\end{equation}

\begin{figure*}
\begin{tabular}{ccc}
 \includegraphics[width=.3\linewidth]{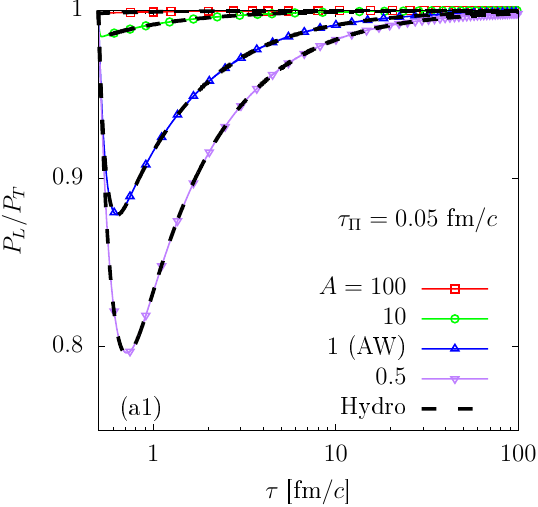} & 
 \includegraphics[width=.3\linewidth]{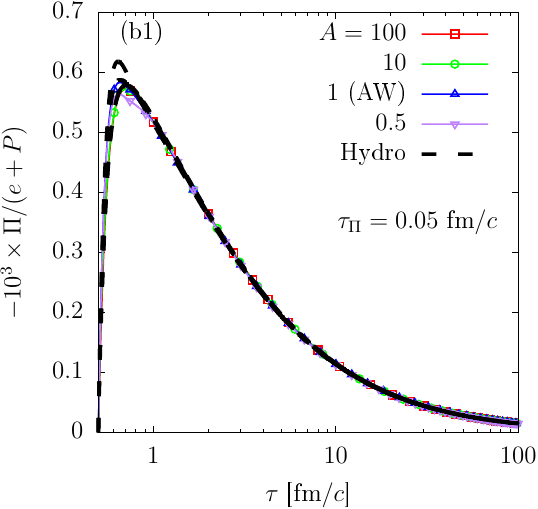} & 
 \includegraphics[width=.3\linewidth]{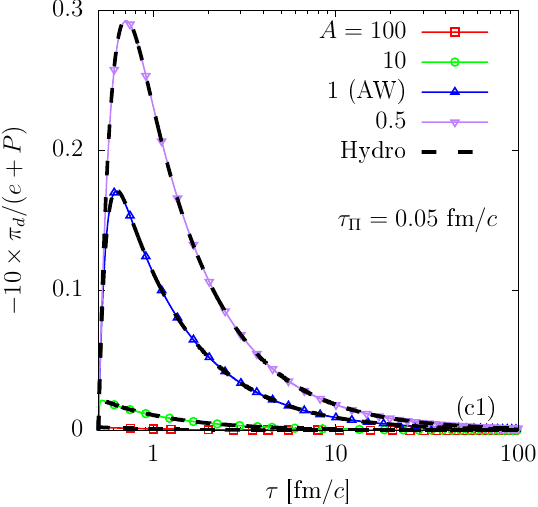} \\
 \includegraphics[width=.3\linewidth]{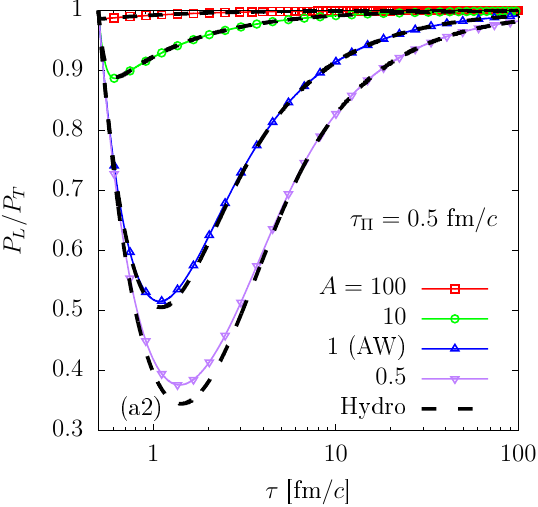} & 
 \includegraphics[width=.3\linewidth]{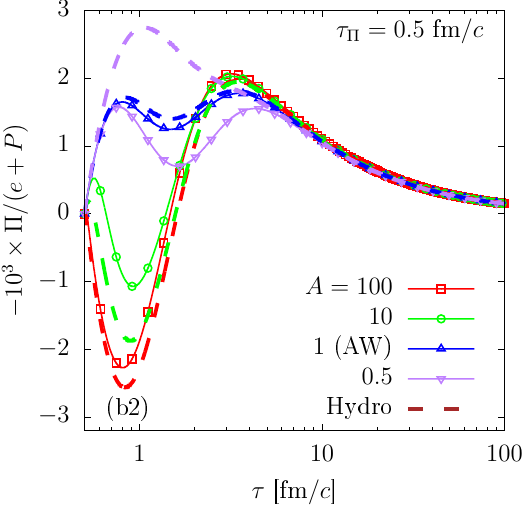} & 
 \includegraphics[width=.3\linewidth]{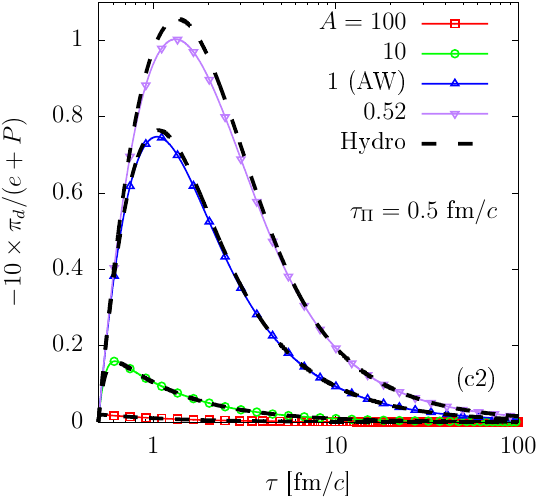} \\
\end{tabular}
\caption{
The ratios
(left) $P_L/P_T$, (middle) $-\Pi / (e + P)$, and (right) $-\pi_d / (e + P)$ in the case of Bjorken flow, obtained using the Shakhov model (solid lines with symbols) and second-order hydrodynamics (dashed lines) for (top) $\tau_\Pi = 0.05$ fm and (bottom) $\tau_\Pi = 0.5$ fm.
In all cases, the parameters of the Shakhov model are taken such that $H A = 1$. The case $A = 1$ corresponds to the Anderson-Witting model. The hydrodynamics results are shown in black in all instances, except for panel (b2), where the colour of the Hydro curves follows that of the kinetic theory curves.
\label{fig:bjork_AH}
}
\end{figure*}

\subsection{The Shakhov collision term}\label{sec:bjork:shk_coll}

\begin{figure*}
\begin{tabular}{ccc}
 \includegraphics[width=.3\linewidth]{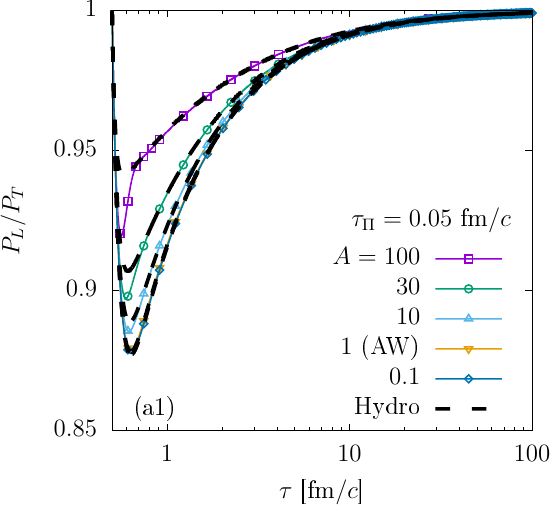} & 
 \includegraphics[width=.3\linewidth]{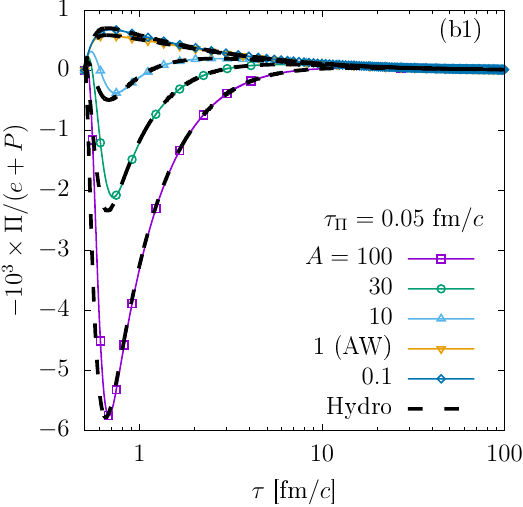} & 
 \includegraphics[width=.3\linewidth]{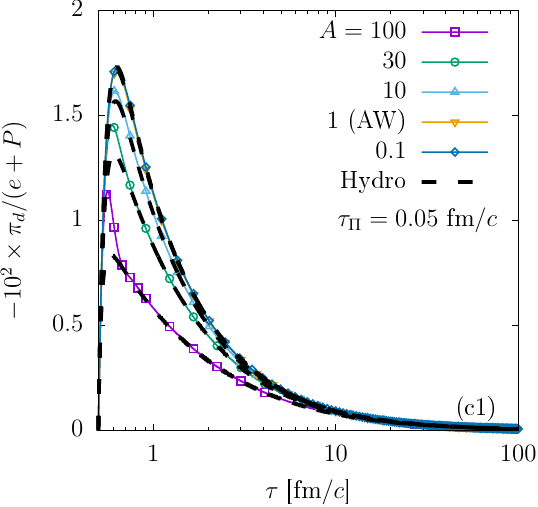} \\
 \includegraphics[width=.3\linewidth]{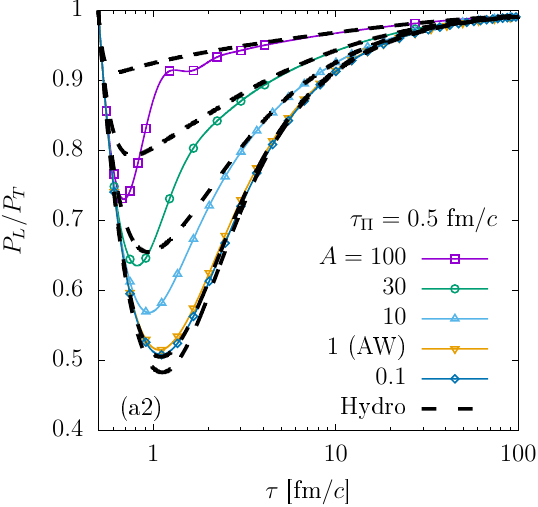} & 
 \includegraphics[width=.3\linewidth]{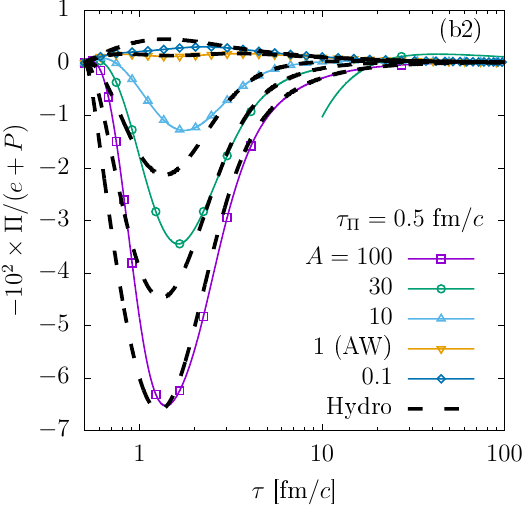} & 
 \includegraphics[width=.3\linewidth]{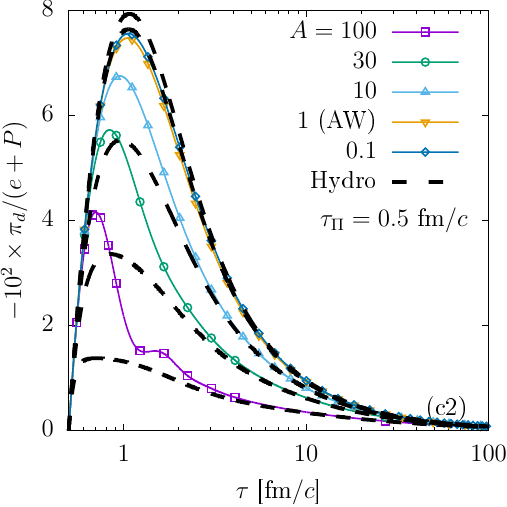} \\
\end{tabular}
\caption{
Same as Fig.~\ref{fig:bjork_AH} for the case when $H = 1$ is fixed and $A$ varies between $0.1$ and $100$ (see legend).\label{fig:bjork_A}
}
\end{figure*}

The construction of the Shakhov collision term relies on the functions $\widetilde{\mathcal{H}}^{(\ell)}_{\bk n}$ that ensure the recovery of the moments $\rho^{\mu_1 \cdots \mu_\ell}_{{\rm S};n}$ of $\delta f_{{\rm S}\bk} = f_{{\rm S}\bk} - f_{0\bk}$ for $-s_\ell \le n \le N_\ell$. Normally, this can be achieved when $\widetilde{\mathcal{H}}^{(\ell)}_{\bk n}$ is taken as a polynomial of order $N_\ell + s_\ell + 1$ in $E_\bk = k^\mu u_\mu$. However, given that the symmetries of the Bjorken flow of a neutral gas (with no conserved charge) imply that even- and odd-order moments do not mix, we only need to ensure the recovery of even-order moments of $\delta f_{{\rm S}\bk}$. To avoid confusion, we shall employ the notation $\tilde{\mathfrak{h}}^{(2)}_{\bk n}$ to denote this basis in even powers of energy, as opposed to the standard basis $\widetilde{\mathcal{H}}^{(\ell)}_{\bk n}$. Specifically, we are employing a Shakhov model with $(N_2, s_2) = (0,2)$, for which we need to recover only $\rho^{\mu\nu}_{{\rm S};-2}$ and $\rho^{\mu\nu}_{{\rm S};0}$, thus $\tilde{\mathfrak{h}}^{(2)}_{\bk n}$ become polynomials of the form $A_n + B_n E_\bk^2$. Then, $\mathbb{S}_\bk = \delta f_{{\rm S}\bk} / f_{0\bk} \tilde{f}_{0\bk}$ becomes 
\begin{equation}
 \mathbb{S}_\bk = \left(\pi_{{\rm S};-2} \tilde{\mathfrak{h}}^{(2)}_{\bk 0} + 
 \pi_{{\rm S};0} \tilde{\mathfrak{h}}^{(2)}_{\bk 2}\right) \left(\frac{k_\eta^2}{\tau^2 k_\tau^2} - \frac{k_\perp^2}{2k_\tau^2}\right),
 \label{eq:bjorken_Sk}
\end{equation}
where the last term follows by noting that, under our assumptions of longitudinal-boost invariance and transverse plane homogeneity, $\rho^{\mu\nu}_{{\rm S};r} = {\rm diag}(0,-\frac{1}{2}, -\frac{1}{2}, \tau^{-2}) \pi_{{\rm S};r}$ and
\begin{equation}
 \rho^{\mu\nu}_{{\rm S};r} k_{\langle \mu} k_{\nu \rangle} = \pi_{{\rm S};r} 
 \left(\frac{k_\eta^2}{\tau^2} - \frac{k_\perp^2}{2}\right).
\end{equation}
The scalar coefficients $\pi_{{\rm S};-2}$ and $\pi_{{\rm S};0}$ are given by
\begin{align}
 \pi_{{\rm S};-2} &= (1 - \tau_R \mathcal{A}^{(2)}_{{\rm S};-2,-2}) \pi_{-2} - \tau_R \mathcal{A}^{(2)}_{{\rm S};-2,0} \pi_0,\nonumber\\
 \pi_{{\rm S};0} &= - \tau_R \mathcal{A}^{(2)}_{{\rm S};0,-2} \pi_{-2} + (1 - \tau_R \mathcal{A}^{(2)}_{{\rm S};0,0}) \pi_0. 
\end{align}

The polynomials $\tilde{\mathfrak{h}}^{(2)}_{\bk n}$ must be constructed to ensure 
\begin{align}
 \int \d K\, \delta f_{{\rm S}\bk} E_\bk^{-2} k^{\langle \mu} k^{\nu \rangle} &= \rho^{\mu\nu}_{{\rm S};-2},\nonumber\\
 \int \d K\, \delta f_{{\rm S}\bk} k^{\langle \mu} k^{\nu \rangle} &= \rho^{\mu\nu}_{{\rm S};0}.
 \label{eq:bjorken_hdef}
\end{align}
Taking into account the integration identity 
\begin{multline}
 \int \d K\, F_\bk k^{\langle \mu_1} \cdots k^{\mu_m \rangle} k_{\langle \nu_1} \cdots k_{\nu_n \rangle}  \\
 = \frac{m! \delta_{mn}}{(2m+1)!!} \Delta^{\mu_1 \cdots \mu_m}_{\nu_1 \cdots \nu_m} \int \d K\, F_\bk (\Delta^{\alpha\beta} k_\alpha k_\beta)^m,
\end{multline}
one may rewrite Eq.~\eqref{eq:bjorken_hdef} as 
\begin{equation}
 2(A_n J_{r, 2} + B_n J_{2+r, 2}) = \delta_{rn},
\end{equation}
valid for $r,n \in \{0, 2\}$. The polynomials $\tilde{\mathfrak{h}}^{(2)}_{\bk n} = A_n + B_n E_\bk^2$ can be obtained as
\begin{align}
 \tilde{\mathfrak{h}}^{(2)}_{\bk 0} &= \frac{J_{42} - J_{22} E_\bk^2}{2(J_{02} J_{42} - J_{22}^2)}, &
 \tilde{\mathfrak{h}}^{(2)}_{\bk 2} &= \frac{-J_{22} + J_{02} E_\bk^2}{2(J_{02} J_{42} - J_{22}^2)}.
 \label{eq:bjorken_h}
\end{align}

\subsection{Numerical results} \label{sec:bjork:res}

In order to validate the kinetic Shakhov model, we performed numerical simulations of both the hydrodynamics and the kinetic theory models, taking constant values for the relaxation time: $\tau_R = \tau_\Pi = 0.05$ fm$/c$ and $0.5$ fm$/c$. The particle mass was set to $m = 1\ {\rm GeV} / c^2$ and at initial time $\tau_0 = 0.5\ {\rm fm} / c$, we set $f_{\bk}(\tau_0) = \exp(-\beta_0 E_\bk)$, with initial temperature $\beta_0^{-1} = 0.6\ {\rm GeV}$. 

In the first set of simulations, shown in Fig.~\ref{fig:bjork_AH}, we took the parameters $A$ and $H$ of the Shakhov model to obey $AH = 1$, such that the $A = H = 1$ case corresponds to the AW model. We considered a wide range of values for $A$, from $0.5$ to $100$.
The kinetic theory results are shown with solid lines and symbols, while the solutions of the corresponding second-order fluid-dynamical equations \eqref{eq:bjork_hydro} are shown with dashed lines.
Panels (a1)--(c1) of Fig.~\ref{fig:bjork_AH} show the results for $\tau_R =\tau_\Pi = 0.05\ {\rm fm}/c$, while panels (a2)--(c2) correspond to $\tau_R =\tau_\Pi = 0.5\ {\rm fm}/c$. 
Panels (a1) and (a2) show the ratio $P_L / P_T$, while panels (c1) and (c2) show the dimensionless ratio $-10\pi_d / (e + P)$. In these panels, kinetic theory and hydrodynamics are always in good agreement. As expected, increasing $A$ and decreasing $H$ has the effect of lowering $|\pi_d|$ and thus leads to ratios $P_L / P_T$ which are closer to unity. Panels (b1) and (b2) show the ratio $-10^3 \Pi / (e + P)$. In the case when $\tau_\Pi = 0.05\ {\rm fm}/c$, shown in panel (b1), the magnitude of $\Pi/ (e +P)$ remains unchanged for all tested values of $A$, as expected since while $\pi_d$ scales with $H$, the contribution $\lambda_{\Pi\pi} \pi_d$ to $\Pi$ scales like $AH = 1$. In panel (b2), $\Pi / (e + P)$ exhibits a clear dependence on $A$, and moreover the agreement between kinetic theory and hydrodynamics deteriorates, especially for $A = 0.5$, indicating a breakdown of the hydrodynamics assumptions.

In Fig.~\ref{fig:bjork_A}, we considered the case when $H = 1$ is fixed and $A$ was varied from $100$ down to $0.1$, with $A = H = 1$ corresponding to the AW model. As before, the kinetic theory and hydrodynamics results for $\tau_R = \tau_\Pi = 0.05\ {\rm fm}/c$ are in good agreement. At $\tau_R = \tau_\Pi = 0.5\ {\rm fm}/c$, visible discrepancies can be seen in the case of large $A$, most notably during the early-time evolution, where hydrodynamics cannot be expected to be valid.

\section{Shear-diffusion coupling: longitudinal waves}\label{sec:long}

In the previous section, we discussed a second-order Shakhov model modifying the cross-coupling coefficient $\bar{\lambda}_{\Pi \pi}$ of a fluid having constituents of mass $m$ and with no conserved charge. We now discuss the second-order Shakhov model which modifies the shear-diffusion cross-coupling coefficients $\ell_{V \pi}$ and $\ell_{\pi V}$, and for simplicity, we focus on an ultrarelativistic ideal gas, whose properties are summarized in Sec.~\ref{sec:shk2:UR}. We start with a brief problem description in Sec.~\ref{sec:long:intro}, while our proposed extended Shakhov model is introduced in Sec.~\ref{sec:long:shk} and validated in Sec.~\ref{sec:long:results}. 

\subsection{Problem description}\label{sec:long:intro}

We consider the propagation of longitudinal waves in a fluid at rest, as discussed in Refs.~\cite{Ambrus:2017keg,Ambrus:2022vif}. Taking the wave vector $\mathbf{k}$ along the $z$ axis, the four velocity reads $u^\mu \partial_\mu \simeq \partial_t + \delta v \partial_z$, where $|\delta v| \ll 1$ is the velocity along the $z$ axis. Denoting via $n_0$, $e_0$ the particle number density and energy density of the background state, we consider the perturbations $\delta n = n - n_0$ and $\delta e = e - e_0$ to be of the same order as $\delta v$. To linear order, the diffusion current $V^\mu$ and shear-stress tensor $\pi^{\mu\nu}$ take the form
\begin{equation}
 V^\mu \partial_\mu \simeq \delta V \partial_z, \quad 
 \pi^{\mu\nu} \simeq \delta \pi {\rm diag}\left(0, -\frac{1}{2}, -\frac{1}{2}, 1\right)\;,
\end{equation}
where the scalar quantities $\delta V$ and $\delta \pi$ are also assumed to be infinitesimal.

The conservation equations $\partial_\mu N^\mu = 0$ and $\partial_\mu T^{\mu\nu} = 0$ reduce to 
\begin{align}
 \partial_t \delta n + n_0 \partial_z \delta v + \partial_z \delta V &= 0\;, \nonumber\\
 \partial_t \delta e + (e_0 + P_0) \partial_z \delta v &= 0\;,\nonumber\\
 (e_0 + P_0)\partial_t \delta v + \partial_z \delta P + 
 \partial_z \delta \pi &= 0\;,
 \label{eq:long_cons}
\end{align}
while the dynamical equations for $\delta V$ and $\delta \pi$ are 
\begin{align}
 \tau_V \partial_t \delta V + \delta V &= -\kappa \partial_z \delta \alpha + \ell_{V\pi} \partial_z \delta \pi\;, \nonumber\\
 \tau_\pi \partial_t \delta \pi + \delta \pi &= -\frac{4\eta}{3} \partial_z \delta v -\frac{2}{3} \ell_{\pi V} \partial_z \delta V\;.
 \label{eq:long_dissipative}
\end{align}
The above equations feature $6$ independent transport coefficients: $\tau_V, \kappa,$ and $\ell_{V\pi}$ determine the behaviour of the diffusion current $\delta V$, while $\tau_\pi, \eta,$ and $\ell_{\pi V}$ control the evolution of the shear-stress $\delta \pi$. These coefficients can be fixed in the extended Shakhov model by employing, e.g., $2 \times 2$ matrices for both the vector and the tensor sectors. In this section, we will focus on changing only the first-order transport coefficients $\kappa$ and $\eta$, as well as the ratios of the cross-coupling coefficients with their respective relaxation times, $\ell_{V\pi} / \tau_V$ and $\ell_{\pi V} / \tau_\pi$. This can be achieved by employing the parameters $(N_1, N_2, s_1, s_2) = (1,0,0,1)$. In the following, we will refer to this model as the $(1001)$ model. For definiteness, we take the overall relaxation time as 
\begin{equation}
 \tau_R = \frac{5 \eta}{4P}.
\end{equation}

\subsection{Shakhov model for the shear-diffusion coupling}
\label{sec:long:shk}

We will now discuss the $(1001)$ extended Shakhov model in more detail.
In the diffusion sector, the matrix $\mathcal{A}^{(1)}_{{\rm S};rn}$ features a single element, which is related to the respective relaxation time as $\tau_V = \tau^{(1)}_{{\rm S};0,0} = 1/\mathcal{A}^{(1)}_{{\rm S};0,0}$.
Considering the diffusion coefficient $\kappa$ fixed, the relaxation time $\tau_V$ is readily obtained as
\begin{equation}
 \tau_V = \frac{12 \kappa}{\beta P}.
 \label{eq:1001_tauV}
\end{equation}

The collision matrix $\mathcal{A}^{(2)}_{{\rm S};rn}$ for the shear sector and its inverse can be written as
\begin{align}
 \mathcal{A}^{(2)}_{{\rm S};rn} =& 
 \begin{pmatrix}
  \mathcal{A}^{(2)}_{{\rm S};-1,-1} & 
  \mathcal{A}^{(2)}_{{\rm S};-1,0} \\
  \mathcal{A}^{(2)}_{{\rm S};0,-1} & 
  \mathcal{A}^{(2)}_{{\rm S};0,0}
 \end{pmatrix}, \nonumber\\
 \tau^{(2)}_{{\rm S};rn} =& 
 \begin{pmatrix}
  \tau^{(2)}_{{\rm S};-1,-1} & 
  \tau^{(2)}_{{\rm S};-1,0} \\
  \tau^{(2)}_{{\rm S};0,-1} & 
  \tau^{(2)}_{{\rm S};0,0}
 \end{pmatrix}.
\end{align}
The first-order transport coefficients of the shear sector are given by
\begin{align}
 \eta_{-1} =& \frac{P}{5}\left(\beta \tau^{(2)}_{{\rm S};-1,-1} + 4 \tau^{(2)}_{{\rm S};-1,0}\right),\nonumber\\
 \eta_0 =& \frac{P}{5}\left(\beta \tau^{(2)}_{{\rm S};0,-1} + 4 \tau^{(2)}_{{\rm S};0,0}\right),
\end{align}
such that 
\begin{equation}
 \mathcal{C}^{(2)}_{-1} = \frac{\eta_{-1}}{\eta_0} = 
 \frac{\tau^{(2)}_{{\rm S};-1,-1} + 4 \beta^{-1} \tau^{(2)}_{{\rm S};-1,0}}
 {\beta \tau^{(2)}_{{\rm S};0,-1} + 4 \tau^{(2)}_{{\rm S};0,0}}\beta.
 \label{eq:C2_-1}
\end{equation}
The relaxation time $\tau_\pi$ is
\begin{equation}
 \tau_\pi = \tau^{(2)}_{{\rm S};0,-1} \mathcal{C}^{(2)}_{-1} + \tau^{(2)}_{{\rm S};0,0},
\end{equation}
while the shear-diffusion cross-coupling coefficients read 
\begin{equation} 
 \ell_{\pi V} = \frac{2}{5} \tau^{(2)}_{{\rm S};0,-1}\;,\quad \ell_{V\pi} = \tau_V \left(\frac{\beta}{4} - \mathcal{C}^{(2)}_{-1}\right) \;.
\end{equation}
We express $\tau^{(2)}_{{\rm S};0,-1}$, $\tau^{(2)}_{{\rm S};0,0}$, and $\mathcal{C}^{(2)}_{-1}$ in terms of $\eta$, $\ell_{V\pi}$, and $\ell_{\pi V}$ as
\begin{gather}
 \tau^{(2)}_{{\rm S};0,-1} = \frac{5\ell_{\pi V}}{2}, \qquad 
 \tau^{(2)}_{{\rm S};0,0} = \frac{5\eta}{4P} - \frac{5\beta}{8} \ell_{\pi V}, \nonumber\\
 \mathcal{C}^{(2)}_{-1} = \frac{\beta}{4} - \frac{\ell_{V\pi}}{\tau_V}.
\end{gather}
Note that knowledge of the transport coefficients cannot fix both entries on the $r = -1$ line of $\tau^{(2)}_{{\rm S}; rn}$, since these entries appear only through the combination shown in $\mathcal{C}^{(2)}_{-1}$. We will take advantage of the second degree of freedom below, in order to optimize the structure of the resulting matrix. For now, we express also the relaxation time $\tau_\pi$ as
\begin{equation}
 \tau_\pi = \frac{5\eta}{4P} - \frac{5 \ell_{V\pi} \ell_{\pi V}}{2\tau_V}.
 \label{eq:1001_taupi}
\end{equation}
Besides the transport coefficients in Eq.~\eqref{eq:UR_tcoeffs_fixed} which are fixed for ultrarelativistic particles, we express the remaining ones using Eqs.~\eqref{eq:UR_tcoeff}:
\begin{gather}
 \lambda_{VV} = \frac{3}{5} \tau_V, \quad 
 \lambda_{V\pi} = \frac{\beta \tau_V}{16} \left(1 - \frac{4\ell_{V\pi}}{\beta \tau_V}\right), \quad 
 \lambda_{\pi V} = 0, \nonumber\\
 \tau_{\pi\pi} = \frac{5}{7}\left(\frac{5\eta}{2P} - \frac{\beta \ell_{\pi V}}{2} - 3\frac{\ell_{V\pi} \ell_{\pi V}}{\tau_V}\right).
 \label{eq:1001_tcoeffs}
\end{gather}

We have now determined all transport coefficients of the $(1001)$ model. In order to assemble the collision matrix and its inverse, we still need to specify the remaining degree of freedom pertaining to the $r = -1$ line of $\tau^{(2)}_{{\rm S};rn}$. To simplify the discussion, we introduce the following notation:
\begin{equation}
 H = \frac{5 \eta}{4\tau_\pi P}, \quad 
 L_{V\pi} = \frac{4\ell_{V\pi}}{\beta \tau_V}, \quad 
 L_{\pi V} = \frac{5\beta \ell_{\pi V}}{8\tau_\pi}.
 \label{eq:long_prelim}
\end{equation}
Considering $L_{V\pi}$ and $L_{\pi V}$ as input parameters, $H$ can be obtained by dividing Eq.~\eqref{eq:1001_taupi} by $\tau_\pi$:
\begin{equation}
 H = 1 + L_{V\pi} L_{\pi V}.
\end{equation}
This allows $\tau_\pi$ to be expressed as
\begin{equation}
 \tau_\pi = \frac{5\eta}{4P(1 + L_{V\pi} L_{\pi V})}.
\end{equation}
Furthermore, the transport coefficients in Eq.~\eqref{eq:1001_tcoeffs} read
\begin{gather}
 \frac{\lambda_{VV}}{\tau_V} = \frac{3}{5}, \quad 
 \frac{\lambda_{V\pi}}{\tau_V} = \frac{\beta}{16}(1 - L_{V\pi}), \quad 
 \lambda_{\pi V} = 0, \nonumber\\
 \frac{\tau_{\pi\pi}}{\tau_\pi} = \frac{2}{7}(5H - 2L_{\pi V} - 3L_{V\pi} L_{\pi V}).
\end{gather}
Moreover, the elements of $\tau^{(2)}_{\rm S}$ satisfy:
\begin{gather}
 \frac{\tau^{(2)}_{{\rm S};0,-1}}{\tau_\pi} = \frac{4}{\beta} L_{\pi V}, \quad 
 \frac{\tau^{(2)}_{{\rm S};0,0}}{\tau_\pi} = H - L_{\pi V}, \nonumber\\
 \frac{\tau^{(2)}_{{\rm S};-1,-1}}{\tau_\pi} + \frac{4\tau^{(2)}_{{\rm S};-1,0}}{\beta \tau_\pi} = H(1 - L_{V\pi}).
\end{gather}

\begin{figure*}[t]
\begin{tabular}{cc}
\includegraphics[width=\columnwidth]{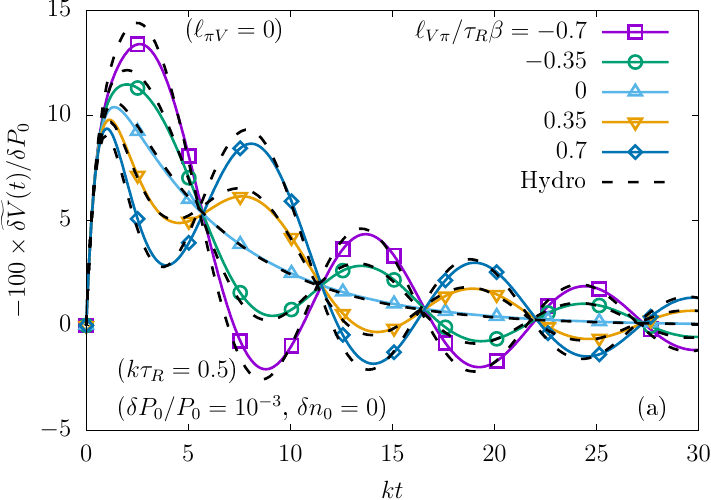} &
\includegraphics[width=\columnwidth]{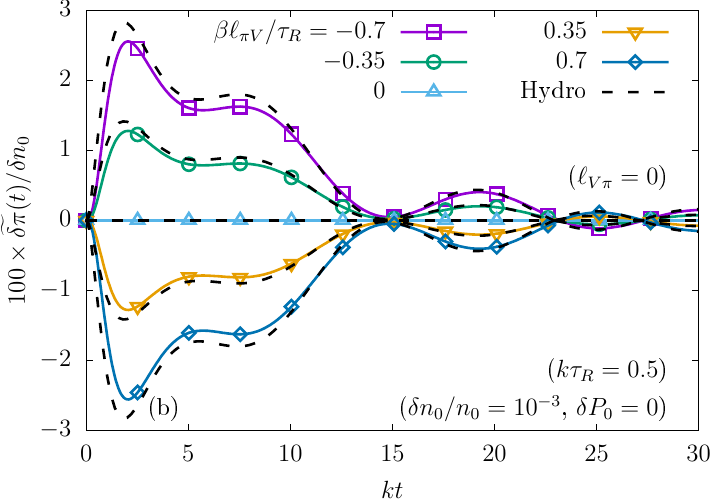} \\
\includegraphics[width=\columnwidth]{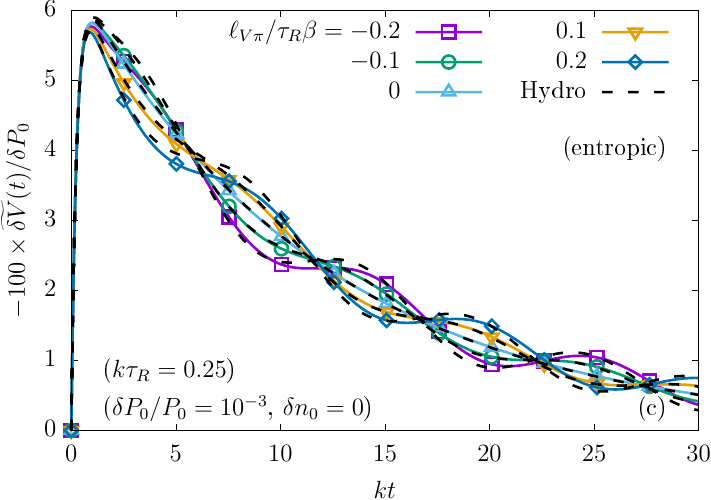} &
\includegraphics[width=\columnwidth]{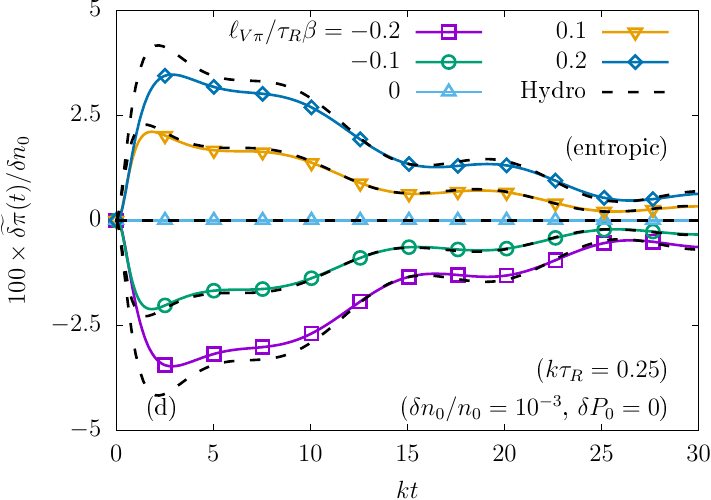}
\end{tabular}
\caption{Time evolution of the amplitudes (a,c) $\widetilde{\delta V}$ and (b,d) $\widetilde{\delta \pi}$, multiplied by the factors $-100/\delta P_0$ and $100 / \delta n_0$. The initial fluctuation amplitudes are $(\delta n_0, \delta P_0) = (0, 10^{-3} P_0)$ and $(10^{-3} n_0, 0)$, respectively. The curves represented with lines and symbol correspond to the Shakhov model results, while the hydro solutions (obtained numerically) are represented with black dashed lines. In panel (a), we varied $\ell_{V\pi} \in \{0, \pm 0.35, \pm 0.7\}$ at vanishing $\ell_{\pi V} = 0$. In panel (b), we varied $\ell_{\pi V} \in \{\pm 0, \pm 0.35, \pm 0.7\}$ with $\ell_{V \pi} = 0$. In panels (c) and (d), we imposed the entropy constraint \eqref{eq:long_entropy} to obtain $\ell_{\pi V}$ for $\ell_{V\pi} \in \{0, \pm 0.1, \pm 0.2\}$. In panels (a)--(b), we set $k \tau_R = 0.5$, while in panels (c)--(d), we employed $k\tau_R = 0.25$.
\label{fig:long}}
\end{figure*}

Coming back to the $r = -1$ line of the $\tau^{(2)}_{{\rm S};rn}$ matrix, one may be tempted to simply set the off-diagonal term to $0$, i.e. $\tau^{(2)}_{{\rm S};-1,0} = 0$. However, this choice fails when $\ell_{V\pi}$ approaches $\beta \tau_V / 4$ and hence $L_{V\pi}$ approaches $1$, since then $\tau^{(2)}_{{\rm S};-1,-1} \rightarrow -\frac{4}{\beta} \tau^{(2)}_{{\rm S};-1,0} \rightarrow 0$ and $\tau^{(2)}_{\rm S}$ becomes singular. To circumvent this problem, we take advantage of the $\tau^{(2)}_{{\rm S}; -1,0}$ degree of freedom to ensure that the eigenvalues of $\tau^{(2)}_{\rm S}$ remain positive. Introducing the notation
\begin{align}
 \tau^{(2)}_{{\rm S};-1,0} = \frac{\beta \tau_\pi}{4} x\;, 
\end{align}
the matrix $\tau^{(2)}_{\rm S}$ can be written as
\begin{equation}
 \tau^{(2)}_{\rm S} = \tau_\pi
 \begin{pmatrix}
  H (1 - L_{V\pi}) - x & \frac{\beta }{4} x \\
  \frac{4}{\beta} L_{\pi V} & H -  L_{\pi V}
 \end{pmatrix}\;,
\end{equation}
and its eigenvalues are given by
\begin{multline}
 \lambda_{1,2} = \frac{\tau_\pi}{2} [H(2 -L_{V\pi}) - L_{\pi V} - x \\
 \mp \sqrt{(H L_{V\pi} + L_{\pi V} + x)^2 - 4 H L_{V\pi} L_{\pi V}}]\;.
\end{multline}
Considering $\lambda_2 > \lambda_1$, we have 
\begin{equation}
 \lim_{x \rightarrow -\infty} \lambda_1 = \lim_{x \rightarrow \infty} \lambda_2 = \tau_\pi H.
\end{equation}
At finite values of $x$, it holds that $\lambda_1 < \tau_\pi H < \lambda_2$. Writing $\lambda_1 = \alpha \tau_\pi H$, with $0 < \alpha < 1$, we have
\begin{equation}
 x = H(1 - \alpha - L_{V\pi}) - L_{\pi V} - \frac{1 - H}{1 - \alpha}\;,
 \label{eq:long_valx}
\end{equation}
while $\lambda_2$ is given by
\begin{equation}
 \lambda_2 = \frac{1 - H \alpha}{1 - \alpha} \tau_\pi\;.
\end{equation}

Noting that $\det \tau^{(2)}_{\rm  S} = \tau_\pi^2 \alpha H (1 - \alpha H) / (1 - \alpha)$, the matrix $\tau^{(2)}_{\rm S}$ can be inverted to obtain the collision matrix $\mathcal{A}^{(2)}_{\rm S}$ as
\begin{equation}
 \mathcal{A}^{(2)}_{\rm S} = \frac{1 - \alpha}{\alpha H \tau_\pi (1 - \alpha H)} 
 \begin{pmatrix}
  H - L_{\pi V} & -\frac{\beta}{4} x \\ 
  -\frac{4}{\beta} L_{\pi V} & H(1 - L_{V\pi}) - x
 \end{pmatrix},
 \label{eq:long_A2}
\end{equation}
where $x$ is fixed by the value of $\alpha$ (taking values between $0$ and $1$) via Eq.~\eqref{eq:long_valx}.
In the following, we will employ for definiteness $\alpha = 1/2$, such that 
\begin{equation}
 \lambda_1 = \frac{H}{2} \tau_\pi, \quad 
 \lambda_2 = (2 - H) \tau_\pi\;,
\end{equation}
with $\det \tau^{(2)}_{\rm S} = \frac{H}{2} (2 - H) \tau_\pi^2$ and 
$x = \frac{5H}{2} - H L_{V\pi} - L_{\pi V} - 2$. 

Before moving on to the numerical results, we remark that, while our proposed Shakhov model allows the coefficients $\ell_{V\pi}, \ell_{\pi V}, \kappa$ and $\eta$ to be controlled independently, they are in principle related through the constraint \cite{Gavassino:2022nff}
\begin{equation}
 \frac{\ell_{V\pi}}{\kappa} + \frac{\ell_{\pi V}}{2\eta T} = 0
 \label{eq:long_entropy}
\end{equation}
which is necessary for the phenomenological entropy current to have a nonnegative divergence. Under this constraint, it holds that 
\begin{equation}
 L_{\pi V} = - 3 H L_{V\pi}, \quad 
 H = \frac{1}{1 + 3 L_{V\pi}^2}.
\end{equation}
Furthermore, the constraint \eqref{eq:long_entropy} forces the couplings $\ell_{V\pi}$ and $\ell_{\pi V}$ to have opposite signs, thereby ensuring that the relaxation time \eqref{eq:1001_taupi} stays positive.

\subsection{Numerical results}
\label{sec:long:results}

We now consider a system of size $L = 2\pi/k$ with periodic boundary conditions, where $k$ is the wave number and $L$ is the wavelength of the propagating wave. We initialize the system in local thermodynamic equilibrium at rest, with pressure and density given by
\begin{align}
 P(t = 0) &= P_0 + \delta P_0 \cos(k z), \nonumber\\
 n(t = 0) &= n_0 + \delta n_0 \cos(k z).
\end{align}
As can be seen from Eq.~\eqref{eq:long_dissipative}, $\ell_{V\pi}$ and $\ell_{\pi V}$ introduce the coupling between the shear and diffusion sectors. In particular, when $\delta n_0 = 0$ and $\delta P_0 \neq 0$, $\ell_{V\pi}$ introduces shear modes into the evolution of $\delta V$, which cause it to oscillate. Conversely, when $\delta P_0 = 0$ and $\delta n_0 \neq 0$, $\ell_{\pi V}$ allows $\delta \pi$ to develop a non-vanishing value through its coupling to $\delta V$. In order to characterize the evolution of $\delta V$ and $\delta \pi$, we consider the Fourier amplitudes
\begin{align}
  \widetilde{\delta V}(t) &= 
  \frac{k}{\pi} \int_0^{2\pi / k} \d z\, \delta V\, \sin(kz)\;,\nonumber\\ 
  \widetilde{\delta \pi}(t) &= 
  \frac{k}{\pi} \int_0^{2\pi / k} \d z\, \delta \pi\, \cos(kz)\;.
\end{align}
The time evolution of $\widetilde{\delta V}(t)$ and $\widetilde{\delta \pi}(t)$ is shown in Fig.~\ref{fig:long} for the cases $(\delta P_0 / P_0, \delta n_0 / n_0) = (10^{-3}, 0)$ and $(0, 10^{-3})$ in panels (a) and (c); and (b) and (d), respectively. 

In all cases, the Shakhov model was constructed using $\mathcal{A}^{(1)}_{\rm S} = \tau_R^{-1}$ and $\alpha = 1/2$ for the $\mathcal{A}^{(2)}_{\rm S}$ matrix. We also considered $\eta = \eta_0 = \frac{4}{5} \tau_R P$, like in the Anderson-Witting model, such that $\tau_\pi = \tau_R / H$, as indicated in Eq.~\eqref{eq:long_prelim}.
In panel (a), we took $\ell_{\pi V} = 0$ and considered non-vanishing values of $\ell_{V\pi}$, in which case $H =1$ and $x = \frac{1}{2} - L_{V\pi}$. Then, the Shakhov matrix \eqref{eq:long_A2} reduces to:
\begin{equation}
 \mathcal{A}^{(2)}_{\rm S} = \frac{1}{\tau_\pi} 
 \begin{pmatrix}
  2 & -\frac{\beta}{4} (1 - 2L_{V\pi}) \\ 
  0 & 1
 \end{pmatrix}.
\end{equation}
For panel (b), we considered the case when $\ell_{V\pi} = 0$, such that $H = 1$ and $x = \frac{1}{2} - L_{\pi V}$ and Eq.~\eqref{eq:long_A2} reduces to
\begin{equation}
 \mathcal{A}^{(2)}_{\rm S} = \frac{2}{\tau_\pi} 
 \begin{pmatrix}
  1 - L_{\pi V} & -\frac{\beta}{4} (\frac{1}{2} - L_{\pi V}) \\
  -\frac{4}{\beta} L_{\pi V} & \frac{1}{2} + L_{\pi V}
 \end{pmatrix}\;.
\end{equation}
Finally, for panels (c) and (d), we enforced the entropy constraint in Eq.~\eqref{eq:long_entropy}, leading to $H = 1 / (1 + 3 L_{V\pi}^2)$, $x = -\frac{H}{2}(12 L_{V\pi}^2 - 4 L_{V\pi} - 1)$ and 
\begin{equation}
 \mathcal{A}^{(2)}_{\rm S} = \frac{2}{\tau_\pi(2 - H)} 
 \begin{pmatrix}
  1 + 3L_{V\pi} & \frac{\beta}{8} (12 L_{V \pi}^2 - 4 L_{V\pi} - 1) \\
  \frac{12}{\beta} L_{V\pi} & 6L_{V\pi}^2 - 3L_{V\pi} + \frac{1}{2}
 \end{pmatrix}\;.
\end{equation}
Finally, we employed a constant relaxation time $\tau_R$, taken such that $k \tau_R = 0.5$ for panels (a) and (b); and $0.25$ for panels (c) and (d). For definiteness, we considered the wavelength to be $L = 2\pi / k = 1\ {\rm fm}$, while the initial temperature was set to $T_0 = 1\ {\rm GeV}$ and the initial chemical potential was set to $0$, such that $n_0 = 212.04\ {\rm fm}^{-3}$ and $P_0 = 212.04\ {\rm GeV} /{\rm fm}^3$. This corresponds to a shear viscosity to entropy density ratio of $4\pi \eta / s = L T/5 \simeq 1.02$ for panels (a) and (b) and half of that ($4\pi \eta / s = LT / 10 \simeq 0.51$) for panels (c) and (d).

All kinetic theory results shown in Fig.~\ref{fig:long} using coloured lines and symbols are in good agreement with the corresponding hydrodynamics results, shown with dashed black lines. As expected, the agreement improves at smaller $k \tau_R$ and it becomes worse when $k \tau_R$ is increased.

\section{Hard-sphere gas: the Riemann problem}\label{sec:riemann}

\begin{table*}
\begin{tabular}{||l||c|c|c|c|c||c|c|c|c|c||}
	\hline\hline 
	Model 
	& $\eta\sigma \beta$ 
    & $\tau_\pi/\lambda_{\rm mfp}$ 
    & $\beta \ell_{\pi V} / \tau_\pi$
	& $\tau_{\pi\pi} / \tau_\pi$
	& $\beta \lambda_{\pi V} / \tau_\pi$
    & $\kappa \sigma$ 
    & $\tau_V / \lambda_{\rm mfp}$ 
	& $\ell_{V\pi} / \beta \tau_V$
	& $\lambda_{VV}/\tau_V$ 
	& $\lambda_{V\pi} / \beta \tau_V$  \\[2pt] \hline
	{\bf HS}
    & $\mathbf{1.2676}$
	& $\mathbf{1.6557}$
	& $\mathbf{-0.56960}$
	& $\mathbf{1.6945}$ 
	& $\mathbf{0.20503}$
    & $\mathbf{0.15892}$
	& $\mathbf{2.0838}$
	& $\mathbf{0.028371}$
	& $\mathbf{0.89862}$ 
	& $\mathbf{0.069273}$ \\[2pt] \hline\hline
    AW
    & $\mathbf{1.2676}$
	& $1.5845$
	& $0$
	& $1.4286$ 
	& $0$
    & $0.13204$
	& $1.5845$
	& $0$
	& $0.6$ 
	& $0.0625$ \\[2pt] \hline 
    $1000$ 
    & $\mathbf{1.2676}$
	& $1.5845$
	& $0$
	& $1.4286$ 
	& $0$
	& $\mathbf{0.15892}$
	& $1.9070$
	& $0$
	& $0.6$ 
	& $0.0625$ \\[2pt] \hline
    $1001$ 
    & $\mathbf{1.2676}$
	& $1.6457$
	& $-0.52446$
	& $1.5946$ 
	& $0$
	& $\mathbf{0.15892}$
	& $1.9070$
	& $\mathbf{0.028371}$
	& $0.6$ 
	& $0.055407$ \\[2pt] \hline
    $1012$ 
    & $\mathbf{1.2676}$
	& $\mathbf{1.6557}$
	& $\mathbf{-0.56960}$
	& $\mathbf{1.6945}$
	& $\mathbf{0.20503}$
	& $\mathbf{0.15892}$
	& $\mathbf{2.0838}$
	& $\mathbf{0.028371}$
	& $0.762023$
	& $0.062933$ \\[2pt] \hline
    $2012$ 
    & $\mathbf{1.2676}$
	& $\mathbf{1.6557}$
	& $\mathbf{-0.56960}$
	& $\mathbf{1.6945}$ 
	& $\mathbf{0.20503}$
    & $\mathbf{0.15892}$
	& $\mathbf{2.0838}$
	& $\mathbf{0.028371}$
	& $\mathbf{0.89862}$ 
	& $\mathbf{0.069273}$ \\[2pt] \hline\hline
\end{tabular}
\caption{
Transport coefficients for the hard-sphere gas of ultrarelativistic particles interacting via an isotropic cross-section $\sigma$, computed using the IReD approach \cite{Wagner:2022ayd,Wagner:2023joq}, as well as for the AW and Shakhov models considered in Sec.~\ref{sec:riemann}.
Besides the transport coefficients shown above, all models have $\delta_{VV} = \tau_V$, $\delta_{\pi\pi} = 4\tau_\pi / 3$, $\tau_{V\pi} = \ell_{V\pi}$, and $\tau_{\pi V} = 4\ell_{\pi V}$.
\label{tbl:riem_tcoeffs}
}
\end{table*}

In this section, we focus on the problem of the propagation of shock waves (the Riemann problem). The Shakhov models presented here will be validated against the values obtained in Refs. \cite{Bouras:2010hm, Denicol:2012vq} using the BAMPS (Boltzmann approach to multiparton scattering) algorithm, where binary elastic scattering of ultrarelativistic (massless) particles via a constant, momentum-independent cross section was considered. In a sense, this section extends the considerations of Sec.~\ref{sec:long} to the case when nonlinear effects become important. Keeping to the notation $(N_1, N_2, s_1, s_2)$ introduced in Sec.~\ref{sec:shk2:dofs}, we will distinguish between Shakhov models of various orders, as described below. Our goal is to develop a Shakhov kinetic model which is able to reproduce the BAMPS data, and in doing so, we will aim to reproduce increasingly more transport coefficients of the hard-sphere (HS) gas, enumerated in Table~\ref{tbl:riem_tcoeffs}.

First, we will consider the Anderson-Witting (AW) model, where the relaxation time $\tau_R$ is taken to match the shear viscosity $\eta$ of the BAMPS model, namely 
\begin{equation}
 \text{AW:} \qquad \tau_R = \frac{5\eta}{4P}, \quad 
 \eta = \frac{1.2676}{\sigma \beta} = 1.2676 P \lambda_{\rm mfp},
 \label{eq:riem_tauR}
\end{equation}
where $\lambda_{\rm mfp} = 1/n \sigma$ is the particle mean free path and $n = \beta P$ is the particle number density. In units of $\lambda_{\rm mfp}$, the relaxation time $\tau_R$ becomes
\begin{equation}
 \tau_R = 1.5845 \lambda_{\rm mfp}.
\end{equation}
For definiteness, we keep the above relation between $\tau_R$ and $\lambda_{\rm mfp}$ for all models discussed below.
One may hope that with the above choice, the AW model can give a reasonable description of shear-driven quantities, such as $\pi^{\mu\nu}$. However, diffusion-related phenomena cannot be accurately described, as the diffusion coefficient of the AW model is incorrect (see Table~\ref{tbl:riem_tcoeffs}). 

We therefore also consider the first-order Shakhov model introduced in Ref.~\cite{Ambrus:2023ilm} and summarized in Sec.~\ref{sec:shk1}, dubbed $(1000)$ according to our convention. For definiteness, we consider $\tau_\pi = \tau_R$ with $\tau_R$ given in Eq.~\eqref{eq:riem_tauR}, while $\tau_V$ is computed based on
\begin{equation}
 \tau_V = \frac{\kappa_{\rm HS}}{\kappa_{\rm R}} \tau_R = 1.2036 \tau_R,
 \label{eq:riem_tauV}
\end{equation}
where the values of $\kappa$ for the hard-sphere and AW models are given in Table~\ref{tbl:riem_tcoeffs}. As it will become clear in the applications subsections, the above $20.4\%$ increase of the diffusion coefficient is insufficient to capture the magnitude of the diffusion current. The resulting Shakhov model has the following collision matrices:
\begin{equation}
 (1000): \quad \mathcal{A}^{(1)}_{\rm S} = \frac{0.52437}{\lambda_{\rm mfp}}, \quad 
 \mathcal{A}^{(2)}_{\rm S} = \frac{0.63111}{\lambda_{\rm mfp}},
\end{equation}
with $\tau_V = 1.2036 \tau_R = 1.9071 \lambda_{\rm mfp}$ and $\tau_\pi = \tau_R = 5 \eta / 4P = 1.5845 \lambda_{\rm mfp}$, as explained above. Please note that $\rho^{\mu\nu}_{0;{\rm S}} = 1 - \tau_R \mathcal{A}^{(2)}_{{\rm S};0,0} = 0$, hence the Shakhov term $\mathbb{S}_\bk$ of the $(1000)$ model consists only of the vector term.

The next model that we employed is the $(1001)$ model discussed in Sec.~\ref{sec:long}. As before, we set $\tau_R$ according to Eq.~\eqref{eq:riem_tauR} and take $\eta = \eta_R$, thus $\tau_\pi = \tau_R / H$. Similarly, $\tau_V$ is set via Eq.~\eqref{eq:riem_tauV}, such that $\kappa$ given in Eq.~\eqref{eq:1001_tauV} takes the HS value. Imposing the HS value for $\ell_{V\pi}/\tau_V = 0.028371 \beta$ and the entropy constraint \eqref{eq:long_entropy} gives $\ell_{\pi V}/\tau_\pi = -0.52446 / \beta$, while $L_{V \pi} = 0.11348$ and $L_{\pi V} = -0.32779$. Thus, $H = 0.96280$ and subsequently all transport coefficients come out as shown in Table~\ref{tbl:riem_tcoeffs}. The Shakhov matrices read
\begin{gather}
 (1001): \qquad 
 \mathcal{A}^{(1)}_{\rm S} = \frac{0.52437}{\lambda_{\rm mfp}}, \nonumber\\
 \mathcal{A}^{(2)}_{\rm S} = \frac{1}{\lambda_{\rm mfp}} \begin{pmatrix}
  1.5706 & -0.19031 \beta\\
  1.5956/\beta & \mathbf{0.27748}
 \end{pmatrix},
\end{gather}
where the element in bold font represents the $\mathcal{A}^{(2)}_{{\rm S};0,0}$ entry of the Shakhov matrix.

We also consider two higher-order models which are derived in Appendix \ref{app:shk_higher}. In the $(1012)$ model, we are able to fix all transport coefficients except $\lambda_{VV}$ and $\lambda_{V\pi}$. The collision matrices are given in Eq.~\eqref{eq:1012_IReD} and reproduced below for convenience:
\begin{gather}
 (1012): \quad 
 \mathcal{A}^{(1)}_{\rm S} = \frac{1}{\lambda_{\rm mfp}}
 \begin{pmatrix}
  0.63419 & 0 \\ 0.22111/ \beta & \mathbf{0.34155}
 \end{pmatrix},\nonumber\\
 \mathcal{A}^{(2)}_{\rm S} = \frac{1}{\lambda_{\rm mfp}}
 \begin{pmatrix}
  0.84927 & 0 & 0 \\
  0 & 0.70961 & 0 \\
  -1.3008/ \beta^2 & 1.5229 / \beta & \mathbf{0.37307}
 \end{pmatrix}\;.
\end{gather}

Finally, the $(2012)$ model allows all transport coefficients to be set to the values obtained from kinetic theory, employing the Shakhov matrices from Eq.~\eqref{eq:2012_IReD}, reproduced below:
\begin{gather}
 (2012): \hspace{\linewidth}\nonumber\\
 \mathcal{A}_{\rm S}^{(1)} = \frac{1}{\lambda_{\rm mfp}}
 \begin{pmatrix}
  0.62732 & 0 &0 \\
  0.11113/\beta & \mathbf{0.59563} & 0.011012 \beta^2 \\
  0& 0& 0.42171
 \end{pmatrix} \;,\nonumber\\
 \mathcal{A}_{\rm S}^{(2)} = \frac{1}{\lambda_{\rm mfp}} \begin{pmatrix}
  0.82802 & 0& 0\\
  0 & 0.70553 & 0\\
  -1.2528/\beta^2 & 1.5120/\beta & \mathbf{0.37256}
 \end{pmatrix} \;.
\end{gather}

\subsection{Problem description} \label{sec:riemann:intro}

\begin{figure}[!ht]
\begin{tabular}{c}
\includegraphics[width=0.95\columnwidth]{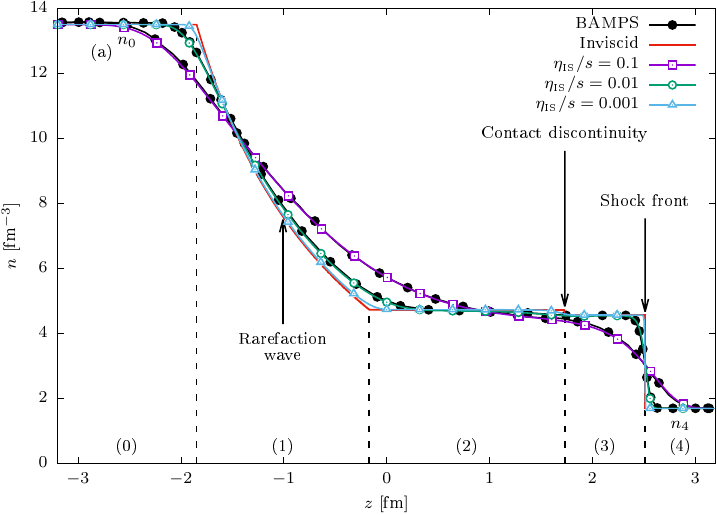} \\
\includegraphics[width=0.95\columnwidth]{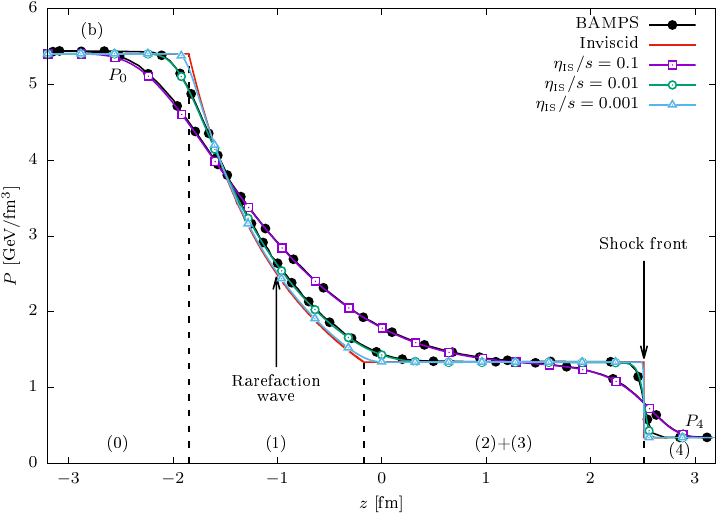} \\
\includegraphics[width=0.95\columnwidth]{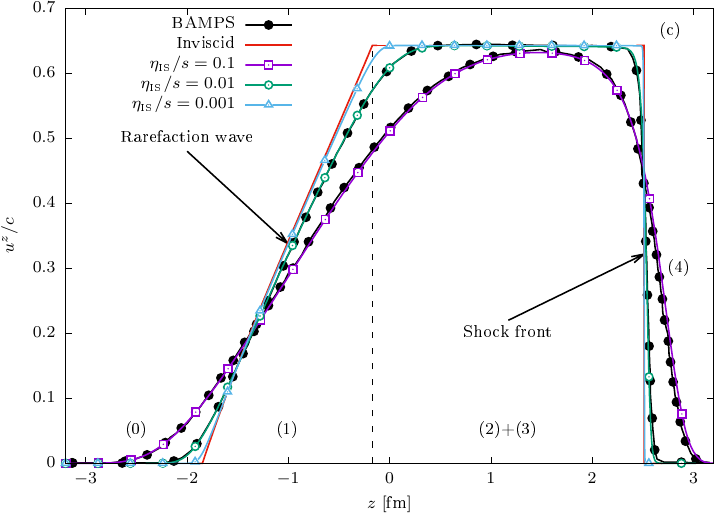} 
\end{tabular}
\caption{
Profiles of (a) density $n$, (b) pressure $P$, and (c) $z$ component of the flow velocity, $u^z$, for $\eta_{\rm IS} / s = 0.1$, $0.01$, and $0.001$, shown with lines and points, as well as the analytical solution for the inviscid case (shown with a solid red line). The dotted black lines help distinguish the various features of the flow structure.
\label{fig:riem}
}
\end{figure}

The Riemann problem constitutes a well-established test for the accuracy of fluid-dynamical codes. It consists in determining the flow of a fluid which is initially separated into distinct regions with different (constant) values for their relevant parameters, such as flow velocity, pressure, and density. 
In the cases considered here, which are equivalent to the setups of Refs.~\cite{Bouras:2010hm, Denicol:2012vq}, we assume the fluid to be homogeneous in the $(x,y)$-plane, such that the system is effectively $(1+1)$-dimensional. The discontinuity is taken to be at $z=0$, thus dividing the fluid at the initial time into two regions: the left region, where $z<0$, and the right region, where $z>0$.
In both regions, the flow velocity is taken to vanish at initial time. 

This setup corresponds to the so-called Sod shock tube \cite{sod1978survey} and the subsequent evolution of the fluid gives rise to 5 distinct regions. Far to the left (region 0), we have the unperturbed initial state characterized by $n = n_0$ and $P = P_0$. A rarefaction wave (region 1) connects the unperturbed fluid to a central plateau (region 2), where $n = n_2$ and $P = P_2$ assume constant values. The central plateau is split in two regions by the contact discontinuity. In region 3, to the right of the contact discontinuity, the density is $n_3  \neq n_2$, while the pressure remains unchanged, $P_2 = P_3$. Region 4, corresponding to the unperturbed fluid with parameters $n = n_4$ and $P = P_4$, lies to the right of the shock front. In the perfect (inviscid) fluid limit, the Sod shock tube problem can be solved analytically \cite{Rezzolla.2013}. The solution for the particular case of an ultrarelativistic, classical ideal gas is presented in detail in Ref.~\cite{Ambrus:2016aub}. 

This analytical solution for the initial conditions considered in Sec.~\ref{sec:riemann:bouras} is represented with a solid red line in Fig.~\ref{fig:riem}, with the particle density $n$, pressure $P$ and four-velocity $u^z$ shown in panels (a), (b) and (c), respectively. The dotted black lines, arrows and inset labels indicate the above-mentioned regions of the flow. We also show here the results obtained using our code for the simple case of the Anderson-Witting model, as well as BAMPS data from Ref.~\cite{Bouras:2010hm} for comparison. It can be seen from the figure that our numerical solution approaches the analytical one as the shear viscosity to entropy density ratio $\eta_{\rm IS}/ s$ decreases (the meaning of the IS subscript will be elucidated in the following subsection). 

\begin{figure*}[t]
\begin{tabular}{cc}
\includegraphics[width=0.95\columnwidth]{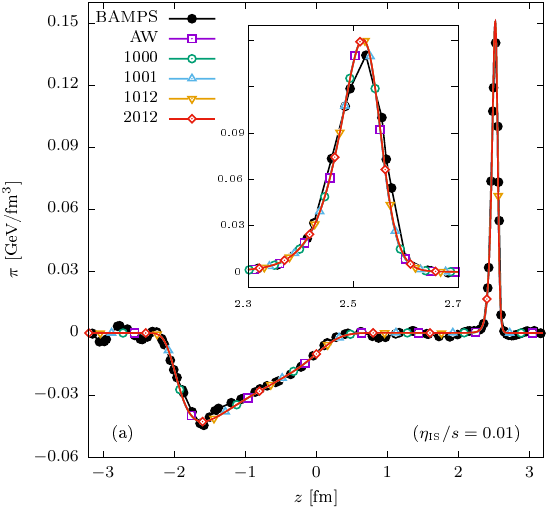} &
\includegraphics[width=0.95\columnwidth]{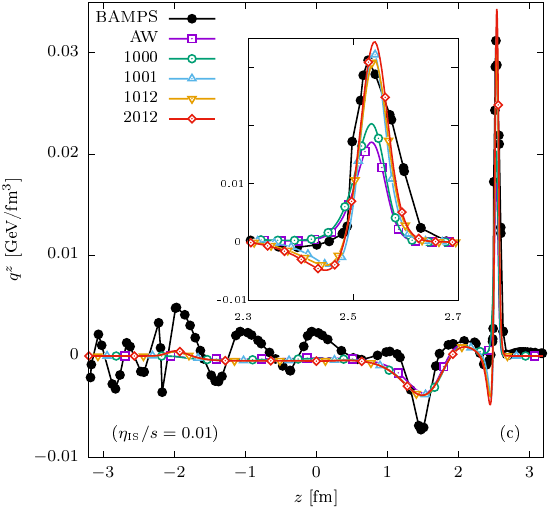} \\
\includegraphics[width=0.95\columnwidth]{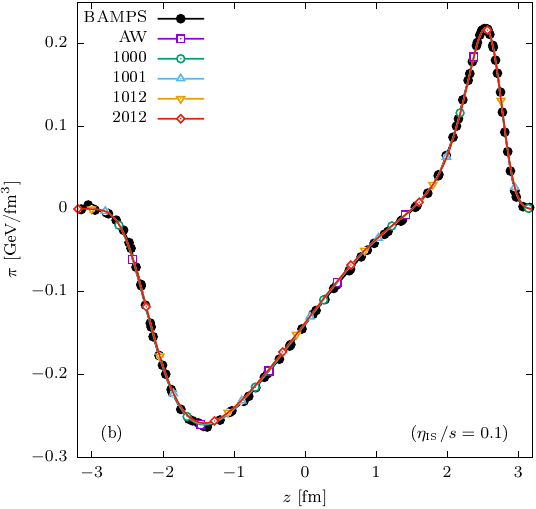} &
\includegraphics[width=0.95\columnwidth]{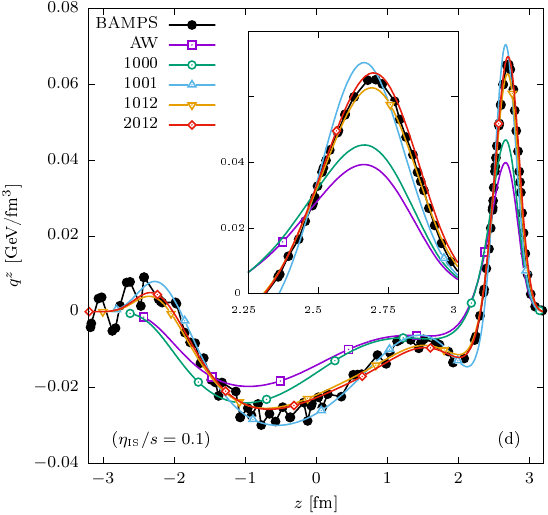}
\end{tabular}
\caption{Profile of (left) $\pi = \pi^{zz} / \gamma^2$ and (right) $q^z = -(e + P) V^z / n$, taken at $t = 3.2\ {\rm fm}$ for $\eta_{\rm IS} / s = 0.01$ (top) and $0.1$ (bottom). The BAMPS results shown using black lines and filled circles are taken from Ref.~\cite{Bouras:2010hm}. The AW model results are shown with the purple line and squares. The Shakhov results obtained using the $(1000)$, $(1001)$, $(1012)$ and $(2012)$ models are shown with lines and empty symbols (circles, upper triangles, lower triangles and rhombi, respectively). \label{fig:bouras}}
\end{figure*}

\subsection{Sod shock tube}\label{sec:riemann:bouras}

Our first test case will be the Sod shock tube setup presented in Ref.~\cite{Bouras:2010hm}, where the system is initialized at vanishing chemical potential, $\mu = 0$. The initial temperatures in the left ($x<0$) and right ($x > 0$) regions are $T_0=0.4\ \mathrm{GeV}$ and $T_4=0.2\ \mathrm{GeV}$, respectively. The shear-stress tensor and the diffusion current are zero at the initial time, but they develop nontrivial profiles during the subsequent evolution of the fluid.
For these quantities, we benchmark the performance of the various Shakhov models by comparison to the results obtained using the BAMPS algorithm.

Before discussing the results, a few remarks regarding our proposed comparison are in order. First, we note that the results presented in Figs.~3--7 of Ref.~\cite{Bouras:2010hm} probe the perfect fluid limit, being obtained at very low shear viscosity to entropy density ratio, $\eta_{\rm IS} / s = 0.001$, or at very large constant cross section, $\sigma = 224.431\ {\rm mb}$. In both cases, reasonable agreement with the analytical solution of the relativistic Euler equations is found, however such large cross-section simulations require a fine spatial resolution, making the simulations computationally very time consuming (see discussion at the start of Sec. IV.C of Ref.~\cite{Bouras:2010hm}). Consequently, as the corresponding BAMPS results are very noisy, we will not consider this limit in this paper and instead we will focus on the cases $\eta_{\rm IS} /s = 0.1$ and $0.01$, shown in Figs.~8--10 of Ref.~\cite{Bouras:2010hm}. 

To understand the comparison, we first note that, in the BAMPS simulations, the desired $\eta / s$ ratio was achieved by employing a local interaction cross-section $\sigma = 1 / n \lambda_{\rm mfp}$ leading to the desired value of $\eta$. Based on the more recent method of moments analysis of DNMR \cite{Denicol:2012cn}, we know that in a hard-sphere gas, the shear viscosity is $\eta \simeq 1.2676 / \beta \sigma$. However, in Ref.~\cite{Bouras:2010hm}, the Israel-Stewart relation $\eta_{\rm IS} = \frac{2}{5} e \lambda_{\rm mfp} = 1.2 / \beta \sigma$ is employed. Since the shear viscosity of the gas is given by the cross-section $\sigma$, we conclude that the results shown in Ref.~\cite{Bouras:2010hm} have a true $\eta / s$ ratio given by 
\begin{equation}
 \frac{\eta}{s} = \frac{\eta_{\rm IS}}{s} \times \frac{1.2676}{1.2} \simeq \frac{\eta_{\rm IS}}{s} \times 1.056,
 \label{eq:riem_etas}
\end{equation}
thus the actual ratio $\eta / s$ is $5.6\%$ larger than the one reported in Ref.~\cite{Bouras:2010hm}. 

The comparison between the Shakhov and BAMPS results is shown in Fig.~\ref{fig:bouras}. Remarkably, $\pi = \pi^{zz} / \gamma^2$ is recovered well in all considered models, showing that fixing the value of the shear viscosity alone is sufficient to capture the dynamics of the shear-stress tensor, as can be seen in panels (a) and (b) of Fig.~\ref{fig:bouras}.

The heat flux $q^z$, shown in panels (c) and (d) of Fig.~\ref{fig:bouras}, exhibits an unexpected sensitivity to the higher-order transport coefficients. For both $\eta_{\rm IS} / s = 0.01$ and $\eta_{\rm IS} / s = 0.1$, $q^z$ has two peaks: the first corresponds to the contact discontinuity, exhibiting a small, negative $q^z$; the second corresponds to the shock front and exhibits a large, positive $q^z$. Furthermore, at $\eta_{\rm IS} / s = 0.1$, the heat flux develops a non-trivial structure inside the rarefaction wave. None of these features are correctly recovered by the AW model. Remarkably, the shock front peak is almost two times taller in BAMPS than in the AW model. Clearly, fixing the diffusion coefficient to match the one in BAMPS is insufficient, as this provides a roughly $20\%$ increase in the height of the AW peak, as indicated by the $(1000)$ curves. Surprisingly, fixing the cross-coupling coefficients $\ell_{V\pi}$ and $\ell_{\pi V}$ within the $(1001)$ model provides a very good match not just at the level of the two peaks, but also throughout the rarefaction wave. 

From these plots, it can be seen that all extended Shakhov models ($1001$, $1012$ and $2012$) perform better than standard RTA, which is not able to reproduce especially the stronger variations seen in the BAMPS data. Interestingly, the first-order Shakhov model that fixes the diffusion coefficient $\kappa$ is still strongly inconsistent with the BAMPS data.

\begin{figure*}[ht]
\begin{tabular}{cc}
\includegraphics[width=\columnwidth]{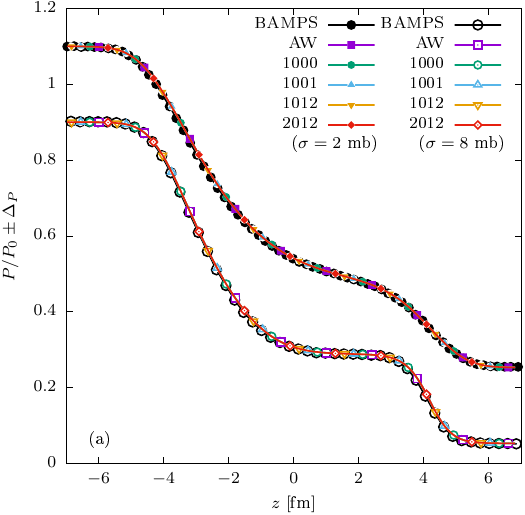} & 
\includegraphics[width=\columnwidth]{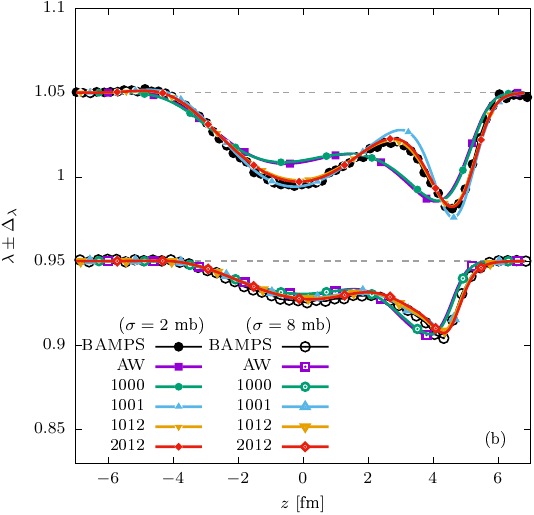} \\
\includegraphics[width=\columnwidth]{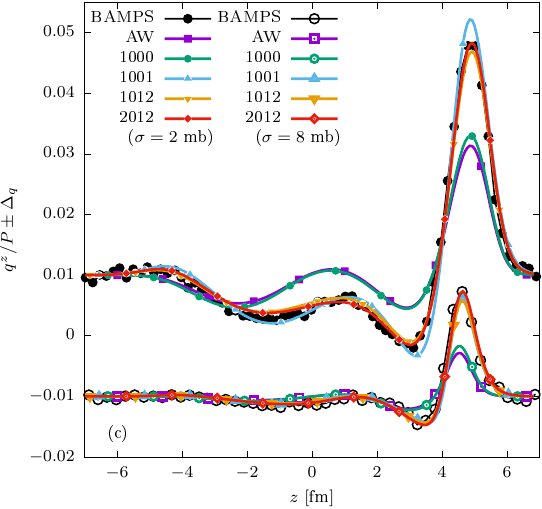} & 
\includegraphics[width=\columnwidth]{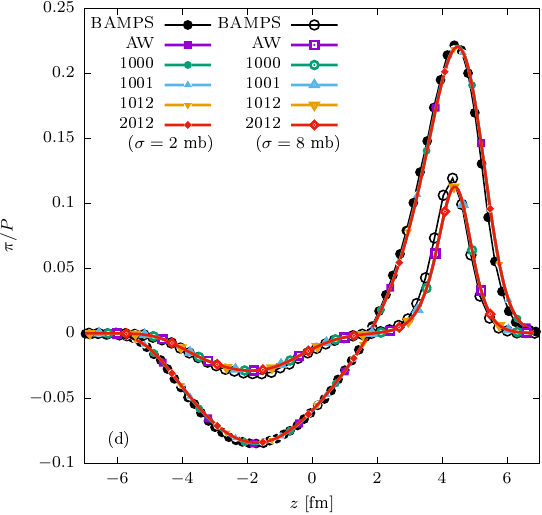}
\end{tabular}
\caption{
Profiles of (a) $P/P_0$, (b) $\lambda = e^{\alpha}$, (c) $q^z / P$ and (d) $\pi / P$ with respect to $z$ taken at $t = 6\ {\rm fm}$ for the initial conditions referred to as ``case 1,'' described in Sec.~\ref{sec:riemann:DNBMXRG}. The datasets corresponding to $\sigma = 2\ {\rm mb}$ (filled symbols) and $8\ {\rm mb}$ (open symbols) are shifted symmetrically by the quantities $+\Delta$ and $-\Delta$, respectively, with $\Delta_P = 0.1$, $\Delta_\lambda = 0.05$, $\Delta_q = 0.01$, and $\Delta_\pi = 0$. 
\label{fig:DNBMXRG1}
}
\end{figure*}

\begin{figure*}[t]
\begin{tabular}{cc}
\includegraphics[width=\columnwidth]{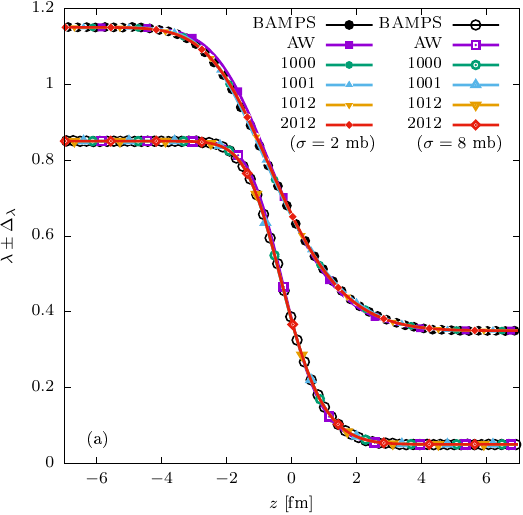} & 
\includegraphics[width=\columnwidth]{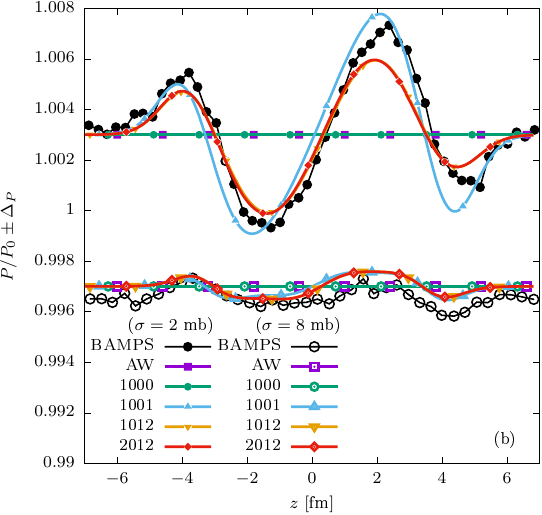} \\
\includegraphics[width=\columnwidth]{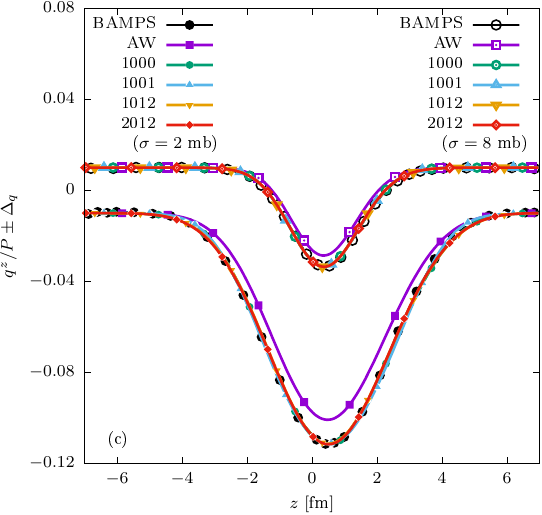} & 
\includegraphics[width=\columnwidth]{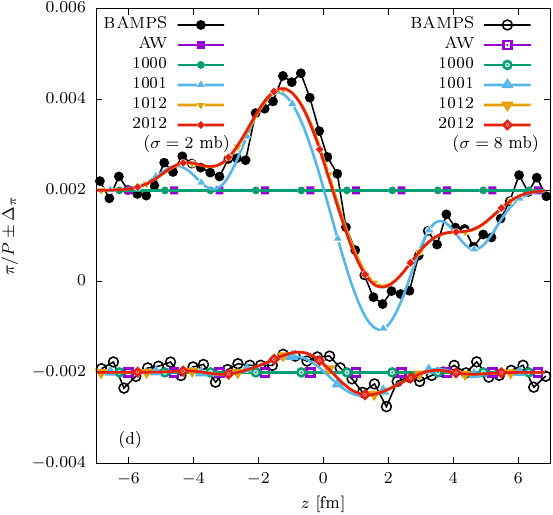}
\end{tabular}
\caption{
Profiles of (a) $\lambda = e^\alpha$, (b) $P/P_0$, (c) $q^z / P$ and (d) $\pi / P$ with respect to $z$ taken at $t = 6\ {\rm fm}$ for the initial conditions referred to as ``case 2.'' As in Fig.~\ref{fig:DNBMXRG1}, we have shifted the datasets with $\Delta_\lambda = 0.15$, $\Delta_P = 0.003$, $\Delta_q = -0.01$, and $\Delta_\pi = 0.002$. 
\label{fig:DNBMXRG2}
}
\end{figure*}

\subsection{Heat flow problem} \label{sec:riemann:DNBMXRG}

The second setup we consider is that of DNBMXRG \cite{Denicol:2012vq}, consisting of two individual cases. In both cases, the asymptotic left state is described by the temperature $T_{-\infty} = 0.4\ {\rm GeV}$ and vanishing chemical potential, $\mu_{-\infty} = 0$, such that the reference pressure becomes $P_0 \equiv P_{-\infty} = g T_{-\infty}^4 / \pi^2 = 5.401\ {\rm GeV}/{\rm fm}^3$ for a degeneracy factor of $g = 16$ and the asymptotic left fugacity is $\lambda_{-\infty} = e^{\mu_{-\infty} / T_{-\infty}} = 1$. The reference particle number density then reads $n_0 \equiv n_{-\infty} = P_{-\infty} / T_{-\infty} = 13.50\ {\rm fm}^{-3}$. The initial conditions are specified at the level of the pressure and fugacity, which are given by the Woods-Saxon profile:
\begin{equation}
 P(z) = P_\infty + \frac{P_0 - P_\infty}{1 + e^{z / D}}, \quad 
 \lambda(z) = \lambda_\infty + \frac{\lambda_0 - \lambda_\infty}{1 + e^{z / D}},
\end{equation}
with $D = 0.3\ {\rm fm}$ being the shock thickness. Considering the relations $P = g T^4 \lambda / \pi^2$ and $n = P / T$, the particle number density $n$ and temperature $T$ can be obtained from $P$ and $\lambda$ via
\begin{equation}
 T = \left(\frac{\pi^2 P}{g \lambda}\right)^{1/4}, \quad 
 n = \left(\frac{g \lambda}{\pi^2} P^3\right)^{1/4}.
\end{equation}
At initial time, the fluid is at rest [$u^\mu = (1,0,0,0)$] and the diffusion current and shear-stress tensor vanish identically, $V^\mu = \pi^{\mu\nu} = 0$. 

\begin{figure*}[ht]
\begin{tabular}{cc}
\includegraphics[width=\columnwidth]{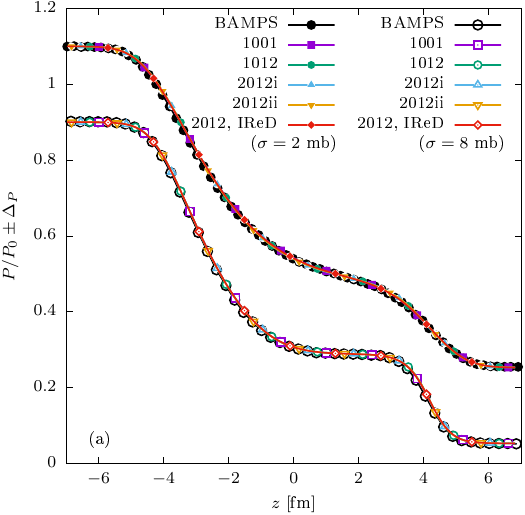} & 
\includegraphics[width=\columnwidth]{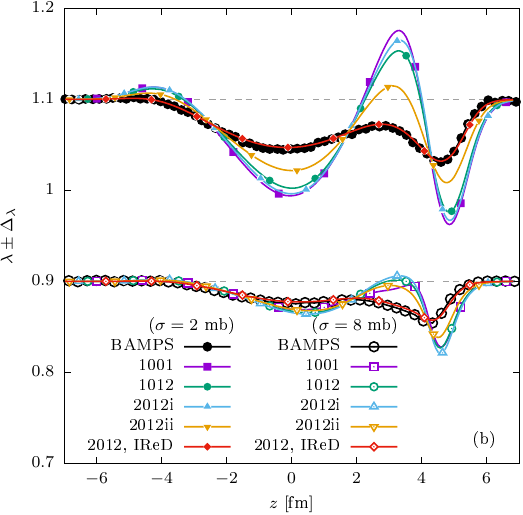} \\
\includegraphics[width=\columnwidth]{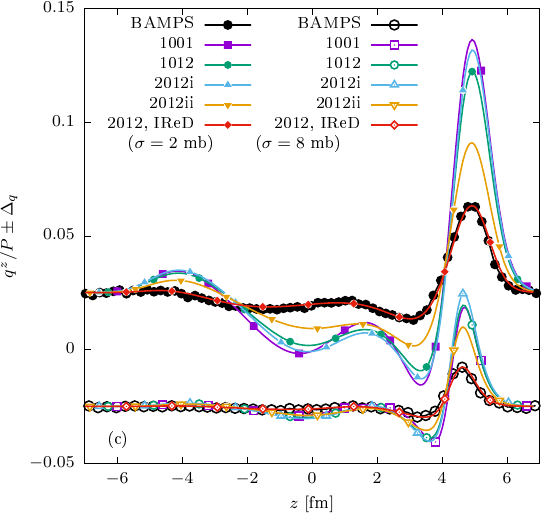} & 
\includegraphics[width=\columnwidth]{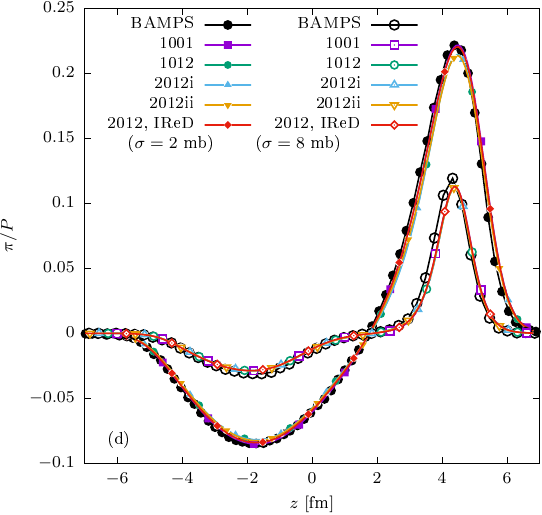}
\end{tabular}
\caption{
Same as Fig.~\ref{fig:DNBMXRG1} for the Shakhov model implementing the DNMR transport coefficients. The ``2012, IReD'' entry shown in red with rhombi corresponds to the results obtained using the IReD transport coefficients, being in excellent agreement with the BAMPS data.
\label{fig:DNBMXRG_DNMR}
}
\end{figure*}

Following Ref.~\cite{Denicol:2012vq}, we consider two sets of initial conditions, labeled as case (i) and case (ii). For case (i), the fugacity stays constant, $\lambda(z) = e^{\mu(z) / T(z)} = 1$, and the pressure drops to $P_\infty = g T_\infty^4 / \pi^2 = 0.824\ {\rm GeV}/{\rm fm}^3$, corresponding to $T_\infty = 0.25\ {\rm GeV}$. In summary, we have
\begin{align}
 (\textrm{i}): \quad 
 \lambda_\infty &= 1, &
 P_\infty &= 0.824\ {\rm GeV}/{\rm fm}^3,\nonumber\\
 T_\infty &= 0.250\ {\rm GeV}, &
 n_\infty &= 3.297\ {\rm fm}^{-3}. 
\end{align}
For case (ii), the pressure stays constant, $P(z) = P_0$, while the fugacity drops to its asymptotic right state, $\lambda_{\infty} = 0.2$, such that
\begin{align}
 (\textrm{ii}):\quad 
 \lambda_\infty &= 0.2, &
 P_\infty &= 5.401\ {\rm GeV}/{\rm fm}^3,\nonumber\\
 T_\infty &= 0.598\ {\rm GeV}, &
 n_\infty &= 9.030\ {\rm fm}^{-3}. 
\end{align}
The simulation domain spans $L = 14\ {\rm fm}$, such that $-L/2 < z < L/2$ and the total simulated time is $T = 6\ {\rm fm}$. 

As in the preceding section, we consider that the fluid is made up of ultrarelativistic hard-sphere particles interacting via an isotropic cross-section $\sigma$. Contrary to the situation in the previous section, $\sigma$ is fixed and we shall consider either $\sigma = 2\ {\rm mb}$ or $8\ {\rm mb}$. This means that the ratio $\eta / s$ is no longer constrained to be a constant. Nevertheless, the relaxation time $\tau_R$ is still fixed via Eq.~\eqref{eq:riem_tauR}, namely
\begin{equation}
 \tau_R = \frac{5\eta}{4P} = \frac{1.5845}{n \sigma},
\end{equation}
where $n \equiv n(z)$ is the local particle number density. The Shakhov model is then implemented exactly as discussed in the preceeding section, using $\lambda_{\rm mfp} = 1/n \sigma$. The numerical results are shown in Figs.~\ref{fig:DNBMXRG1} and \ref{fig:DNBMXRG2} and, before discussing them in detail below, we mention that for presentation purposes, we have chosen to show the results for both values of $\sigma$ together on the same canvas. For this purpose, we have shifted each quantity $A$ by an offset $\Delta_A$ for $\sigma = 2\ {\rm mb}$ and by $-\Delta_A$ for $\sigma = 8\ {\rm mb}$. 

Case (i) above is a milder version of the Sod shock tube problem considered in the previous subsection. First, the temperature jump is smaller ($T_\infty = 0.25\ {\rm GeV}$ compared to $0.2\ {\rm GeV}$ considered in Sec.~\ref{sec:riemann:bouras}), and second, the initial discontinuity is smoothed out by the Woods-Saxon profile. Fig.~\ref{fig:DNBMXRG1} shows the comparison between the BAMPS data and our simulation results. Panels (a) and (d) show the normalized pressure $P / P_0$ and the shear-stress tensor coefficient $\pi \equiv \pi^{zz} /\gamma^2$. All models (including the AW model) are in good agreement with the BAMPS data, confirming that the dynamics of these quantities are dominated by the shear viscosity $\eta$, which is correctly recovered by all considered models. 

Panels (b) and (c) of Fig.~\ref{fig:DNBMXRG1} show the fugacity, $\lambda = e^{\mu / T}$, and the heat flux $q^z = -(e + P) V^z / n$. For the larger cross-section, $\sigma = 8\ {\rm mb}$, shown with empty symbols, all model results seem to be consistent with the BAMPS data at the level of the fugacity. However, panel (c) shows that the AW and $(1000)$ models significantly underestimate the peak values of the heat flux, while the higher-order models ($1001$, $1012$ and $2012$) capture the BAMPS data quite accurately. For the lower cross-section, $\sigma = 2\ {\rm mb}$, shown with full symbols, the AW and $(1000)$ models deviate significantly from the BAMPS data for both $\lambda$ and $q^z$. The $(1001)$ model provides a significant improvement, however one may observe a slight discrepancy in the $\lambda$ profile before the onset of the shock front, as well as a slight overestimation of the heat-flux peak. On the other hand, both high-order models ($1012$ and $2012$) are in excellent agreement with BAMPS.

The setup considered as ``case (ii),'' shown in Fig.~\ref{fig:DNBMXRG2}, favours a flow pattern dominated by the heat flux. While the fugacity $\lambda$, shown in panel (a), is excellently captured by all models, the pressure fluctuations, arising at a subpercent level around the initial value, are completely missed by the AW and $1000$ models. The higher-order models give a reasonable representation of the general trend of these fluctuations, with the $1001$ model performing (surprisingly) marginally better than the $1012$ and $2012$ models, however the small amplitude of the oscillations may indicate that transport phenomena beyond second order may play an important role. The same remarks hold equally for the shear-stress coefficient $\pi$, shown in panel (d).

Surprisingly, the heat flux profile shown in panel (c) of Fig.~\ref{fig:DNBMXRG2} is recovered by all Shakhov models, while the AW model underestimates its magnitude by roughly $11\%$. The first-order Shakhov model $(1000)$, implementing the correct value of the diffusion coefficient, already provides an excellent match to the BAMPS data. 
Furthermore, the $(1001)$ model already captures the main features of the pressure and shear-stress tensor fluctuations, while the higher order models ($1012$ and $2012$) provide a marginal improvement over the $(1001)$ model.

\begin{table*}
\begin{tabular}{||l||c|c|c|c|c||c|c|c|c|c||}
	\hline\hline 
	Model 
	& $\eta\sigma \beta$ 
    & $\tau_\pi/\lambda_{\rm mfp}$ 
    & $\beta \ell_{\pi V} / \tau_\pi$
	& $\tau_{\pi\pi} / \tau_\pi$
	& $\beta \lambda_{\pi V} / \tau_\pi$
    & $\kappa \sigma$ 
    & $\tau_V / \lambda_{\rm mfp}$ 
	& $\ell_{V\pi} / \beta \tau_V$
	& $\lambda_{VV}/\tau_V$ 
	& $\lambda_{V\pi} / \beta \tau_V$  \\[2pt] \hline
	{\bf HS}
    & $\mathbf{1.2676}$
	& $\mathbf{2}$
	& $\mathbf{-0.68317}$
	& $\mathbf{1.6888}$ 
	& $\mathbf{0.24188}$
    & $\mathbf{0.15892}$
	& $\mathbf{2.5721}$
	& $\mathbf{0.11921}$
	& $\mathbf{0.92095}$ 
	& $\mathbf{0.051709}$ \\[2pt] \hline\hline
    $1001$ 
    & $\mathbf{1.2676}$
	& $1.98958$
	& $\mathbf{-0.68317}$
	& $1.43987$
	& $0$
	& $\mathbf{0.15892}$
	& $1.9070$
	& $\mathbf{0.11921}$
	& $0.6$ 
	& $0.032698$ \\[2pt] \hline
    $1012$ 
    & $\mathbf{1.2676}$
	& $\mathbf{2}$
	& $\mathbf{-0.68317}$
	& $\mathbf{1.6888}$ 
	& $\mathbf{0.24188}$
    & $\mathbf{0.15892}$
	& $\mathbf{2.5721}$
	& $\mathbf{0.11921}$
	& $0.76998$
	& $0.043070$ \\[2pt] \hline
    $2012$ 
    & $\mathbf{1.2676}$
	& $\mathbf{2}$
	& $\mathbf{-0.68317}$
	& $\mathbf{1.6888}$ 
	& $\mathbf{0.24188}$
    & $\mathbf{0.15892}$
	& $\mathbf{2.5721}$
	& $\mathbf{0.11921}$
	& $\mathbf{0.92095}$ 
	& $\mathbf{0.051709}$ \\[2pt] \hline\hline
\end{tabular}
\caption{
Transport coefficients for the hard-sphere gas of ultrarelativistic particles interacting via an isotropic cross-section $\sigma$, computed using the original DNMR approach \cite{Denicol:2012cn,Wagner:2023joq}, as well as for the high-order Shakhov models $(1001)$, $(1012)$, and $(2012)$ considered in Sec.~\ref{sec:riemann:DNMR}.
Besides the transport coefficients shown above, all models have $\delta_{VV} = \tau_V$, $\delta_{\pi\pi} = 4\tau_\pi / 3$, $\tau_{V\pi} = \ell_{V\pi}$, and $\tau_{\pi V} = 4\ell_{\pi V}$.
\label{tbl:riem_tcoeffs_DNMR}
}
\end{table*}

\subsection{Using the DNMR transport coefficients}\label{sec:riemann:DNMR}

Before ending this section, we remark that the point of the original DNBMXRG paper \cite{Denicol:2012vq} was to demonstrate that one can obtain agreement with the BAMPS data within the framework of second-order fluid dynamics by increasing the number of dynamical moments. Surprisingly, Figs.~2 and 3 of Ref.~\cite{Denicol:2012vq} indicate that keeping the original $14$ dynamical moments -- for ultrarelativistic particles, just $13$ moments, namely $n$, $u^\mu$, $e$, $V^\mu$, and $\pi^{\mu\nu}$ -- and increasing the accuracy for the computation of the transport coefficients within the DNMR method of moments leads to worse agreement with BAMPS. 

The reason for this apparently divergent behaviour lies in the way the second-order hydrodynamics scheme is set up. In the original DNMR framework, one encounters second-order terms of the type ${\rm Re}^{-2}$ (not considered in our present work), ${\rm Re}^{-1} {\rm Kn}$ (the $\mathcal{J}$, $\mathcal{J}^\mu$, and $\mathcal{J}^{\mu\nu}$ terms), and ${\rm Kn}^2$ (the $\mathcal{K}$, $\mathcal{K}^\mu$, and $\mathcal{K}^{\mu\nu}$ terms). The latter ${\rm Kn}^2$ terms are parabolic and must thus be omitted from a hydrodynamics solver, as is done also in Ref.~\cite{Denicol:2012vq}. Within the IReD framework of Ref.~\cite{Wagner:2022ayd}, these ${\rm Kn}^2$ terms are consistently absorbed in the ${\rm Re}^{-1} {\rm Kn}$ terms, which leads to a restoration of the second-order accuracy by a modification of the second-order transport coefficients appearing in the ${\rm Re}^{-1} {\rm Kn}$ terms. Hence, employing the values of the transport coefficients for the ${\rm Re}^{-1} {\rm Kn}$ terms derived in the DNMR framework while discarding the $O({\rm Kn}^2)$ terms cannot lead to a hydrodynamic model which is second-order accurate, hence the persistent discrepancy to the BAMPS data.

To test this, we consider again the Shakhov models discussed until now, tuned to recover the DNMR transport coefficients for hard-sphere interactions, shown in Table~\ref{tbl:riem_tcoeffs_DNMR}. The equivalent AW and $(1000)$ models are evidently identical to the IReD ones, since they fix only the first-order transport coefficients, which are identical between the IReD and DNMR approaches. For the $(1001)$ model, we employed $\ell_{V\pi} = 0.11921 \beta \tau_V$ and $\ell_{\pi V} = -0.68317 \tau_\pi / \beta$. The higher order $(1012)$ and $(2012)$ models are derived in Sections~\ref{sec:1012_DNMR} and \ref{sec:2012_DNMR} of the appendix. In all cases, we summarize the resulting transport coefficients in Table~\ref{tbl:riem_tcoeffs_DNMR}

Our results shown in Fig.~\ref{fig:DNBMXRG_DNMR} confirm the interpretation that the transport coefficients derived within the DNMR framework are ill-suited for hydrodynamical simulations. While the pressure [panel (a)] and shear-stress coefficient $\pi$ [panel (d)] profiles are recovered well by all models, showing that the dynamics of these quantities is dominated just by the value of the shear viscosity $\eta$, the fugacity $\lambda$ [panel (b)] and most of all, the heat flux [panel (c)], are strongly sensitive to the second-order transport coefficients and the Shakhov model results with DNMR coefficients exhibit strong disagreement to the BAMPS data. One can see again that employing the IReD transport coefficients within the high-order $(2012)$ model gives excellent agreement with the BAMPS data.


\section{Conclusions} \label{sec:conc}

In this work, we presented a general method to extend the standard Anderson-Witting (AW) relaxation-time approximation (RTA) for the Boltzmann equation via a high-order Shakhov-like construction. Depending on the chosen truncation, the model allows for a varying number of transport coefficients to be fixed from the hydrodynamic limit of kinetic theory.
To validate our construction, we first considered two canonical setups: the Bjorken flow for massive particles and longitudinal waves for massless particles, where we compared simulation results between the Shakhov model and second-order Israel-Stewart-like hydrodynamics with transport coefficients taken from the Shakhov model. In the context of the Bjorken flow, we demonstrated that the Shakhov model can be used to simultaneously tweak the shear viscosity $\eta$, bulk viscosity $\zeta$, and the bulk-shear coupling coefficient $\lambda_{\Pi \pi}$. In the longitudinal waves setup, we demonstrated that the Shakhov model is able to simultaneously tweak the shear viscosity $\eta$, particle diffusivity $\kappa$, and the shear-diffusion cross-coupling coefficients, $\ell_{V\pi}$ and $\ell_{\pi V}$. The results of the kinetic Shakhov model were in good agreement with those from hydrodynamical simulations with the transport coefficients expected from the Shakhov model, validating the hydrodynamic limit of the Shakhov model.

The third example that we considered consisted of several incarnations of the Riemann problem, where we employed the Shakhov model to reproduce the solution of the full Boltzmann equation obtained using the BAMPS (Boltzmann approach to multi-parton scattering) code for massless particles interacting via an isotropic cross-section $\sigma$. The transport coefficients of such a hard-sphere gas were computed using the method of moments in the famous DNMR paper \cite{Denicol:2012cn}, and we took them in the IReD formulation that completely avoids parabolic, $O({\rm Kn}^2)$ terms \cite{Wagner:2022ayd} (see also Ref.~\cite{Wagner:2023joq} for a discussion on the analytical structure of the hard-sphere collision matrix). Here we considered several models, in increasing order of complexity:
\begin{itemize}
\item The AW model, which recovers just the shear viscosity $\eta$.
\item The first-order Shakhov model, labeled $(1000)$, recovering both $\eta$ and the particle diffusivity $\kappa$;
\item The $(1001)$ model, discussed in the context of longitudinal waves, recovering $\eta$, $\kappa$, as well as the diffusion-shear coupling coefficients $\ell_{V\pi}$ and $\ell_{\pi V}$;
\item The $(1012)$ model, recovering all transport coefficients except $\lambda_{VV}$ and $\lambda_{V\pi}$;
\item The $(2012)$ model, recovering all first- and second-order transport coefficients of the hard-sphere model.
\end{itemize}
In the above, we employed the $(N_1, N_2, s_1, s_2)$ models introduced in Sec.~\ref{sec:shk2:dofs}. As explained in Eq.~\eqref{eq:dofs_UR}, such models benefit from $2(N_1 + N_2 + s_1 + s_2)$ independent degrees of freedom, plus the overall relaxation time $\tau_R$. For simplicity, we took $\tau_R$ to be related to the model's shear viscosity $\eta$ through the standard RTA relation, $\tau_R = 5\eta / 4P$. Then, the $(1000)$ model has $2$ parameters, used to fix $\eta$ and $\kappa$; the $(1001)$ model has $4$ parameters, fixing $\eta$, $\kappa$, $\ell_{V\pi}$, and $\ell_{\pi V}$; the $(1012)$ model has $8$ parameters, fixing $\eta$, $\kappa$, $\tau_\pi$, $\tau_V$, $\ell_{V\pi}$, $\ell_{\pi V}$, $\tau_{\pi\pi}$, and $\lambda_{\pi V}$; finally, the $(2012)$ model has $10$ free parameters, fixing also $\lambda_{VV}$ and $\lambda_{V\pi}$.

Our first conclusion was that the dynamics of both the pressure $P$ and the shear-stress tensor $\pi^{\mu\nu}$ is dominated by the shear viscosity $\eta$ and is thus accurately recovered by all considered models, including the AW mode. Contrary to expectations, the dynamics of the heat flux $q^\mu$ requires more than just fixing the diffusion coefficient $\kappa$. While performing in principle better than the AW model, the first-order $(1000)$ Shakhov model is still in visible disagreement with the BAMPS data. A notable exception is that of ``case 2'' from DNBMXRG \cite{Denicol:2012vq}, where both $P$ and $\pi^{\mu\nu}$ exhibit subleading fluctuations and the dynamics of $q^\mu$ is dominated by $\kappa$, such that the $(1000)$ model is in excellent agreement with the BAMPS data. In a more general setting, cross-couplings are important and the $(1001)$ model provides a significant improvement over the AW and $(1000)$ models by correctly capturing the $\ell_{V\pi}$ and $\ell_{\pi V}$ cross-coupling coefficients. The small discrepancies with respect to the BAMPS data are almost entirely removed when considering the $(1012)$ and $(2012)$ models. 

We must remark that in this paper, we have focused entirely on the linearized part of the collision term. For a generic $2 \rightarrow 2$ scattering, the moments of the collision term receive both a linear and a quadratic contribution \cite{Molnar.2014}. Since our models already display very good agreement to the BAMPS data, we can only conclude that for the instances of the Riemann problem considered in this paper, these missing ${\rm Re}^{-2}$ terms make subleading contributions, however it is not inconceivable that such terms may become important in more general settings.

The setups presented here can be applied to more complex $(3+1)$-dimensional problems, in particular related to the study of heavy-ion collisions. While usually in simulations a hydrodynamic stage is followed by an afterburner which is based on kinetic theory, a Shakhov-type model can cover both regimes, eliminating the need for particlization. The Shakhov extension of the RTA provides the means to determine realistic models for, e.g., the shear and bulk viscosities, by employing Bayesian analysis on numerical and experimental data, as discussed in Ref.~\cite{Bernhard:2019bmu}. This should bring an important improvement to the results reported in Ref.~\cite{Bernhard:2019bmu}, where pure hydrodynamics was used to obtain the numerical data, since hydrodynamics is known to be inaccurate during preequilibrum, when the quark-gluon plasma fluid is in a far-from-equilibrium state \cite{Ambrus:2022qya}.

\begin{acknowledgments}
We thank A. Palermo, D. Rischke and P. Aasha for fruitful 
discussions.  
The authors gratefully acknowledge support through a grant of the Ministry of Research, Innovation and Digitization, CNCS - UEFISCDI, project number PN-III-P1-1.1-TE-2021-1707, within PNCDI III; support by the Deutsche Forschungsgemeinschaft (DFG, German Research Foundation) through the CRC-TR 211 'Strong-interaction matter under extreme conditions'– project number 315477589 – TRR 211; and support by the European Union's Horizon 2020 research and innovation program under grant agreement No.~824093 (STRONG-2020). We also acknowledge Dr. Flotte for hospitality and fruitful discussions.
\end{acknowledgments}

\appendix

\section{Numerical scheme}\label{app:numsch}

In this section of the appendix, we discuss the method employed to obtain numerical solutions of the Boltzmann equation with the Shakhov-like collision model. The scheme extends that introduced in Refs.~\cite{Ambrus:2023qcl} and \cite{Ambrus:2023ilm} and employs the so-called rapidity-based moments, summarized in Sec.~\ref{app:numsch:beq}. Sections~\ref{app:numsch:bjork} and \ref{app:numsch:riem} summarize the strategy employed for the Bjorken flow and Riemann problems, respectively. Section~\ref{app:numsch:disc} summarizes the discretization scheme and numerical algorithm. Finally, we provide a note on code availability in Sec.~\ref{app:numsch:code}.

\subsection{Boltzmann Equation in \texorpdfstring{1+1}{1+1}-D Minkowski space and \texorpdfstring{0+1}{0+1}-D Milne space} \label{app:numsch:beq}

The applications considered in the validation sections concern flows which are homogeneous with respect to the transverse plane spanned by $\mathbf{x}_\perp$, such that the Boltzmann equation reads
\begin{equation}
 k^t \partial_t \fk + k^z \partial_z \fk = -\frac{E_\bk}{\tau_R} (\fk - \fS).
 \label{eq:app_boltz}
\end{equation}
We parametrize the momentum space as
\begin{equation}
 \begin{pmatrix} k^t \\ k^z \end{pmatrix} = m_\perp \begin{pmatrix} \cosh y \\ \sinh y \end{pmatrix},\quad 
 \begin{pmatrix} k^x \\ k^y \end{pmatrix} = k_\perp \begin{pmatrix} \cos \varphi_k \\ \sin \varphi_k \end{pmatrix},
\end{equation}
where $m_\perp = \sqrt{\bk_\perp^2 + m^2}$ is the transverse mass, $y= \tanh^{-1} v^z$ is the rapidity, and $v^z = k^z / k^t$. Dividing Eq.~\eqref{eq:app_boltz} by $k^t$, we arrive at
\begin{equation}
 \partial_t \fk + v^z \partial_z \fk = -\frac{\gamma(1 - \beta^z v^z)}{\tau_R} (\fk - \fS),
 \label{eq:app_boltz_shock}
\end{equation}
where we assumed that the transverse components of the four-velocity vanish, $u^\mu  \partial_\mu = \gamma(\partial_t + \beta^z \partial_z)$.
Equation \eqref{eq:app_boltz_shock} is appropriate to analyze the sound and shock wave propagation problems considered in Sections~\ref{sec:long} and \ref{sec:riemann}.

In the case of the Bjorken flow considered in Sec.~\ref{sec:bjork}, invariance with respect to longitudinal boosts imposes the velocity profile $u^\mu \partial_\mu = \tau^{-1} (t \partial_t + z \partial_z)$, where $\tau = \sqrt{t^2 - z^2}$ is the Bjorken time. Introducing the space-time rapidity $\eta = \tanh^{-1} (z / t)$, such that 
\begin{equation}
 t = \tau \cosh \eta, \quad 
 z = \tau \sinh \eta, 
\end{equation}
Eq.~\eqref{eq:app_boltz} becomes
\begin{equation}
 \frac{\partial f_\bk}{\partial \tau} + \tanh(y - \eta) \frac{1}{\tau} \frac{\partial \fk}{\partial \eta} = -\frac{1}{\tau_R}(f_\bk - \fS),\label{eq:app_boltz_bjork_aux}
\end{equation}
where the functional dependence of the distribution function is $f_\bk \equiv f_\bk(\tau, \mathbf{x}_\perp, \eta; m_\perp, \varphi_k, \tanh y)$. 
Boost invariance dictates that $f_\bk$ depends on $\eta$ and $y$ only through their difference. It is therefore convenient to parametrize the momentum space using $(m_\perp, \varphi_k, v^z)$ with $v^z \equiv \tanh(y - \eta)$, such that Eq.~\eqref{eq:app_boltz_bjork_aux} finally reads \cite{Ambrus:2023qcl,Ambrus:2023ilm}
\begin{equation}
 \frac{\partial f_\bk}{\partial \tau} - \frac{v^z(1 - v_z^2)}{\tau} \frac{\partial \fk}{\partial v^z} = -\frac{1}{\tau_R}(f_\bk - \fS),\label{eq:app_boltz_bjork}
\end{equation}
where in the above, $f_\bk \equiv f_\bk(\tau; m_\perp, \varphi_k, v_z)$. This parametrization can be related to that arising when the curvilinear coordinates $(\tau, \eta)$ are employed, when it is convenient to introduce
\begin{align}
 k^\tau &= \frac{\partial \tau}{\partial t} k^t + \frac{\partial \tau}{\partial z} k^z = m_\perp \cosh (y - \eta),\nonumber\\
 k^{\eta} &= \frac{\partial \eta}{\partial t} k^t + \frac{\partial \eta}{\partial z} k^z = \frac{m_\perp}{\tau} \sinh (y - \eta),
\end{align}
such that $v^z =\tau k^{\eta}/k^\tau$.

The macroscopic moments of the distribution function are obtained after integration over the momentum space using the integration measure $\d^3k / k^0$, which reads $\d k^x \d k^y \d k^z / k^t$ on Minkowski space and $\tau \d k^x \d k^y \d k^{\eta} / k^\tau$ on Milne space. In both cases, when employing the appropriate parametrization $(m_\perp, \varphi_k, v^z)$, this integration measure becomes 
\begin{equation}
 \int \frac{\d^3k}{k^0} \rightarrow  
 \int_{-1}^1 \frac{\d v^z}{1-v_z^2} \int_0^{2\pi} \d \varphi_k \int_m^\infty \d m_\perp m_\perp\;.
\end{equation}

In the case of the $0+1$-D and $1+1$-D flows considered in this paper,
the parametrization of the momentum space using the rapidity-based degrees of freedom $(m_\perp, \varphi_k, v^z)$ allows the dimensionality of the momentum space to be reduced from $3$ degrees of freedom to a single one, namely $v^z$. Introducing the rapidity-based moments 
\begin{equation}
 F_n(v^z) = \frac{g}{(2\pi)^3} \int_0^{2\pi} \d\varphi_k \int_{m}^\infty \frac{\d m_\perp\, m_\perp^{n+1}}{(1-v_z^2)^{(n+2)/2}} f_\bk,
 \label{eq:Fn_def}
\end{equation}
it is possible to recast Eq.~\eqref{eq:app_boltz_shock} as
\begin{subequations}
\begin{equation}
 \frac{\partial F_n}{\partial t} + v^z \frac{\partial F_n}{\partial z} = -\frac{\gamma(1 - \beta^z v^z)}{\tau_R} (F_n - F^{\rm S}_n), \label{eq:bjorken_eq_Fn}
\end{equation}
while Eq.~\eqref{eq:app_boltz_bjork} becomes:
\begin{multline}
 \frac{\partial F_n}{\partial\tau } + \frac{1}{\tau} [1 + (n - 1) v_z^2] F_n - \frac{1}{\tau} \frac{\partial [v^z(1 - v_z^2) F_n]}{\partial v^z} \\
 = -\frac{1}{\tau_R}(F_n - F_n^{\rm S}). \label{eq:riemann_eq_Fn}
\end{multline}
\end{subequations}
The functions $F_n^{\rm S}$ corresponding to the Shakhov distribution $f_{{\rm S}\bk}$ must be computed using Eq.~\eqref{eq:Fn_def}. A general expression is cumbersome and uninsightful, such that we defer the details of this computation to the following subsections, where we will separately make use of the particular symmetries of the $0+1$-D Bjorken flow of massive particles and the $1+1$-D Riemann problem for massless particles.


\subsection{Strategy for Bjorken flow} \label{app:numsch:bjork}

Due to the symmetries of the $0+1$-D Bjorken flow, $T^{\mu\nu} = {\rm diag}(e, P_T, P_T, \tau^{-2} P_L)$ has a diagonal form, while $u^\mu \partial_\mu = \partial_\tau$. The quantities $e = T^{\tau\tau}$, $P_L = \tau^2 T^{\eta \eta}$ and $P_T = \frac{1}{2}(e - P_L - T^\mu{}_\mu)$ can be obtained from the moments $F_n$ using
\begin{align}
 T^\mu{}_\mu &= m^2 \int_{-1}^1 \d v^z F_0,\nonumber\\ 
 \begin{pmatrix}
  T^{\tau\tau} \\ \tau^2 T^{\eta\eta} 
 \end{pmatrix} &= \int_{-1}^1 \d v^z 
 \begin{pmatrix}
     1 \\ v_z^2
 \end{pmatrix} F_2.
 \label{eq:T_tt_zz}
\end{align}
In the Bjorken flow setup, we consider only the case of a gas in chemical equilibrium ($\alpha = 0$), for which the vector moments are not relevant. 

The requirements of orthogonality to $u^\mu$, tracelessness and transverse-plane isotropy constrains the form of the irreducible rank-2 tensors, written with respect to the Bjorken coordinates $(\tau,x,y,\eta)$, to
\begin{align}
 \rho_r^{\mu\nu} &= {\rm diag}\left(0, -\frac{1}{2}, -\frac{1}{2}, \frac{1}{\tau^2}\right) \pi_r,
\end{align} 
such that their dynamics can be described by the scalar quantities $\pi_r$, with $\pi_0 = \frac{2}{3}(P_L - P_T)$. In terms of the functions $F_n$, the quantities $\pi_r$ can be written as:
\begin{equation}
 \pi_r = \frac{1}{3} \int_{-1}^1 \d v^z [(3v_z^2 - 1) F_{r+2} + m^2 F_r].
 \label{eq:bjorken_pir}
\end{equation}
Similarly, the non-equilibrium contributions $\rho_r$ to the scalar moments can be obtained via
\begin{equation}
 \rho_r = \int_{-1}^1 \d v^z (F_r - F_r^{\rm eq}),
\end{equation}
with the bulk viscous pressure given by $\Pi = -\frac{m^2}{3} \rho_0 = \frac{1}{3} (e - 3P - T^\mu{}_\mu)$.

We finally discuss the construction of the Shakhov functions $F^S_n$. In the case of the ideal gas considered in this section, the Shakhov term $f_{{\rm S}\bk}$ becomes  
\begin{equation}
 f_{{\rm S}\bk} = f_{0\bk}(1 + \mathbb{S}_\bk),
\end{equation}
with $\mathbb{S}_\bk$ given by Eq.~\eqref{eq:bjorken_Sk}. Using Eq.~\eqref{eq:bjorken_h}, $\mathbb{S}_\bk$ becomes 
\begin{multline}
 \mathbb{S}_\bk = \begin{pmatrix} 
  \pi_{{\rm S};-2} & \pi_{{\rm S};0} 
\end{pmatrix} \begin{pmatrix}
    j_4 & -j_2 \\
    -j_2 & j_0
\end{pmatrix} \begin{pmatrix}
    1 \\
    m_\perp^2 / (1 - v_z^2)
\end{pmatrix} \\\times 
\left(\frac{3v_z^2 - 1}{2}+ \frac{m^2}{2} \frac{1 - v_z^2}{m_\perp^2}\right),
\end{multline}
where $j_n = J_{n2} / [2(J_{02} J_{22} - J_{42}^2)]$.
Specifically, we find
\begin{multline}
 F^{\rm S}_n = F^{\rm eq}_n + \begin{pmatrix} 
  \pi_{{\rm S};-2} & \pi_{{\rm S};0} 
\end{pmatrix} \begin{pmatrix}
    j_4 & -j_2 \\
    -j_2 & j_0
\end{pmatrix} \\\times 
\left[\frac{3v_z^2 - 1}{2} \begin{pmatrix}
 F_n^{\rm eq} \\ 
 F_{n+2}^{\rm eq}  
\end{pmatrix} + \frac{m^2}{2} \begin{pmatrix}
 F_{n-2}^{\rm eq} \\ 
 F_{n}^{\rm eq}
\end{pmatrix}\right],
\end{multline}
where the functions $F_n^{\rm eq}$ introduced in Eq.~\eqref{eq:Fn_def} evaluate to
\begin{equation}
 F_n^{\rm eq} = \frac{g e^\alpha}{4\pi^2} T^{n+2} \,
 \Gamma\left(n + 2, \frac{m / T}{\sqrt{1 - v_z^2}}\right) \; ,
 \label{eq:Fn_eq}
\end{equation}
with $\Gamma(n,x) = \int_{x}^{\infty} t^{n-1} e^{-t} \d t$ being the incomplete
Gamma function. The coefficients $\pi_{{\rm S};-2}$ and $\pi_{{\rm S};0}$ characterizing the Shakhov distribution can be calculated from the coefficients $\pi_{-2}$ and $\pi_0$ obtained from $f_\bk$ using the following matrix equation:
\begin{equation}
 \begin{pmatrix}
     \pi_{{\rm S};-2} \\ \pi_{{\rm S};0} 
 \end{pmatrix} = \begin{pmatrix}
     1 - \tau_R \mathcal{A}^{(2)}_{{\rm S};-2,-2} & -\tau_R \mathcal{A}^{(2)}_{{\rm S};-2,0} \\
     - \tau_R \mathcal{A}^{(2)}_{{\rm S};0,-2} & 1-\tau_R \mathcal{A}^{(2)}_{{\rm S};0,0}
 \end{pmatrix} \begin{pmatrix}
     \pi_{-2} \\ \pi_0
 \end{pmatrix},
\end{equation}
with the matrix elements $\mathcal{A}^{(2)}_{{\rm S};rn}$ given in Eq.~\eqref{eq:bjorken_A2}. 
Taking into account that $\pi_r$ can be expressed in terms of $F_{r+2}$ and $F_r$, as shown in Eq.~\eqref{eq:bjorken_pir}, the above relation shows that the Boltzmann equation for the rapidity-based moments, Eq.~\eqref{eq:bjorken_eq_Fn}, can be closed by tracking the evolution of $F_{-2}$, $F_0$ and $F_2$.

\subsection{Strategy for \texorpdfstring{$\mathbf{1+1}$}{1+1}-D flows} \label{app:numsch:riem}

We now move to the case of $1+1$-D flows. Due to the symmetries of a system with transverse-space homogeneity, the only non-vanishing components of $T^{\mu\nu}$ are its diagonal ones, as well as $T^{tz} = T^{zt}$. The components $T^{tt}$ and $T^{zz}$ are as in Eq.~\eqref{eq:T_tt_zz}, the trace cancels ($T^\mu{}_\mu = 0$), while $T^{tz}$ is given by
\begin{align}
 T^{tz} &= \int_{-1}^1 \d v^z \, v^z F_2.
\end{align}
Due to isotropy in the transverse plane, the transverse components satisfy $T^{xx} = T^{yy} =\frac{1}{2} (T^{tt} -  T^{zz})$. 
The Landau frame can be obtained by solving the eigenvalue equation $T^\mu{}_\nu u^\nu = e u^\mu$, i.e.
\begin{equation}
 \begin{pmatrix}
  T^{tt} & -T^{tz} \\ 
  T^{tz} & -T^{zz} 
 \end{pmatrix}
 \begin{pmatrix}
     1 \\ \beta^z
 \end{pmatrix} =  e
 \begin{pmatrix}
     1 \\ \beta^z
 \end{pmatrix}.
\end{equation}
This leads to the solution
\begin{equation}
 e = \frac{1}{2} [T^{tt} - T^{zz} + \sqrt{(T^{tt} + T^{zz})^2 - 4(T^{tz})^2}],
\end{equation}
while $\beta^z = T^{tz} / (e + T^{zz})$ \cite{Ambrus:2017keg}. The particle number density $n = u_\mu N^\mu = \gamma(N^t - \beta^z N^z)$ can be obtained using
\begin{align}
 \begin{pmatrix}
  N^t \\ N^{z}
 \end{pmatrix} &= \int_{-1}^1 \d v^z 
 \begin{pmatrix}
     1 \\ v^z
 \end{pmatrix} F_1.
\end{align}

As in the case of the Bjorken flow, the vector and tensor off-equilibrium moments $\rho^\mu_r$ and $\rho^{\mu\nu}_r$ are each characterized by a single number, $V_r$ and $\pi_r$, which we define as follows:
\begin{align}
 \rho^\mu_r \partial_\mu &= V_r (\beta^z \partial_t + \partial_z),\nonumber\\
 \rho^{\mu\nu}_r &= \pi_r 
 \begin{pmatrix}
  \beta_z^2 \gamma^2 & 0 & 0 & \beta^z\gamma^2 \\ 
  0 & -\frac{1}{2} & 0 & 0 \\
  0 & 0 & -\frac{1}{2} & 0 \\
  \beta^z\gamma^2 & 0 & 0 & \gamma^2
 \end{pmatrix}.
 \label{eq:rho_munu_expl}
\end{align}

Defining the macroscopic quantities $N^\mu_r = \langle k^\mu E_\bk^r \rangle$ and $T^{\mu\nu}_r = \langle k^\mu k^\nu E_\bk^r \rangle$, we have 
\begin{align}
 \begin{pmatrix}
  N^t_r \\ N^{z}_r
 \end{pmatrix} &= \int_{-1}^1 \d v^z 
 \begin{pmatrix}
     1 \\ v^z
 \end{pmatrix} (u \cdot v)^r F_{r+1}\;, \nonumber\\
 \begin{pmatrix}
  T^{tt}_r \\  T^{tz}_r \\ T^{zz}_r 
 \end{pmatrix} &= \int_{-1}^1 \d v^z 
 \begin{pmatrix}
     1 \\ v^z \\ v_z^2
 \end{pmatrix} (u \cdot v)^r F_{r+2}\;,
\end{align}
where $v^\mu = k^\mu / k^0$ and $u \cdot v = \gamma (1 - \beta^z v^z)$.
Apart from the components $T_r^{ij} = \frac{1}{2} \delta^{ij} (T^{tt}_r -  T^{zz}_r - T^\mu_{r;\mu})$ with $i,j \in \{1, 2\}$ being transverse indices, all other transverse components $N^i_r$, $T^{t i}_r$ and $T^{iz}_r$ can be taken to vanish. The non-equilibrium quantities $V_r$ and $\pi_r$ can be obtained via 
\begin{align}
 V_r &= \Delta^z_\nu N^\nu_r = \gamma^2 \int_{-1}^1 dv^z (u \cdot v)^r F_r(v^z - \beta^z), \nonumber\\
 \pi_r &= \frac{1}{\gamma^2} \Delta^{zz}_{\mu\nu} T^{\mu\nu}_r 
 = \int_{-1}^1 dv^z (u \cdot v)^r F_r \nonumber\\
 &\qquad\qquad \times \left[\frac{2}{3} \gamma^2(v^z - \beta^z)^2 -\frac{1 - v_z^2}{3} \right]\;.
\end{align}

We now discuss the moments $F^{\rm S}_n$ of the Shakhov distribution function. Considering the generic split $f_{{\rm S}\bk} = f_{0\bk} + \delta f_{{\rm S}\bk}$, we have 
\begin{eqnarray}
 F^{\rm S}_n = F^{\rm eq}_n + \delta F^{\rm S}_n,
\end{eqnarray}
where $F_n^{\rm eq}$ reduces to
\begin{align}
 F_n^{\rm eq} &= \frac{g e^\alpha}{4\pi^2} \int_0^\infty \frac{\d m_\perp\, m_\perp^{n+1}}{(1 - v_z^2)^{(n+2)/2}} \exp\left(-\frac{\beta m_\perp (u \cdot v)}{\sqrt{1 - v_z^2}} \right) \nonumber\\
 &= \frac{I_{n0}}{2(u \cdot v)^{n+2}},
\end{align}
with $I_{n0} = g e^\alpha (n+1)! / 2\pi^2 \beta^{n+2}$, as shown in Eq.~\eqref{eq:Inq_m0}. Furthermore, considering that 
$\delta f_{{\rm S}\bk} = f_{0\mathbf{k}}\widetilde{f}_{0\mathbf{k}}\sum_{\ell = 0}^L \sum_{r = -s_\ell}^{N_\ell} \rho^{\mu_1 \cdots \mu_\ell}_{{\rm S};r} k_{\langle \mu_1} \cdots k_{\mu_\ell \rangle} E_\bk^{-s_\ell} \widetilde{\mathcal{H}}^{(\ell)}_{\bk,r + s_\ell}$, we have 
\begin{multline}
 \delta F^{\rm S}_n = \frac{1}{2} \sum_{\ell = 0}^L \frac{(2\ell + 1)!!}{(-1)^\ell \ell!} \\\times 
 \sum_{r = -s_\ell}^{N_\ell} \frac{\rho^{\mu_1 \cdots \mu_\ell}_{{\rm S};r} v_{\langle \mu_1} \cdots v_{\mu_\ell \rangle}}{(u \cdot v)^{\ell + n + 2}} \widetilde{\mathcal{F}}^{(\ell)}_{\ell - s_\ell - n, r + s_\ell}\;,
 \label{eq:delta_Fn_S}
\end{multline}
where $\widetilde{\mathcal{F}}^{(\ell)}_{rn}$ was introduced in Eq.~\eqref{eq:F_def} and was evaluated in Eq.~\eqref{eq:F_m0} for the massless case. 
In order to arrive at Eq. \eqref{eq:delta_Fn_S}, we inserted $k^t=m_\perp/\sqrt{1-v_z^2}$, performed the $\varphi_k$-integral and employed that in the massless limit
\begin{multline}
    \widetilde{\mathcal{F}}_{rn}^{(\ell)}=\frac{(-1)^\ell \ell!}{(2\ell+1)!!} \beta^{-2(\ell-s_\ell)-2 +r} \frac{g}{2\pi^2}\nonumber\\
    \times\int_0^\infty \mathrm{d}x x^{2(\ell-s_\ell)+1-r} f_0(x)\widetilde{f}_0(x) \widetilde{\mathcal{H}}_{\mathbf{k} n}^{(\ell)}(x)\;,
\end{multline}
where $x=\beta k^0$ and we evaluated the integral in the rest frame of the fluid.
Specializing the above to the case when $1 \le \ell \le 2$ gives
\begin{multline}
 F^{\rm S}_n = F^{\rm eq}_n - \frac{3(\beta^z - v^z)}{2(u \cdot v)^{n + 3}} 
 \sum_{r = -s_1}^{N_1} V_{{\rm S};r} \widetilde{\mathcal{F}}^{(1)}_{1 - s_1 - n, r + s_1} \\
 + \frac{15}{4(u \cdot v)^{n+4}} \left[\gamma^2(\beta^z - v^z)^2 - \frac{1 - v_z^2}{2} \right] \\\times 
 \sum_{r = -s_2}^{N_2} \pi_{{\rm S};r} \widetilde{\mathcal{F}}^{(2)}_{2 - s_2 - n, r + s_2}\;,
\end{multline}
where we employed Eq. \eqref{eq:rho_munu_expl} and used that $k^\mu k_\mu =0$.
The coefficients $V_{{\rm S};r}$ and $\pi_{{\rm S};r}$ corresponding to the Shakhov distribution can be obtained from the coefficients $V_r$ and $\pi_r$ corresponding to $f_\bk$ via 
\begin{align}
 V_{{\rm S};r} &= \sum_{n = -s_1}^{N_1} (\delta_{rn} - \tau_R \mathcal{A}^{(1)}_{{\rm S};rn}) V_n\;, \nonumber\\
 \pi_{{\rm S};r} &= \sum_{n = -s_2}^{N_2} (\delta_{rn} - \tau_R \mathcal{A}^{(2)}_{{\rm S};rn}) \pi_n\;,
\end{align}
cf. Eq. \eqref{eq:connection_rhoS_rho}.

\subsection{Space, time and momentum space discretization} \label{app:numsch:disc}

The numerical algorithms employed in this paper are identical to those described in the supplementary material of Ref.~\cite{Ambrus:2023ilm} and fall in the category of the discrete velocity methods (DVM) (see Refs.~\cite{Ambrus:2016aub,Ambrus:2021sjg,Ambrus:2021fej,Ambrus:2022adp,Ambrus:2023ilm,Ambrus:2023qcl} for previous developments of this code).
Due to the symmetries of the flows considered in this paper, the momentum space becomes one dimensional. The remaining $v^z$ degree of freedom is discretized according to the Gauss-Legendre prescription, such that the continuous functions $F_n(v^z)$ are replaced by
\begin{equation}
 F_{n;j} = w_j F_n(v^z_j), \quad 
 w_j = \frac{2(1 - v_{z;j}^2)}{[(K+1) P_{K+1}(v^z_j)]^2},
\end{equation}
where $v^z_j$ ($1 \le j \le K$) are the $K$ roots of the Legendre polynomial $P_K(z)$ and $w_j$ are the associated weights. The derivative with respect to $v^z$ appearing in Eq.~\eqref{eq:app_boltz_bjork} is replaced by the finite sum
\begin{equation}
 \left[\frac{\partial[v^z(1 - v_z^2) F_n]}{\partial v^z}\right]_j = \sum_{j' = 1}^K \mathcal{K}_{j,j'} F_{n;j'}, 
\end{equation}
where the kernel $\mathcal{K}_{j,j'}$ is obtained by projecting the left-hand side expression onto the space of Legendre polynomials \cite{Ambrus:2016aub}:
\begin{multline}
 \mathcal{K}_{j,j'} = w_j \sum_{m = 1}^{K - 3} \frac{m(m+1)(m+2)}{2(2m+3)} P_m(v^z_j) P_{m+2}(v^z_{j'}) \\
 - w_j \sum_{m = 1}^{K - 1} \frac{m(m+1)}{2} P_m(v^z_j) \left[\frac{(2m+1) P_m(v^z_{j'})}{(2m-1)(2m+3)} \right.\\
 \left. + \frac{m-1}{2m-1} P_{m-2}(v^z_{j'})\right].
\end{multline}

The time stepping is performed using the third-order total variation diminishing (TVD) Runge-Kutta algorithm of Refs.~\cite{Shu:1988,Gottlieb:1998}. Considering the equation $\partial_t f = L[t, f]$, the scheme advances $f_n \equiv f(t_n)$ from time $t_n$ to $f_{n+1} \equiv f(t_{n+1})$ at time $t_{n+1} = t_n + \delta t_n$ via two intermediate stages:
\begin{align}
 f_{n+1} &= \frac{1}{3} f_n + \frac{2}{3} f^{(2)}_n + \frac{2}{3} \delta t_n L\left[t_n + \frac{1}{2} \delta t_n, f^{(2)}_n\right] \;, \nonumber\\
 f^{(2)}_n &= \frac{3}{4} f_n + \frac{1}{4} f^{(1)}_n + \frac{1}{4} \delta t_n L[t_n + \delta t_n, f^{(1)}_n] \;,\nonumber\\
f^{(1)}_n &= f_n + \delta t_n L[t_n, f_n] \;.
\label{eq:RK}
\end{align}
In the case of Bjorken flow, the time step is taken adaptively, $\delta \tau_n = {\rm min} (10^{-3} \tau_n, \tau_R / 2)$, while in the case of the longitudinal waves and Riemann problems, we employ equal time steps.

Finally, the spatial advection for the $1+1$-D problems is implemented using a flux-based scheme. The domain of length $L$ is discretized using $S$ equal intervals of size $\delta z = L / S$. The spatial nodes are taken as the centers of these cells, having coordinates $z_s =(s - \frac{1}{2}) \delta z - \frac{L}{2}$ ($1 \le s \le S)$. The spatial derivative at point $z_s$ is calculated using
\begin{equation}
 \left(v^z \frac{\partial F_n}{\partial z}\right)_s = \frac{\mathbf{F}_{s+1/2} - \mathbf{F}_{s-1/2}}{\delta z},
\end{equation}
where the flux $\mathbf{F}_{s + 1/2}$ at the interface between cells $s$ and $s - 1$ is computed using the upwind-biased WENO-5 (fifth-order weighted essentially non-oscillatory) scheme of Refs.~\cite{Jiang:1996,Rezzolla.2013}.

\subsection{Note on code availability} \label{app:numsch:code}

The numerical code, raw data, and scripts
to generate the plots shown in this paper are available on Code
Ocean \cite{codeoceanshk2}. The code consists of two separate programs, one for the Bjorken flow and the other for the Riemann problem, extending the code in Ref.~\cite{Ambrus:2023ilm} (the original version for the Bjorken flow code was introduced in Ref.~\cite{Ambrus:2023qcl}). 
We remark that the evaluation of the modified Bessel functions $K_n(z)$ and of the Bickley function $\mathrm{Ki}_1(z)$, required for the Bjorken flow problem, is performed using the algorithms designed by D.E. Amos in Refs.~\cite{Amos609,Amos644, Amos644comment1, Amos644comment2}, openly available through the OpenSpecfun project.\footnote{Source files downloaded from
\texttt{https://github.com/JuliaMath/openspecfun}, commit number
\texttt{70239b8d1fe351042ad3321e33ae97923967f7b9}.}

\section{Shakhov matrices for ultarelativistic hard spheres}\label{app:shk_higher}

In this appendix, we illustrate the procedure for the implementation of the ($1012$) and ($2012$) high-order Shakhov matrices, tuned to recover the transport coefficients of a classical gas of hard-sphere particles, interacting via the constant cross section $\sigma$, shown on the first lines of Tables~\ref{tbl:riem_tcoeffs} and \ref{tbl:riem_tcoeffs_DNMR}. In sections~\ref{sec:1012} and \ref{sec:2012}, we focus on recovering the transport coefficients obtained in the IReD (inverse-Reynolds dominance) approach \cite{Wagner:2022ayd,Wagner:2023joq}, summarized in Table~\ref{tbl:riem_tcoeffs}. Sections~\ref{sec:1012_DNMR} and \ref{sec:2012_DNMR} present the collision matrices recovering the values of the transport coefficients calculated using the standard DNMR approach \cite{Denicol:2012cn,Wagner:2023joq}.

\subsection{IReD (\texorpdfstring{$\mathbf{N_1}$, $\mathbf{N_2}$, $\mathbf{s_1}$, $\mathbf{s_2}$}{N\textoneinferior, N\texttwoinferior, s\textoneinferior, s\texttwoinferior})=(1012) model}
\label{sec:1012}
In this truncation, the basis is shifted downwards as far as possible in the ultrarelativistic case ($s_1 = 1$ and $s_2 = 2$). 
The interesting submatrices and their inverses for $\ell=1$ and $\ell=2$ are given by
\begin{eqnarray}
    \mathcal{A}_{{\rm S};rn}^{(1)}&=&\begin{pmatrix}
     \mathcal{A}_{{\rm S};-1,-1}^{(1)} & \mathcal{A}_{{\rm S};-1,0}^{(1)}\\
     \mathcal{A}_{{\rm S};0,-1}^{(1)} & \boldsymbol{\mathcal{A}}_{\mathbf{\text{\bf S};0,0}}^{\mathbf{(1)}}
    \end{pmatrix}\;,\nonumber\\
    \tau_{{\rm S};rn}^{(1)}&=&\begin{pmatrix}
     \tau_{{\rm S};-1,-1}^{(1)} & \tau_{{\rm S};-1,0}^{(1)}\\
     \tau_{{\rm S};0,-1}^{(1)} & \boldsymbol{\tau}_{\mathbf{\text{\bf S};0,0}}^{\mathbf{(1)}}
    \end{pmatrix}\;,
\end{eqnarray}
and 
\begin{eqnarray}
    \mathcal{A}_{{\rm S};rn}^{(2)}&=&\begin{pmatrix}
     \mathcal{A}_{{\rm S};-2,-2}^{(2)} & \mathcal{A}_{{\rm S};-2,-1}^{(2)} & \mathcal{A}_{{\rm S};-2,0}^{(2)}\\
     \mathcal{A}_{{\rm S};-1,-2}^{(2)} & \mathcal{A}_{{\rm S};-1,-1}^{(2)} & \mathcal{A}_{{\rm S};-1,0}^{(2)}\\
     \mathcal{A}_{{\rm S};0,-2}^{(2)} & \mathcal{A}_{{\rm S};0,-1}^{(2)} & \boldsymbol{\mathcal{A}}_{\mathbf{\text{\bf S};0,0}}^{\mathbf{(2)}}
    \end{pmatrix}\;,\nonumber\\
    \tau_{{\rm S};rn}^{(2)}&=&\begin{pmatrix}
     \tau_{{\rm S};-2,-2}^{(2)} & \tau_{{\rm S};-2,-1}^{(2)} & \tau_{{\rm S};-2,0}^{(2)}\\
     \tau_{{\rm S};-1,-2}^{(2)} & \tau_{{\rm S};-1,-1}^{(2)} & \tau_{{\rm S};-1,0}^{(2)}\\
     \tau_{{\rm S};0,-2}^{(2)} & \tau_{{\rm S};0,-1}^{(2)} & \boldsymbol{\tau}_{\mathbf{\text{\bf S};0,0}}^{\mathbf{(1)}}
    \end{pmatrix}\;,
\end{eqnarray}
respectively. In the above and in what follows, we employed bold font to highlight the $(0,0)$ entry in each matrix.
The first-order transport coefficients read in this case
\begin{align}
    \kappa_{0}&=\frac{P\beta}{12}\left( \tau_{{\rm S};0,-1}^{(1)} \beta+\tau_{{\rm S};0,0}^{(1)}\right)\;,\nonumber\\
    \eta_{0}&=\frac{P}{15}\left( \tau_{{\rm S};0,-2}^{(2)}\beta^2 +3\tau_{{\rm S};0,-1}^{(2)}\beta +12 \tau_{{\rm S};0,0}^{(2)}  \right)\;.
\end{align}
The relaxation times are given by
\begin{eqnarray}
    \tau_V &=& \tau_{{\rm S};0,-1}^{(1)} \mathcal{C}_{-1}^{(1)}+\tau_{{\rm S};0,0}^{(1)}\;,\nonumber\\
    \tau_\pi &=& \tau_{{\rm S};0,-2}^{(2)}\mathcal{C}_{-2}^{(2)}+\tau_{{\rm S};0,-1}^{(2)}\mathcal{C}_{-1}^{(2)}+\tau_{{\rm S};0,0}^{(2)}\;,
\end{eqnarray}
while the second-order transport coefficients for the particle diffusion read
\begin{multline}
    \delta_{VV}=\tau_V \;,\quad \lambda_{VV}=\frac{1}{5}\left( \mathcal{C}_{-1}^{(1)}\tau_{{\rm S};0,-1}^{(1)}+3\tau_{{\rm S};0,0}^{(1)} \right)\;,\\
    \ell_{V\pi}= \tau_{V\pi}= \left[ \tau_{{\rm S};0,-1}^{(1)} \left(\frac{\beta^2}{12}-\mathcal{C}_{-2}^{(2)}  \right)+\tau_{{\rm S};0,0}^{(1)} \left(\frac{\beta}{4}-\mathcal{C}_{-1}^{(2)}  \right) \right]\;,\\
    \lambda_{V\pi}=\frac14 \left(2\tau_{{\rm S};0,-1}^{(1)}\mathcal{C}_{-2}^{(2)}+\tau_{{\rm S};0,0}^{(1)}\mathcal{C}_{-1}^{(2)}  \right)\;.
\end{multline}
Lastly, the second-order transport coefficients for the shear-stress tensor are given by
\begin{align}
    \delta_{\pi\pi} &= \frac43 \tau_\pi\;, \quad  \lambda_{\pi V}= \frac{1}{10} \tau^{(2)}_{{\rm S};0,-2} \mathcal{C}^{(1)}_{-1}  \;, \nonumber\\
    \tau_{\pi\pi} &= \frac27\left[\mathcal{C}_{-2}^{(2)}\tau_{{\rm S};0,-2}^{(2)}+3\mathcal{C}_{-1}^{(2)} \tau_{{\rm S};0,-1}^{(2)}+5\tau_{{\rm S};0,0}^{(2)}\right]\;,\nonumber\\
    \ell_{\pi V} &= \frac14\tau_{\pi V} = \frac25 \left(\mathcal{C}_{-1}^{(1)}\tau_{{\rm S};0,-2}^{(2)}+\tau_{{\rm S};0,-1}^{(2)}\right)\;.
\end{align}

Let us briefly discuss how a matching procedure may look like in this case. Firstly, we may fix $\tau_{{\rm S};0,0}^{(1)}$ and $\tau_{{\rm S};0,0}^{(2)}$ via $\kappa_0$ and $\eta_0$, respectively. 
Then, we fix $\tau_{{\rm S};0,-1}^{(1)}$ and $\tau_{{\rm S};0,-1}^{(2)}$ via $\tau_V$ and $\tau_\pi$.
The last remaining coefficient in the zeroth row of $\tau^{(2)}_{\rm S}$, i.e., $\tau_{{\rm S};0,-2}^{(2)}$, is fixed by means of $\ell_{\pi V}$, after which $\mathcal{C}_{-1}^{(2)}$ is expressed via $\tau_{\pi\pi}$.
Having used all nonvanishing second-order coefficients of the shear-stress equation, we turn to the diffusion current and fix $\mathcal{C}_{-2}^{(2)}$ through $\ell_{V \pi}$. 
By now we have left only $\mathcal{C}_{-1}^{(1)}$ as a degree of freedom, but three free coefficients, namely $\lambda_{V\pi}$, $\lambda_{VV}$ and $\lambda_{\pi V}$. Thus, we have to choose one of these coefficients to fix $\mathcal{C}_{-1}^{(1)}$ and compare the results for the other two with the reference values from Ref.~\cite{Wagner:2023joq}, where they have been computed to high precision recently.

Putting in the converged values from Ref. \cite{Wagner:2023joq} (considering the ``IReD''-values, which are based on the approach presented in Ref. \cite{Wagner:2022ayd}) and fixing $\lambda_{\pi V}$, we follow the procedure outlined above numerically to obtain the values of $\tau^{(1)}_{\rm S}$ and $\tau^{(2)}_{\rm S}$. Note that, since only the values of $\mathcal{C}^{(2)}_{-2}$, $\mathcal{C}^{(2)}_{-1}$ and $\mathcal{C}^{(1)}_{-1}$ are fixed, we may set $\tau^{(2)}_{{\rm S};-2,-1}=\tau^{(2)}_{{\rm S};-2,0}=\tau^{(2)}_{{\rm S};-1,-2}=\tau^{(2)}_{{\rm S};-1,0}=\tau^{(1)}_{{\rm S};-1,0}=0$.
Furthermore, since there are two solutions for the matrices, we may classify them by the agreement between the (unfixed) values for $\lambda_{VV}, \,\lambda_{V\pi}$ and the reference values $\lambda_{VV}=1.8725/(P\beta \sigma), \, \lambda_{V\pi}=0.14435/(P\sigma)$ \cite{Wagner:2023joq}. 
The two solutions yield
\begin{align}
    \lambda_{VV}^{(i)}=2.3185/(P\beta \sigma)\;,\quad \lambda_{V\pi}^{(i)}=0.16508/(P \sigma)\;,\nonumber\\
    \lambda_{VV}^{(ii)}=1.5879/(P\beta \sigma)\;,\quad \lambda_{V\pi}^{(ii)}=0.13114/(P \sigma)\;,
\end{align}
where the second solution fits slightly better. Choosing this solution, we obtain 
\begin{align}
    \tau^{(1)}_{\rm S} =& \lambda_{\rm mfp} \begin{pmatrix}
     1.5768 & 0\\
     -1.0208/\beta & \mathbf{2.9279}
    \end{pmatrix}\;,\nonumber\\
    \tau^{(2)}_{\rm S} =& \lambda_{\rm mfp} \begin{pmatrix}
     1.1775 & 0 & 0\\
     0 & 1.4092 & 0\\
     4.1056/\beta^2 & -5.7524/\beta & \mathbf{2.6805}
    \end{pmatrix}\;,
\end{align}
where $\lambda_{\rm mfp} = 1 / P \beta \sigma$ and we have represented in bold font the $(0,0)$ component of each matrix.
The actual matrices needed for the Shakhov term are then
\begin{align}
 \mathcal{A}^{(1)}_{\rm S} =& \frac{1}{\lambda_{\rm mfp}}
 \begin{pmatrix}
  0.63419 & 0 \\ 0.22111\beta^{-1} & \mathbf{0.34155}
 \end{pmatrix},\nonumber\\
 \mathcal{A}^{(2)}_{\rm S} =& \frac{1}{\lambda_{\rm mfp}}
 \begin{pmatrix}
  0.84927 & 0 & 0 \\
  0 & 0.70961 & 0 \\
  -1.3008 \beta^{-2} & 1.5229 \beta^{-1} & \mathbf{0.37307}
 \end{pmatrix}\;.\label{eq:1012_IReD}
\end{align}

\subsection{IReD (\texorpdfstring{$\mathbf{N_1}$, $\mathbf{N_2}$, $\mathbf{s_1}$, $\mathbf{s_2}$}{N\textoneinferior, N\texttwoinferior, s\textoneinferior, s\texttwoinferior})=(2012) model}
\label{sec:2012}

We now consider the truncation where the basis is shifted downwards as far as possible and an additional vector moment is included, i.e., the submatrices are given by
\begin{eqnarray}
    \mathcal{A}_{{\rm S};rn}^{(1)}&=&\begin{pmatrix}
     \mathcal{A}_{{\rm S};-1,-1}^{(1)} & \mathcal{A}_{{\rm S};-1,0}^{(1)} & \mathcal{A}_{{\rm S};-1,2}^{(1)}\\
     \mathcal{A}_{{\rm S};0,-1}^{(1)} & \boldsymbol{\mathcal{A}}_{\mathbf{\text{\bf S};0,0}}^{\mathbf{(1)}} & \mathcal{A}_{{\rm S};0,2}^{(1)} \\
     \mathcal{A}_{{\rm S};2,-1}^{(1)} & \mathcal{A}_{{\rm S};2,0}^{(1)} & \mathcal{A}_{{\rm S};2,2}^{(1)}
    \end{pmatrix}\;,\nonumber\\
    \tau_{{\rm S};rn}^{(1)}&=&\begin{pmatrix}
     \tau_{{\rm S};-1,-1}^{(1)} & \tau_{{\rm S};-1,0}^{(1)} &
     \tau_{{\rm S};-1,2}^{(1)}\\
     \tau_{{\rm S};0,-1}^{(1)} & \boldsymbol{\tau}_{\mathbf{\text{\bf S};0,0}}^{\mathbf{(1)}} &
     \tau_{{\rm S};0,2}^{(1)} \\
     \tau_{{\rm S};2,-1}^{(1)} & \tau_{{\rm S};2,0}^{(1)} &
     \tau_{{\rm S};2,2}^{(1)}
    \end{pmatrix}\;,
\end{eqnarray}
and 
\begin{eqnarray}
    \mathcal{A}_{{\rm S};rn}^{(2)}&=&\begin{pmatrix}
     \mathcal{A}_{{\rm S};-2,-2}^{(2)} & \mathcal{A}_{{\rm S};-2,-1}^{(2)} & \mathcal{A}_{{\rm S};-2,0}^{(2)}\\
     \mathcal{A}_{{\rm S};-1,-2}^{(2)} & \mathcal{A}_{{\rm S};-1,-1}^{(2)} & \mathcal{A}_{{\rm S};-1,0}^{(2)}\\
     \mathcal{A}_{{\rm S};0,-2}^{(2)} & \mathcal{A}_{{\rm S};0,-1}^{(2)} & \boldsymbol{\mathcal{A}}_{\mathbf{\text{\bf S};0,0}}^{\mathbf{(2)}}
    \end{pmatrix}\;,\nonumber\\
    \tau_{{\rm S};rn}^{(2)}&=&\begin{pmatrix}
     \tau_{{\rm S};-2,-2}^{(2)} & \tau_{{\rm S};-2,-1}^{(2)} & \tau_{{\rm S};-2,0}^{(2)}\\
     \tau_{{\rm S};-1,-2}^{(2)} & \tau_{{\rm S};-1,-1}^{(2)} & \tau_{{\rm S};-1,0}^{(2)}\\
     \tau_{{\rm S};0,-2}^{(2)} & \tau_{{\rm S};0,-1}^{(2)} & \boldsymbol{\tau}_{\mathbf{\text{\bf S};0,0}}^{\mathbf{(2)}}
    \end{pmatrix}\;.
\end{eqnarray}

The first-order transport coefficients are given by
\begin{align}
    \kappa_0&= \frac{P\beta}{12}\left(\beta \tau^{(1)}_{{\rm S};0,-1} +\tau_{{\rm S};0,0}^{(1)}-12\beta^{-2}\tau_{{\rm S};0,2}^{(1)}  \right)\;,\nonumber\\
    \eta_{0}&= \frac{P\beta}{15}\left( \beta \tau_{{\rm S};0,-2}^{(2)}+3\tau_{{\rm S};0,-1}^{(2)}+12\beta^{-1}\tau_{{\rm S};0,0}^{(2)}  \right)\;,
\end{align}
while the relaxation times read
\begin{eqnarray}
    \tau_V &=&\tau_{{\rm S};0,-1}^{(1)}\mathcal{C}_{-1}^{(1)}+ \tau_{{\rm S};0,0}^{(1)} +\tau_{{\rm S};02}^{(1)}\mathcal{C}_2^{(1)}\;,\nonumber\\
    \tau_\pi &=& \tau_{{\rm S};0,-2}^{(2)} \mathcal{C}_{-2}^{(2)}+\tau_{{\rm S};0,-1}^{(2)}\mathcal{C}_{-1}^{(2)}+\tau_{{\rm S};0,0}^{(2)}\;.
\end{eqnarray}
The remaining second-order transport coefficients for the diffusion current are
\begin{align}
    \delta_{VV}&=\tau_V ,\,\lambda_{VV}=\frac{1}{5}\left(\mathcal{C}_{-1}^{(1)}\tau_{{\rm S};0,-1}^{(1)}+ 3\tau_{{\rm S};0,0}^{(1)}+7\mathcal{C}_2^{(1)}\tau_{{\rm S};0,2}^{(1)} \right)\,,\nonumber\\
    \ell_{V\pi} &= \tau_{V\pi}= \left[ \tau_{{\rm S};0,-1}^{(1)} \left(\frac{\beta^2}{12}-\mathcal{C}_{-2}^{(2)}  \right)\right.\nonumber\\
    &\left.+\tau_{{\rm S};0,0}^{(1)} \left(\frac{\beta}{4}-\mathcal{C}_{-1}^{(2)}  \right)+\tau_{{\rm S};0,2}^{(1)} \left(\frac{5}{\beta}-\mathcal{C}_{1}^{(2)}  \right) \right]\;,\nonumber\\
    \lambda_{V\pi}&=\frac{1}{4} \left(2\tau_{{\rm S};0,-1}^{(1)}\mathcal{C}_{-2}^{(2)}+\tau_{{\rm S};0,0}^{(1)}\mathcal{C}_{-1}^{(2)}-\tau_{{\rm S};0,2}^{(1)}\mathcal{C}_{1}^{(2)}  \right)\;,
\end{align}
and the second-order coefficients for the shear-stress tensor read
\begin{align}
    \delta_{\pi\pi} &= \frac43 \tau_\pi\;, \quad  \lambda_{\pi V}= \frac{1}{10} \tau^{(2)}_{{\rm S};0,-2} \mathcal{C}^{(1)}_{-1}  \;,\nonumber\\
    \tau_{\pi\pi} &= \frac27\left[\mathcal{C}_{-2}^{(2)} \tau_{{\rm S};0,-2}^{(2)}+3\mathcal{C}_{-1}^{(2)} \tau_{{\rm S};0,-1}^{(2)}+5\tau_{{\rm S};0,0}^{(2)}\right]\;,\nonumber\\
    \ell_{\pi V} &= \frac14\tau_{\pi V}= \frac25 \left(\mathcal{C}_{-1}^{(1)}\tau_{{\rm S};0,-2}^{(2)}+\tau_{{\rm S};0,-1}^{(2)}\right)\;.
\end{align}
Note that we gain two additional parameters compared to the last section, such that we can fix all free quantities to the hydrodynamic second-order transport coefficients.
While $\mathcal{C}^{(1)}_{-1}$, $\mathcal{C}^{(1)}_2$, $\mathcal{C}^{(2)}_{-1}$, and $\mathcal{C}^{(2)}_{-2}$ represent degrees of freedom of the model, the coefficient $\mathcal{C}^{(2)}_1 = \eta_1 / \eta_0$ corresponds to a transport coefficient that lies outside the basis. Since our tensor basis is shifted, we can use Eq.~\eqref{eq:UR_etar_shift} to replace
\begin{equation}
 \mathcal{C}^{(2)}_1 = \sum_{n = -s_2 = -2}^{N_2 = 0} \widetilde{\mathcal{F}}^{(2)}_{-3, n+2} \mathcal{C}^{(2)}_n = \frac{24}{\beta^3} \mathcal{C}^{(2)}_{-2} - \frac{36}{\beta^2} \mathcal{C}^{(2)}_{-1} + \frac{12}{\beta},
\end{equation}
where we employed $\widetilde{\mathcal{F}}^{(2)}_{-3,0} = 24 / \beta^3$, $\widetilde{\mathcal{F}}^{(2)}_{-3,1} = -36 / \beta^2$, and $\widetilde{\mathcal{F}}^{(2)}_{-3,2} = 12 / \beta$.

The set of solutions for the inverse matrices reads in this case
\begin{align}
    \tau_{\rm S}^{(1)} &= \lambda_{\rm mfp} \begin{pmatrix}
     1.5941 & 0 &0 \\
     -0.29743\beta & \mathbf{1.6789} & -0.043798 \beta^2 \\
     0& 0& 2.3713
    \end{pmatrix} \;,\nonumber\\
    \tau_{\rm S}^{(2)} &= \lambda_{\rm mfp} \begin{pmatrix}
     1.2077 & 0& 0\\
     0 & 1.4174 & 0\\
     4.0612/\beta^2 & -5.7524/\beta & \mathbf{2.6842}
    \end{pmatrix} \;,
\end{align}
whereas the actual matrices are given by
\begin{align}
    \mathcal{A}_{\rm S}^{(1)} &= \frac{1}{\lambda_{\rm mfp}} \begin{pmatrix}
     0.62732 & 0 &0 \\
     0.11113/\beta & \mathbf{0.59563} & 0.011012 \beta^2 \\
     0& 0& 0.42171
    \end{pmatrix} \;,\nonumber\\
    \mathcal{A}_{\rm S}^{(2)} &= \frac{1}{\lambda_{\rm mfp}} \begin{pmatrix}
     0.82802 & 0& 0\\
     0. & 0.70553 & 0\\
     -1.2528/\beta^2 & 1.5120/\beta & \mathbf{0.37256}
    \end{pmatrix} \;.
    \label{eq:2012_IReD}
\end{align}
In this case, only one viable solution with real entries exists.

\subsection{DNMR (\texorpdfstring{$\mathbf{N_1}$, $\mathbf{N_2}$, $\mathbf{s_1}$, $\mathbf{s_2}$}{N\textoneinferior, N\texttwoinferior, s\textoneinferior, s\texttwoinferior})=(1012) model}
\label{sec:1012_DNMR}

Taking exactly the same steps as in Sec.~\ref{sec:1012}, we find two solutions which yield for the unfixed parameters $\lambda_{VV}$ and $\lambda_{V\pi}$
\begin{align}
    \lambda_{VV}^{(i)}=2.9200/(P\beta \sigma)\;,\qquad \lambda_{V\pi}^{(i)}=&0.16811/(P\sigma)\;,\nonumber\\
    \lambda_{VV}^{(ii)}=1.9805/(P\beta \sigma)\;,\qquad \lambda_{V\pi}^{(ii)}=&0.11078/(P\sigma)\;,
\end{align}
which we compare to the true values of the DNMR theory: $\lambda_{VV}=2.3688/(P\beta \sigma)$ and $\lambda_{V\pi}=0.13300/(P\sigma)$. Since the second solution fits better, we obtain the Shakhov matrices as
\begin{align}
 \mathcal{A}^{(1)}_{\rm S} &= \frac{1}{\lambda_{\rm mfp}}
 \begin{pmatrix}
  0.84344 & 0 \\ 0.40458 \beta^{-1} & \mathbf{0.27285}
 \end{pmatrix},\nonumber\\
 \mathcal{A}^{(2)}_{\rm S} &= \frac{1}{\lambda_{\rm mfp}}
 \begin{pmatrix}
  4.6173 & 0 & 0 \\
  0 & 1.1968 & 0 \\
  -11.978 \beta^{-2} & 3.2933 \beta^{-1} & \mathbf{0.33340}
 \end{pmatrix}\;.
 \label{eq:1012_DNMR}
\end{align}

\subsection{DNMR (\texorpdfstring{$\mathbf{N_1}$, $\mathbf{N_2}$, $\mathbf{s_1}$, $\mathbf{s_2}$}{N\textoneinferior, N\texttwoinferior, s\textoneinferior, s\texttwoinferior})=(2012) model}
\label{sec:2012_DNMR}

In order to derive the $(2012)$ model that recovers all of the DNMR transport coefficients, we follow the same steps as in Sec.~\ref{sec:2012}. 
In contrast to the IReD case, in the DNMR one we find two real solutions, which are given by
\begin{align}
    \mathcal{A}_{\rm S}^{(1),i} &= \frac{1}{\lambda_{\rm mfp}} \begin{pmatrix}
     0.86786 & 0 &0 \\
     0.26578/\beta & \mathbf{0.46905} & 0.0044985 \beta^2 \\
     0& 0& 0.26889
    \end{pmatrix} \;,\nonumber\\
    \mathcal{A}_{\rm S}^{(2),i} &= \frac{1}{\lambda_{\rm mfp}} \begin{pmatrix}
     6.0275 & 0& 0\\
     0. & 1.2396 & 0\\
     -16.191/\beta^2 & 3.4324/\beta & \mathbf{0.33550}
    \end{pmatrix} 
    \label{eq:2012i_DNMR}
\end{align}
and
\begin{align}
    \mathcal{A}_{\rm S}^{(1),ii} &= \frac{1}{\lambda_{\rm mfp}} \begin{pmatrix}
     0.66178 & 0 &0 \\
     0.44072/\beta & \mathbf{0.13535} & -0.0020283 \beta^2 \\
     0& 0& 0.32059
    \end{pmatrix} \;,\nonumber\\
    \mathcal{A}_{\rm S}^{(2),ii} &= \frac{1}{\lambda_{\rm mfp}} \begin{pmatrix}
     1.4065 & 0& 0\\
     0. & 0.95234 & 0\\
     -2.7355/\beta^2 & 2.5039/\beta & \mathbf{0.31856}
    \end{pmatrix} \;,
    \label{eq:2012ii_DNMR}
\end{align}
respectively.

\bibliography{bib_Shakhov}

\end{document}